\documentclass{aa}
\usepackage{graphicx}
\usepackage{txfonts}
\usepackage{natbib}
\usepackage{hyperref}
\hypersetup{colorlinks=true, citecolor=blue}
\usepackage{graphicx}
\usepackage{rotating}
\usepackage{listings}
\usepackage[usenames,dvipsnames]{xcolor}
\usepackage{placeins}
\bibpunct{(}{)}{;}{a}{}{,}

\begin{document}

\title{\texttt{CIGALE}: a python Code Investigating GALaxy Emission}
\author{M. Boquien\inst{1} \and D. Burgarella\inst{2} \and Y. Roehlly\inst{2} \and V. Buat\inst{2} \and L. Ciesla\inst{2} \and D. Corre\inst{2} \and A. K. Inoue\inst{3} \and H. Salas\inst{1}}
\institute{Centro de Astronom\'ia (CITEVA), Universidad de Antofagasta, Avenida Angamos 601, Antofagasta, Chile
  \and Aix Marseille Universit\'e, CNRS, LAM (Laboratoire d'Astrophysique de Marseille) UMR 7326, 13388, Marseille, France
  \and Department of Environmental Science and Technology, Faculty of Design Technology, College of General Education, Osaka Sangyo University, 3-1-1 Nakagaito, Daito, Osaka 574-8530, Japan}
\date{}
\abstract
  {Measuring how the physical properties of galaxies change across cosmic times is essential to understand galaxy formation and evolution. With the advent of numerous ground--based and space--borne instruments launched over the past few decades we now have exquisite multi--wavelength observations of galaxies from the far--ultraviolet (FUV) to the radio domain. To tap into this mine of data and obtain new insight into the formation and evolution of galaxies, it is essential that we are able to extract information from their spectral energy distribution (SED).}
  {We present a completely new implementation of Code Investigating GALaxy Emission (\texttt{CIGALE}). Written in \texttt{python}, its main aims are to easily and efficiently model the FUV to radio spectrum of galaxies and estimate their physical properties such as star formation rate, attenuation, dust luminosity, stellar mass, and many other physical quantities.}
  {To compute the spectral models, \texttt{CIGALE} builds composite stellar populations from simple stellar populations combined with highly flexible star formation histories, calculates the emission from gas ionised by massive stars, and attenuates both the stars and the ionised gas with a highly flexible attenuation curve. Based on an energy balance principle, the absorbed energy is then re--emitted by the dust in the mid-- and far--infrared domains while thermal and non--thermal components are also included, extending the spectrum far into the radio range. A large grid of models is then fitted to the data and the physical properties are estimated through the analysis of the likelihood distribution.}
  {\texttt{CIGALE} is a versatile and easy--to--use tool that makes full use of the architecture of multi--core computers, building grids of millions of models and analysing samples of thousands of galaxies, both at high speed. Beyond fitting the SEDs of galaxies and parameter estimations, it can also be used as a model-generation tool or serve as a library to build new applications.}
  {}
\keywords{methods: data analysis, methods: numerical, methods: statistical, galaxies: general}

\maketitle

\section{Introduction\label{sec:introduction}}

The multi--wavelength emission of galaxies from $\gamma$--rays to the radio domain is the outcome of the complex physical interplay between their main baryonic components: stars of all ages and their remnants; molecular, atomic, and ionised gas; dust; and supermassive black holes. This means that the spectral energy distribution (SED) of a galaxy contains the imprint of the baryonic processes that drove its formation and evolution along cosmic times. In other words, to understand galaxy formation and evolution we need to extract the information tightly woven into the SED of galaxies across a broad range of redshifts.

Over the past decade, major efforts have been undertaken to develop and strengthen two of the main pillars upon which rest our studies of galaxy formation and evolution: panchromatic observations and panchromatic models. On the observational side, large multi--wavelength surveys of galaxies have been carried out to measure the SED of galaxies across space and time, yielding a treasure trove of data that provide us with outstanding insight across the different baryonic components of galaxies. In turn, to interpret these observations and measure the fundamental physical properties of galaxies (e.g. star formation rate (SFR) and history (SFH), stellar mass, attenuation, dust mass and properties, presence and characteristics of an active nucleus, etc.), important investments have been made towards creating ever more precise and accurate models of the emission of galaxies over multiple orders of magnitude in wavelength.

Modelling the SED of galaxies is a heavily intricate problem. Galaxies with very different properties can have broadly similar SEDs. This is particularly the case when considering restricted wavelength ranges rather than the full SED, which is seldomly available. Therefore, estimating the physical properties of galaxies precisely and accurately with only limited data is a considerable challenge. In practice, different avenues can be taken to build physically motivated SED models and attempt to determine their intrinsic physical properties.

A popular approach consists in modelling galaxies using simple dust--attenuated templates representative of the diversity of galaxies at different redshifts. Such an approach is generally adopted by photometric redshift codes that fit only the FUV (far--ultraviolet) to NIR (near--infrared) part of the SED. Although this method is fast and works remarkably well for determining redshifts, as long as spectral breaks are sampled, it shows important limits when it comes to estimating the physical properties of galaxies beyond the stellar mass. In particular it can suffer heavily from degeneracies between the age and the metallicity \citep[e.g.][]{worthey1994a} or between the age and the attenuation \citep[e.g.][]{papovich2001a}: a galaxy can appear red either because it is strongly attenuated, because it does not form stars anymore, or because it has a high metallicity.

A more accurate but much more demanding approach in terms of computational resources is to solve the radiative transfer equation of the emission of stellar populations through a dusty gaseous medium with an arbitrary geometry. While this allows for an exquisitely detailed modelling, the required computing time can be extremely large and the effort required to construct large grids of models rapidly becomes prohibitively expensive even on relatively small samples of galaxies. So far this constraint has largely confined radiative transfer models to theoretical studies \citep[e.g.][]{gordon2001a,tuffs2004a,trayford2017a} and to only a handful of in--depth observational case studies, generally on edge-on galaxies \citep[e.g.][]{xilouris1999a,popescu2000a,bianchi2008a,delooze2012b}, with the modelling of face-on galaxies being a fairly recent development \citep[e.g.][]{delooze2014a, viaene2017a}.

An increasingly popular compromise in terms of speed, precision, and accuracy is to rely on an energy balance principle: the energy emitted by dust in the mid-- and far--IR exactly corresponds to the energy absorbed by dust in the UV--optical range. Such a method has been adopted by modern SED modelling codes such as \texttt{CIGALE} \citep{burgarella2005a,noll2009a}, \texttt{MAGPHYS} \citep{dacunha2008a}, and \texttt{FSPS} \citep{conroy2009a,conroy2010b} for instance. Such codes are very versatile and have been applied to study a wide variety of issues: why quiescent galaxies do not follow the starburst IRX--$\beta$ relation \citep{boquien2012a}, the attenuation properties of galaxies \citep[][Buat et al. (in press), Decleir et al. (submitted)]{buat2011b,buat2012a,boquien2013a,lofaro2017a,salim2018a}, SFR estimators \citep{buat2014a,boquien2014a,boquien2016a}, the separation of the emission of active galactic nuclei (AGNs) from their host galaxy \citep{ciesla2015a}, the imprint of the environment on the SED of galaxies \citep{bitsakis2016a,ciesla2016a}, or more generally the properties of nearby and distant galaxies \citep[e.g.][Małek et al. (in press), Burgarella et al. 2018 (submitted)]{burgarella2011a,giovannoli2011a,johnston2015a,malek2014a,alvarez2016a,pappalardo2016a,hirashita2017a,vika2017a}, to cite but a few of the studies carried out with \texttt{CIGALE}. If this approach is so successful, this is largely owing to the efficiency of the method, which allows to obtain good results, breaking the aforementioned degeneracies with the help of dust emission, while doing so rapidly with relatively modest computing requirements.

A key aspect of many modern models is their use of a Bayesian--like approach. The physical properties are then not evaluated from the best--fit model but by weighting all the models depending on their goodness--of--fit, with the best--fit models having the heaviest weight. This naturally takes into account the uncertainties on the observations while also including the effect of intrinsic degeneracies between physical parameters (different models, sometimes with widely different physical parameters, can yield very similar SEDs over some wavelength ranges, making it difficult to favour one model in particular). By doing so we are able to not only convincingly reproduce the observations but also to obtain more reliable estimates of the physical properties and their related uncertainties.

With ever larger and deeper surveys spanning ever broader wavelength ranges, it is especially important that we have equally more efficient, reliable, and versatile tools to model galaxies and estimate their physical properties. We present in this paper the new \texttt{python} version of \texttt{CIGALE}. While it shares the name, the ``energy balance'' principle, and the Bayesian--like strategy of the original \texttt{FORTRAN} implementation presented in \cite{noll2009a}, it is a completely new code that benefits from years of experience developing, maintaining, and using the original \texttt{CIGALE} \texttt{FORTRAN}, while addressing some of the new challenges and usages that have surfaced over the last few years. The aim of this article is to present the new architecture of \texttt{CIGALE}, its different modules, and various examples of its application. For conciseness, we do not dwell on the more theoretical aspects of the Bayesian strategy that have been presented in \cite{noll2009a} and many other articles. Similarly, we do not give excessive details on the precision and accuracy of the results as the topic has already been covered extensively in several papers from the same authors \citep[e.g.][]{boquien2012a,boquien2016a,buat2014a,ciesla2017a}. Finally, \texttt{CIGALE} being in constant evolution and development, this article describes its status as of version 2018.1. Further developments will be presented in separate publications.

The article is structured as follows. We present the guiding principles and the architecture of this new code in Sect.~\ref{sec:principles}. The modules used to construct the SED and carry out the analysis are presented in Sects.~\ref{sec:SED-creation} and \ref{sec:analysis-modules}. The versatility of \texttt{CIGALE} is shown through various examples of its application in Sect.~\ref{sec:versatility}. We conclude in Sect.~\ref{sec:summary}. For reference we provide additional technical details, performance benchmarks, and various examples in the appendices.

\section{Architecture\label{sec:principles}}

To interpret the results of the modelling of galaxies and avoid a detrimental black--box effect, it is important to understand the whys and wherefores of the model. We present here the broad design principles that we have followed, the reasons for which we have chosen the \texttt{python} language, a high--level overview of the architecture to compute the models and estimate the physical properties of galaxies, and finally some important implementation choices.

\subsection{Design principles}

While the intrinsic scientific quality of a model is certainly one of the most important criteria determining its impact, other factors also play a role. Ideally, a model should provide clear and meaningful results to a wide population of astronomers from Masters students learning galaxy modelling to highly experienced modellers without requiring a detailed knowledge of the internal mechanics and of the implementation. At the same time a model should remain clear in what it is doing, and how it does it, so that it is flexible and easily adaptable even by inexperienced users, allowing it to easily evolve. Last but not least, the model should not require extraordinary resources. Desktop computers or small departmental servers should be sufficient to analyse large samples of galaxies. To reach these overarching goals, we have designed the \texttt{python} version of \texttt{CIGALE} following three major guiding principles: modularity, clarity, and efficiency both for the users and the developers.

\begin{itemize}
 \item Modularity: the code must be split into different blocks that are as independent as can be from one another. Each of the four main stages: input handling (e.g. reading and processing the input files), computation of the models (e.g. the fluxes and the physical properties of each model), analysis (e.g. fit of observations and estimation of physical properties), and output handling (e.g. saving the physical properties, the best--fit spectrum, the $\chi^2$ of each model, the probability distribution function, etc.), must be entirely independent. Each physical component (stellar populations, nebular emission, attenuation by dust, dust emission, active nucleus, etc.) must be dealt with separately in individual modules, and each module must be able to be substituted as transparently as possible from the point--of--view of upstream and downstream modules. For instance it must be possible to change the attenuation law without affecting the rest of the code in any way. Finally, relying on this modularity, it must be possible to use \texttt{CIGALE} not only as initially intended but also as a library to build new tools.
 \item Clarity: the code must be as easy to understand as can be not only for the developers but also for the users in order to avoid a black--box effect and facilitate the development of community--driven extensions. This is very important to keep the evolution of \texttt{CIGALE} in phase with the evolution of knowledge and the creation of new emission models for any physical component existing or newly developed.
 \item Efficiency: large surveys yield increasingly larger multi--wavelength catalogues. We must use computer resources as efficiently as possible in terms of power and memory usage. We aim at being able to model the SED of thousands of galaxies across the universe using  millions of models in a matter of a few hours on a typical multi--core computer readily available off--the--shelf.
\end{itemize}

\subsection{Choice of programming language}

With these guiding principles in mind, we have chosen to develop the new version of \texttt{CIGALE} using the \texttt{python} language. We have made this choice based on three main arguments.

\begin{enumerate}
 \item With its clear syntax and its low barrier of entry, \texttt{python} has become an increasingly popular language in Astronomy. It is often the language of choice for teaching programming and has even become the \textit{de facto} standard for many new developments. For \texttt{CIGALE}, this means a large fraction of the community is readily able to develop and adapt it to their needs, increasing its potential beyond its original design.
 \item A direct cause and consequence of this popularity is that unlike languages more closely tailored for numerical computations such as \texttt{FORTRAN} or \texttt{idl}, \texttt{python} is versatile and has a broad and rich set of specialised and general--purpose libraries. We have relied as much as possible on such libraries, and in particular on \texttt{sqlalchemy}\footnote{\url{http://www.sqlalchemy.org/}} for storing models and filters in a database, \texttt{numpy} and \texttt{scipy} \citep{numpy,scipy} for numerical computations, \texttt{matplotlib} \citep{matplotlib} for plotting, and \texttt{astropy} \citep{astropy2013a,astropy2018a} for astronomically related tasks such as computing cosmology--dependent quantities like the luminosity distances or more generally to handle data input and output in a variety of formats, including \texttt{FITS} and \texttt{VO--table}. This has allowed us to focus our efforts on the scientific challenges rather than on the low level strata of the software.
 \item Even though it is a scripting language, the aforementioned scientific modules allow fast numerical computations in \texttt{python}, minimising the impact on the performance compared to a compiled language such as \texttt{C++} or \texttt{FORTRAN}. In addition to this, the language comes with a built--in module for parallel programming, allowing for efficient use of multiple cores and processors.
 \item Last but not least, \texttt{python} is published under a Free license. This means that users do not have to acquire a license to run \texttt{CIGALE}, unlike with \texttt{idl} for instance. Numerous \texttt{python} distributions are available at   zero cost with all the required libraries to install and run \texttt{CIGALE} easily.
\end{enumerate}

\subsection{High--level overview of \texttt{CIGALE}}

The primary purposes of \texttt{CIGALE} are to generate theoretical models and,  optionally, to use them to estimate the physical properties of galaxies. As the latter case is in fact probably the most common situation, we provide here a high--level overview of how this is achieved, only noting the handful of major differences when \texttt{CIGALE} is simply used to generate theoretical models.

\subsubsection{The division of labour\label{sssec:labour}}

The \texttt{CIGALE} package provides three executable files, each dedicated to a specific task:
\begin{enumerate}
  \item \texttt{pcigale} carries out the computation of the models and if needed the estimation of the physical properties of galaxies. 
  \item \texttt{pcigale-plots} generates plots from the output of \texttt{pcigale}: best SED, $\chi^2$ distribution, probability distribution function, and physical properties estimations from mock catalogues.
  \item \texttt{pcigale-filters} allows to list, delete, add, or plot a filter in the database.
\end{enumerate}
In practice, only \texttt{pcigale} is required to create the models, fit them to observations, and estimate the physical properties. The \texttt{pcigale-plots} and \texttt{pcigale-filters} executables are only provided for convenience and to facilitate the interpretation of the results provided by \texttt{pcigale}.

In more detail, \texttt{pcigale} handles the optional guided construction of the configuration file (\texttt{pcigale init}, which produces a configuration file template where the user then indicates the list of the physical modules to be used, and \texttt{pcigale genconf}, which fills the file with the configuration section for each of the user--requested modules), as well as the computation itself (\texttt{pcigale run}).

The computation is internally divided into four main stages:
\begin{enumerate}
 \item Input handling. First the configuration file is read and interpreted: name of the input data files, number of cores, fluxes and properties to fit, parameters for each module, and so on. Then the input data for each object to be analysed are also read: names, redshifts, distances (optional)\footnote{If the luminosity distance is explicitly provided, it overrides the distance computed from the redshift. The difference can be particularly important for nearby galaxies where peculiar motions dominate over the Hubble flow.}, fluxes, and physical properties (optional). These data are then processed and normalised (e.g. eliminating invalid data, adding missing uncertainties, etc.). If \texttt{CIGALE} is only used to generate models, the data file is only used to extract the list of bands.
 \item Model generation. For every combination of the input parameters, compute the physical properties of the model (SFR, stellar mass, attenuation, etc.) and its fluxes in a given set of bands.
 \item Analysis. For each object: (a) fit all the models to the data, (b) estimate the likelihoods for all the models, and (c) estimate the physical properties from the likelihoods. If only the generation of models is requested, then this step is skipped altogether.
 \item Output handling. For each object, save (a) the physical properties estimated from the likelihood, (b) the fluxes and the physical properties of the best-fitting model, and (c), optionally, additional information such as the spectrum of the best model, the $\chi^2$ of all the models, the probability distribution functions, and so on. If only the generation of models is requested, then only the computed fluxes are saved as well as the individual spectra.
\end{enumerate}

After \texttt{pcigale run} has completed, it is possible to generate a plot of the best model along with the observations as well as a range of plots related to the evaluation of the parameters with \texttt{pcigale-plots}.

\section{Model creation modules\label{sec:SED-creation}}

The physical processes at play in galaxies provide us with a natural path to build models and compute their physical properties. In \texttt{CIGALE}, the models are progressively computed by a series of independent modules called successively, each corresponding to a unique physical component or process. The typical sequence to build each model is the following:
\begin{enumerate}
 \item Computation of the SFH of the galaxy.
 \item Computation of the stellar spectrum from the SFH and single stellar population models.
 \item Computation of the nebular emission (lines and continuum) from the Lyman continuum photons production.
 \item Computation of the attenuation of the stellar and nebular emission assuming an attenuation law; computation of the luminosity absorbed by the dust.
 \item Computation of dust emission in the mid-infrared (mid-IR) and far-IR based an energy balance principle: the energy absorbed by the dust at short wavelengths, which has been computed in the previous step is re--emitted at longer wavelengths.
 \item Computation of the emission of an active nucleus.
 \item Redshifting of the model and computation of the absorption by the  intergalactic medium (IGM).
\end{enumerate}
In practice, the models are progressively computed by successively applying these different modules, each adding a different physical component (spectrum and associated physical parameters). For each model these individual spectral components and the combined spectrum are stored individually to ease the subsequent computation (e.g. to account for the differential reddening between younger and older stellar populations, we need to store these populations separately) and allow the user to easily retrieve the contribution from each physical component. For quantities that are more conveniently computed from the full rest-frame spectrum, in particular those that are directly measured observationally from the spectrum (e.g. line equivalent widths, UV slope $\beta$, colours, etc.), a special module can be added prior to redshifting to calculate them on the rest-frame spectrum. We describe here how we have modelled and parametrised each of these different physical components. As new modelling needs appear in the future, we will keep on improving these modules as well as adding new modules whenever necessary, a unique feature derived from the architecture and modularity of \texttt{CIGALE}. We should note that in addition to the modules we present here, \texttt{CIGALE} also provides unofficial modules that expand its capabilities even further and we support users to develop new modules and encourage them to make them available to the \texttt{CIGALE} community.

\subsection{Star formation history\label{ssec:modules-sfh}}

\begin{figure}[!htbp]
 \includegraphics[width=\columnwidth]{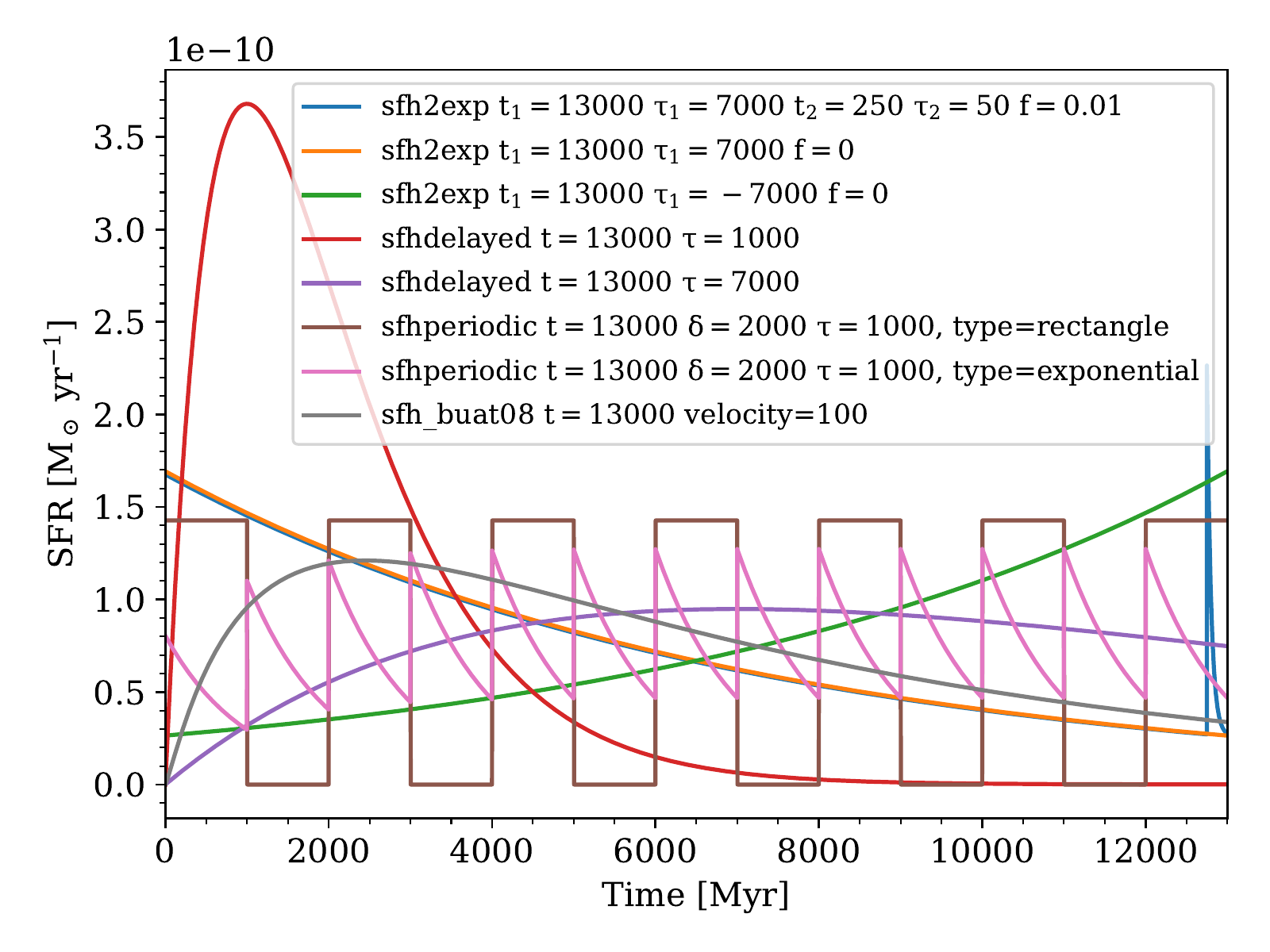}
 \caption{SFH generated with the \texttt{sfh2exp}, \texttt{sfhdelayed}, and \texttt{sfhperiodic} modules. These represent the cases of two decreasing exponentials (blue), a single decreasing exponential (orange), one increasing exponential (green), a delayed SFH with different timescales (red and purple), a periodic rectangular SFH (brown), a periodic exponential SFH (pink), and the rotation velocity--dependent SFH of \cite{buat2008a} (grey). We point out the transitory phase for the periodic exponential as each of the decaying exponentials combine. The exact parameters are indicated in the box. All SFHs have been normalised to have formed 1~M$_\odot$ over 13 Gyr. The diversity of generated SFHs allows for an important flexibility in the modelling.\label{fig:sfh}}
\end{figure}

As galaxies evolve secularly, accrete and expel gas, or interact with one another over cosmic times, their SFR is expected to vary considerably in non--trivial ways, from episodes of intense star formation to very quiescent phases. Constraining the SFH of galaxies is a tremendously difficult task. Not only because these variations are so complex, but also because dramatically different SFHs can sometimes yield remarkably similar SEDs. This difficulty to constrain the SFH of galaxies from broadband data has led most studies to adopt relatively simple SFH prescriptions aimed at reproducing the broad variations of the SFR with time: decaying or rising exponentials, delayed, or \`a la \cite{sandage1986a} for instance. However, with increasingly detailed numerical simulations it is now also possible to adopt more realistic SFH directly derived from such simulations or semi--analytic models \citep[e.g.][]{pacifici2012a,boquien2014a}. To encompass these two approaches, \texttt{CIGALE} handles both analytical SFH depending on several parameters, and arbitrary SFH. In Fig.~\ref{fig:sfh} we present some SFHs obtained from these modules, using a set of parameters representative of their versatility. We note that only one type of SFH (e.g. exponential, delayed, etc.) can be used in a single run. This means that different runs are needed to compare different parametrisations.

\subsubsection{Basic assumptions on the SFH}

Even though star formation is often modelled as mathematically continuous, it is a fundamentally discrete process, with stars being stochastically born one at a time. Building spectra from the ages of individual stars would rapidly become overwhelming computationally (notwithstanding the fact that we do not know this sort of information beyond the local group), it is therefore reasonable to assume some level of discretisaton on the SFH. In \texttt{CIGALE} we introduce two levels of discretisation.

First, we adopt a sampling period of 1~Myr for the SFH. Considering the time $t$ and assuming the age of the galaxy $t_0$, the sampling grid starts at $t=0$~Myr and the last sample is at $t=t_0-1$~Myr. It is important to note that the SFR is computed at the beginning of an age bin, but the contribution of stars to the spectrum is computed at the end of that age bin.

A sampling of 1~Myr is however too long to capture some brief but important stellar evolutionary phases. We therefore assume that in any given bin, star formation occurs in ten instantaneous episodes separated by 0.1 Myr. For instance if the SFH sampling indicates an SFR of 1 M$_\sun$ yr$^{-1}$ for a given bin, we distribute equally in time ten bursts of $10^5$~M$_\sun$. To limit the computation cost of this approach, the single stellar populations presented in Sect.~\ref{ssec:stellar-pops} are stored with a sampling over an age grid of 1~Myr but already precomputed assuming ten smaller bursts.

Finally, each SFH is automatically normalised so that the total mass of stars formed from the onset of star formation to the last time step is always 1~M$_\odot$. This definition does not correspond to the stellar mass because it does not take into account the return fraction, which depends on the specifics of the stellar populations. We see in Sect.~\ref{ssec:analysis} how the models are scaled to the observations, which in effect is equivalent to scaling the SFH to the proper level for each observation.

\subsubsection{\texttt{sfh2exp}, \texttt{sfhdelayed}, and \texttt{sfhperiodic} modules}

We present here three modules defining analytic SFH covering three different general cases: SFH defined by single or double exponentials (\texttt{sfh2exp}), delayed SFH with an optional exponential burst(\texttt{sfhdelayed}), and periodic SFH (\texttt{sfhperiodic}).

\paragraph{\texttt{sfh2exp}} One of the simplest ways to model the SFH of a galaxy is to model it with one or two decaying exponentials. Conceptually, the first exponential models the long-term star formation that has formed the bulk of the stellar mass, whereas the second one models the most recent burst of star formation. The combination can be expressed in the following way:
\begin{equation}
\text{SFR}\left(t\right)\propto
  \begin{cases}
    \exp\left(-t/\tau_0\right)&\text{if } t < t_0-t_1\\
    \exp\left(-t/\tau_0\right)+k\times\exp\left(-t/\tau_1\right)&\text{if } t \ge t_0-t_1,
  \end{cases}
\end{equation}
with $t_1$ being the age of the onset of the second episode of star formation relative to $t_0$ (i.e. if the galaxy started forming stars 13~Gyr ago and had a burst of star formation 100~Myr ago, $t_0=13$~Gyr and $t_1=100$~Myr), $\tau_0$ and $\tau_1$ the e--folding times of the populations modelling the older stellar populations and the most recent episode of star formation, and $k$ the relative amplitude of the second exponential, which is computed from the burst strength $f$ defined as the fraction of stars formed in the second burst relative to the total mass of stars ever formed. As (a) the SFH is sampled with a period of 1~Myr, (b) we assume a constant SFR between two samples, and (c) by convention we assign the first timestep a time of 0~Myr, $f$ can be expressed in the form of discrete integrals:

\begin{equation}
  f = \frac{k\sum_{t=t_0-t_1-1}^{t_0-1}\exp\left(-t/\tau_1\right)}{\sum_{t=0}^{t_0-1}\exp\left(-t/\tau_0\right)+k\sum_{t=t_0-t_1-1}^{t_0-1}\exp\left(-t/\tau_1\right)},\label{eq:sf2exp-f}
\end{equation}
which means that $k$ can be easily computed from the following relation:
\begin{equation}
  k = \frac{f}{1-f}\times\frac{\sum_{t=0}^{t_0-1}\exp\left(-t/\tau_0\right)}{\sum_{t=t_0-t_1-1}^{t_0-1}\exp\left(-t/\tau_1\right)}.\label{eq:sf2exp-k}
\end{equation}

Such a formulation, despite its apparent simplicity, is very versatile:
\begin{itemize}
 \item Very large values of $\tau$ compared to $t_0$ can be used to model a nearly constant SFR.
 \item Rising exponentials are obtained setting $\tau$ to a negative value.
 \item The classical case of a single exponential can be obtained setting $f=0$.
\end{itemize}
This allows for an efficient modelling of elliptical galaxies (case of a single exponential) or of galaxies having had a recent episode of star formation for instance. However, a clear weakness of this module is that it is not adapted for galaxies which have had a recent drop in their SFR, such as galaxies being quenched due to an infall on a cluster or galaxies over a large range in redshift where we successively increase and decrease the amplitude of SFHs.

\paragraph{\texttt{sfhdelayed}}
The sudden onset of star formation and burst episodes in a double--exponential parametrisation may be too extreme in many practical cases where we expect the variation of the SFH to be smoother. An increasingly popular way to model the SFH of galaxies is the so--called ``delayed'' SFH:

\begin{equation}
 \text{SFR}\left(t\right)\propto \frac{t}{\tau^2}\times\exp\left(-t/\tau\right) \text{for } 0\le t \le t_o,\label{eq:sfhdelayed}
\end{equation}
with $t_o$ the age of the onset of star formation, and $\tau$ the time at which the SFR peaks. Such a functional form has the advantage of providing a nearly linear increase of the SFR from the onset of star formation rather than an abrupt one in the case of \texttt{sfh2exp}. After peaking at $t=\tau$, it smoothly decreases.

To allow for more flexibility, the module also allows for an exponential burst representing the latest episode of star formation (Ma\l{}ek et al., in press). The burst strength is defined following the same concept as for \texttt{sfh2exp}, substituting the exponential for the older stellar populations (indices 0) in Eqs.~\ref{eq:sf2exp-f} and \ref{eq:sf2exp-k} for the delayed SFH from Eq.~\ref{eq:sfhdelayed}.

While a delayed SFH allows us to efficiently model early--type (for small $\tau$) and late--type (for large $\tau$) galaxies, one obvious limitation of this functional form is that it does not allow for a recent quenching of the SFR. To address this issue, \cite{ciesla2017a} expanded \texttt{sfhdelayed} allowing for an instantaneous recent variation of the SFR, upward or downward, and setting it to a constant until the last time step. This module is provided as \texttt{sfhdelayedbq}. This approach was used to successfully model  the broad range of KINGFISH galaxies in Hunt et al. (submitted).

\paragraph{\texttt{sfh\_buat08}} Rather than relying a priori on pure analytical functions, an alternative approach has been to tie the SFH to an observed physical quantity of the galaxy. This was done for example by \cite{boissier2003a,buat2008a} who related the SFH to the rotational velocity of the galaxy. Their SFH is parametrised as:
\begin{equation}
 \text{SFR}\left(t\right)\propto 10^{a+b\times\log(t)+c\times t^{1/2}},
\end{equation}
with $t$ ranging from 1 to $t_0$ in units of gigayears, and $a$, $b$, and $c$ being constants that depend on the rotational velocity of the galaxy. We adopt an extended version of the constants presented in Table 2 of \cite{buat2008a}\footnote{Private communication from Samuel Boissier.}.
\paragraph{\texttt{sfhperiodic}}

This module provides periodic SFH. The star formation episodes can be of three forms: exponential, delayed, or rectangular. There are four input parameters: 1. the shape of the star formation episodes, 2. $\delta$, the elapsed time between the beginning of each episode of star formation, 3. $\tau$, the duration of each star formation episode, and 4. $t_o$, the age of the onset of the first star formation episode (i.e. the age of the oldest stars).

\subsubsection{\texttt{sfhfromfile} module}

To build models with arbitrarily complex SFH and combine with hydrodynamical simulations or semi--analytic models, the \texttt{sfhfromfile} module allows to read and process SFH read from files. The first column of the file contains the age, starting from 0 and with a step of 1~Myr and each subsequent column contains the SFH with the SFR in M$_\odot$~yr$^{-1}$ and a step of 1~Myr for each line. The file can be provided indifferently in \texttt{ASCII}, \texttt{FITS}, or \texttt{VO--table} formats.

To use these SFH, besides the file name, the module requires the indices of the columns to consider, and the ages in millions of years at which the model should be computed. This allows to one compute the SED of a given SFH at different time steps, for instance to investigate its variation with respect to time \citep{boquien2014a}. Optionally, it is also possible to normalise the SFH so that the total stellar mass formed is 1~M$_\odot$ in a similar way as for the \texttt{sfh2exp}, \texttt{sfhdelayed}, and \texttt{sfhperiodic} modules.

\subsection{Stellar populations\label{ssec:stellar-pops}}

With the SFH having been computed with one of the modules described in Sect.~\ref{ssec:modules-sfh}, the next step is to compute the intrinsic stellar spectrum. To do so, in addition to the SFH, we need to adopt a library of single stellar populations (SSPs). We rely on two popular libraries of SSPs, that of \cite{bruzual2003a} (module \texttt{bc03}) and that of \cite{maraston2005a} (module \texttt{m2005}).

Each SSP library is available for a broad range of metallicities (0.0001, 0.0004, 0.004, 0.008, 0.02, and 0.05 for \cite{bruzual2003a}, and 0.001, 0.01, 0.02, and 0.04 for \cite{maraston2005a}) and for two initial mass functions (IMFs) (\cite{salpeter1955a} and \cite{chabrier2003b} for \cite{bruzual2003a}, and \cite{salpeter1955a} and \cite{kroupa2001a} for \cite{maraston2005a}).

The \cite{bruzual2003a} SSPs come in low and high resolution versions, both of which are provided with \texttt{CIGALE}. By default the low-resolution models are used as they are generally sufficient for use with broadband data. An option is provided to build the database with high-resolution models, which are useful for instance when dealing with narrow features such as absorption or emission lines.

\begin{figure}[!htbp]
 \includegraphics[width=\columnwidth]{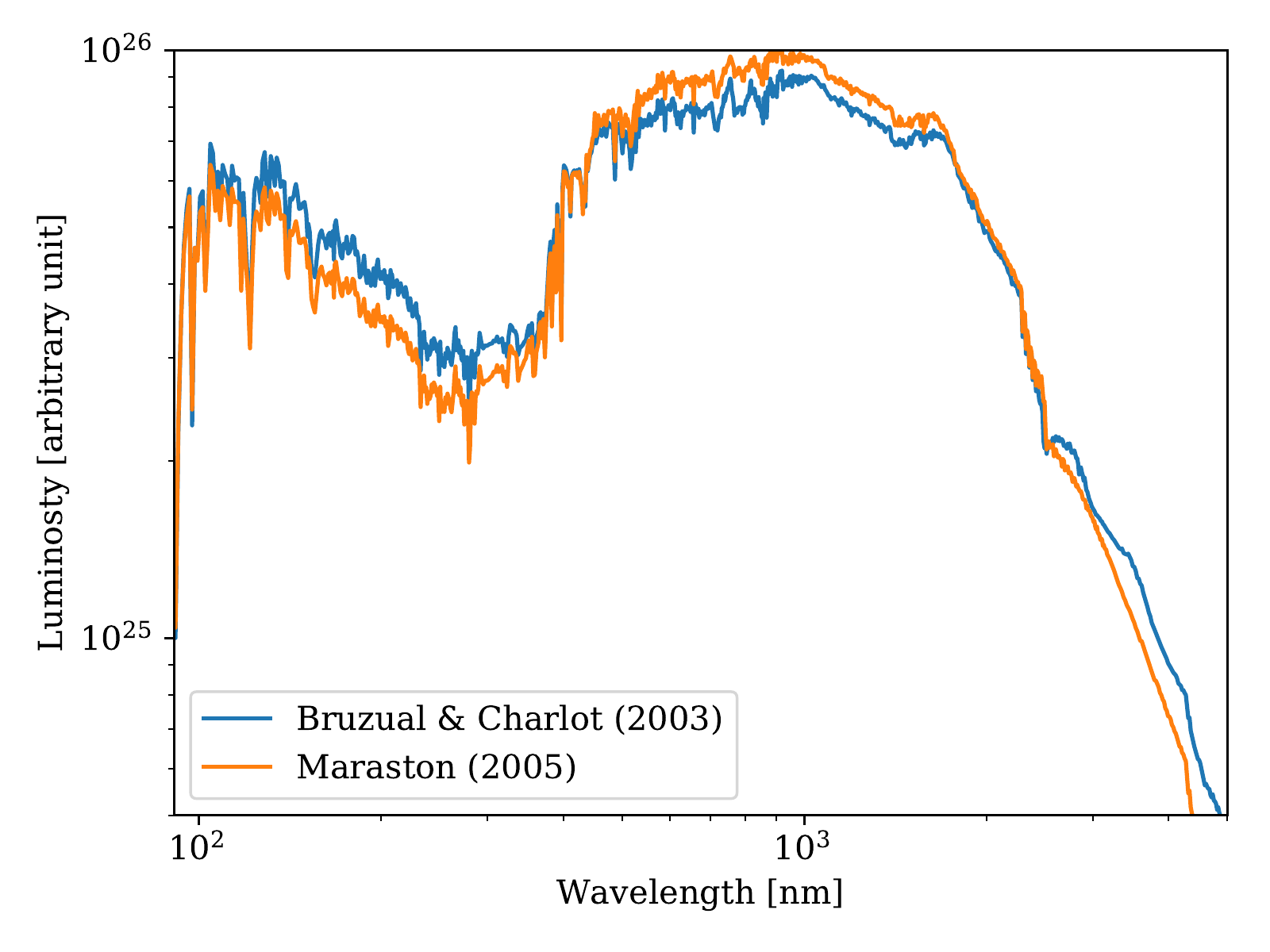}
 \caption{Spectra of the composite stellar populations for the \cite{bruzual2003a} (blue) and \cite{maraston2005a} models (orange). Both models have an identical SFH generated with the \texttt{sfh2exp} module ($t_0=13000$~Myr, $\tau_0=7000$~Myr, $t_1=250$~Myr, $\tau_1=50$~Myr, $f=0.01$), a \cite{salpeter1955a} IMF, and a metallicity $Z=0.02$. Even if the \cite{maraston2005a} models have been developed to handle the contribution of thermally pulsating asymptotic giant branch stars, there are also clear differences at all wavelengths. Such differences are strongly dependent on the actual SFH and the differences seen in this plot are not necessarily representative of what would be obtained with another SFH.\label{fig:comp-bc03-m05}}
\end{figure}

To compute the spectrum of the composite stellar populations, we calculate the dot product of the SFH with the grid containing the evolution of the spectrum of an SSP with steps of 1 Myr\footnote{Taking $\mathrm{SFH(t)}$ to be the SFR at age t, and $\mathrm{SSP(\lambda, t)}$ to be the spectrum of a single population at wavelength $\lambda$ and of age t, as the age grid is regular and identical, the object spectrum $\mathrm{S(\lambda)}$ is $\mathrm{\sum_tSSP(\lambda, t)\times SFH(t)}$.}. In Fig.~\ref{fig:comp-bc03-m05} we show the results of this computation using the \texttt{bc03} and \texttt{m2005} modules.

At this stage the stellar populations are dust--free. We need however to anticipate that there can be a differential reddening between young stellar populations embedded in their dust clouds and older populations that have escaped and are therefore less reddened \citep[e.g.][]{charlot2000a}. To account for this, we compute and store separately the spectra of old and young stars so they can be attenuated independently in a downstream module (Sect.~\ref{ssec:modules-attenuation}). Following the prescription of \cite{charlot2000a}, the default age of separation between these populations is 10~Myr but it can be configured freely.

\subsection{Nebular emission}

The most massive stars emit a significant fraction of their light in the Lyman continuum. These high-energy photons ionise the surrounding gas which re--emits the energy in the form of a series of emission lines and a continuum (free--free, free--bound, and two--photon) that extends far into the radio regime. This emission is important as it provides us with excellent probes into the most recent star formation through hydrogen lines and radio continuum as well as the gas metallicity from metal lines. While the nebular emission generally contributes little to the broadband fluxes of quiescent star--forming galaxies, and is therefore ignored, this is not the case at the local scale when considering starbursting dwarf and very young star--forming regions \citep[e.g.][]{anders2003a,boquien2010c} as well as some high--redshift galaxies \citep[e.g.][]{stark2013a,debarros2014a}. This has a direct impact on SED modelling and the nebular emission needs to be carefully taken into account.

To model the nebular emission in \texttt{CIGALE} we have adopted nebular templates based on \cite{inoue2011a}, which have been generated using \texttt{CLOUDY 13.01} \citep{ferland1998a,ferland2013a}. They predict the relative intensities of 124 lines from H\,\textsc{ii} regions from He\,\textsc{ii} at 30.38~nm to [N\,\textsc{ii}] at 205.4~$\mu$m. These templates are parametrised according to ionisation parameter $U$, and the metallicity $Z$, which is assumed to be the same as the stellar one. The electron density is assumed to be constant and is set to 100~cm$^{-3}$. Important improvements compared to the original templates of \cite{inoue2011a} include a refinement of the sampling in $\log U$ to steps of 0.1~dex, the extension down to $\log U=-4$, and changes in the abundances. The abundance set is based on the Orion nebula. The helium and nitrogen abundances are scaled by metallicity following \cite{nagao2011a}. This is motivated by the fact that helium has a primordial abundance floor and the nitrogen production is dominated by the secondary nucleosynthesis process through the CNO cycle at high metallicity.

In practice the computation of the nebular emission in \texttt{CIGALE} follows several steps. First of all, after having selected a given template (based on $U$, and $Z$), which gives line luminosities normalised to the ionizing photon luminosity, the spectrum of emission lines is computed. Each line has a Gaussian shape with a user--defined line width to take gas motion into account, which can be especially important for narrow--band filters and high--redshift objects due to the apparent line broadening with redshift in the observed frame. While this gives the normalised nebular emission line spectrum which could be rescaled to the appropriate level by multiplying by the ionizing photon luminosity which was computed along with the composite stellar population, this would ignore the fact that not all Lyman-continuum photons ionize the surrounding gas. Two main processes can affect the ionisation rate of the surrounding gas. First of all, a fraction of the Lyman continuum can simply escape from the galaxy. Even though the escape fraction is generally low in the nearby universe, it may reach much higher values at high redshift to reionise the universe \citep[e.g.][]{inoue2006a,hayes2011a}. The other process that can prevent Lyman continuum photons from ionising the surrounding gas is absorption by dust \citep{inoue2001a,inoue2001b}. In this case it contributes to the general dust heating and is accounted for in the dust emission models presented in Sect.~\ref{ssec:dust-emission}. To take these two processes into account, we downscale the nebular line spectrum by the following factor from \cite{inoue2011a}:
\begin{equation}
 k=\frac{1-f_\text{esc}-f_\text{dust}}{1+\alpha_1\left(T_e\right)/\alpha_B\left(T_e\right)\times\left(f_\text{esc}+f_\text{dust}\right)},\label{eqn:correct-fesc-fdust}
\end{equation}
with $\alpha_B$ being the case B recombination rate in m$^3$~s$^{-1}$, $\alpha_1=\alpha_A-\alpha_B$, the recombination rate to the ground level, $T_e$ the electron temperature in K, $f_\text{esc}$ the Lyman continuum escape fraction and $f_\text{dust}$ the  partial absorption by dust before ionisation. Numerically, for $T_e=10^4$~K, we take $\alpha_B=2.58\times10^{-19}$~m$^3$~s$^{-1}$ $\alpha_1=1.54\times10^{-19}$~m$^3$~s$^{-1}$ \citep{ferland1980a}.

The nebular continuum is computed following the prescription by \cite{inoue2010a} with the same parameters as the emission line templates and is computed in a similar fashion, including normalisation to the Lyman continuum photon luminosity and correction for the escape fraction and absorption by dust. Only the hydrogen continuum is taken into account as helium and other metal element continua are weak and negligible.

It is important to note that the nebular emission does not consider the emission of metal and CO lines in photo--dissociation regions and molecular clouds. In effect, [C\,\textsc{ii}] at 158~$\mu$m or [O\,\textsc{i}] 63~$\mu$m/145~$\mu$m are seriously underestimated for galaxies with photo--dissociation regions. This will be considered for a future version of the code.

\begin{figure*}[!htbp]
 \includegraphics[width=\columnwidth]{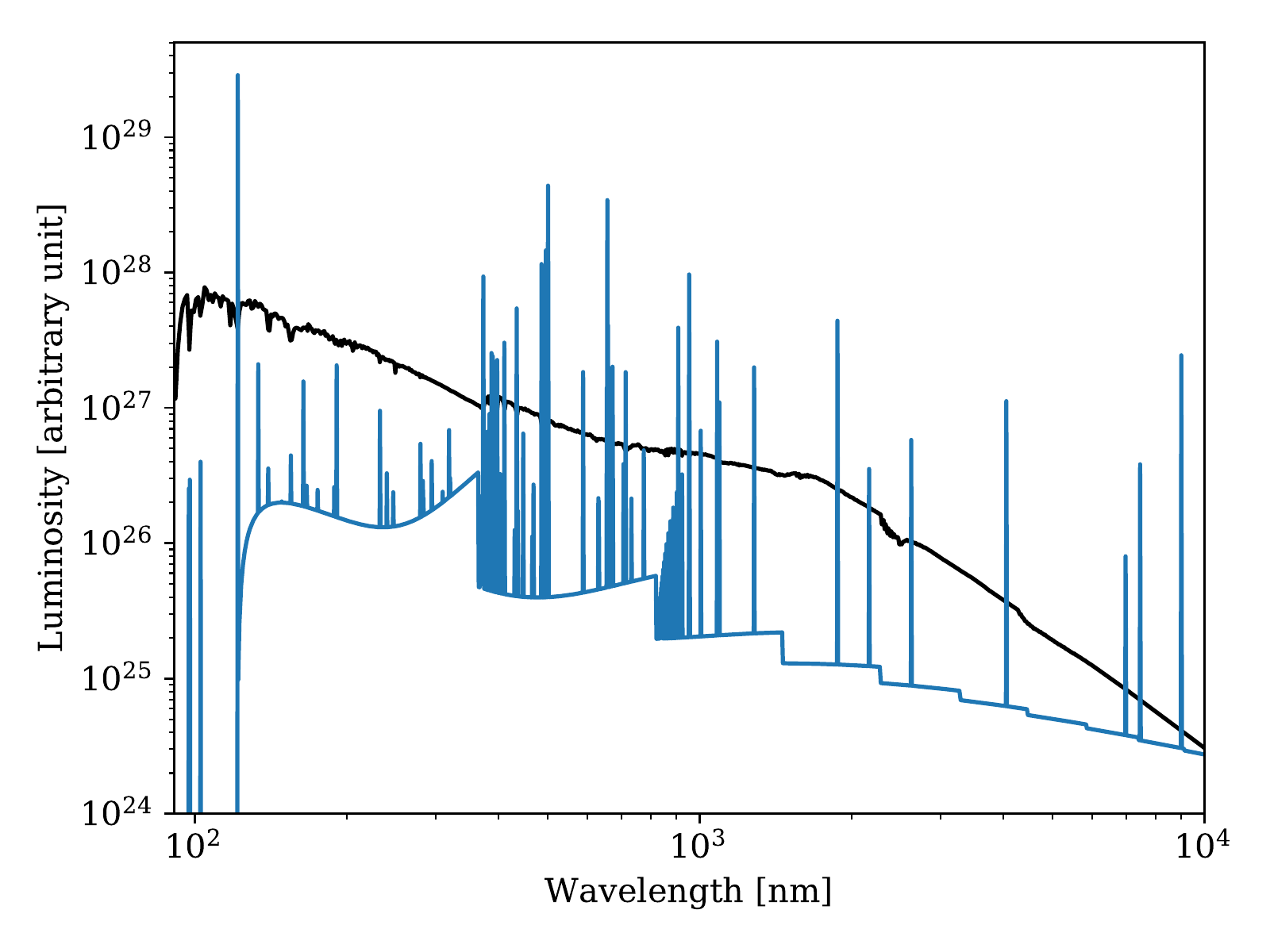}
 \includegraphics[width=\columnwidth]{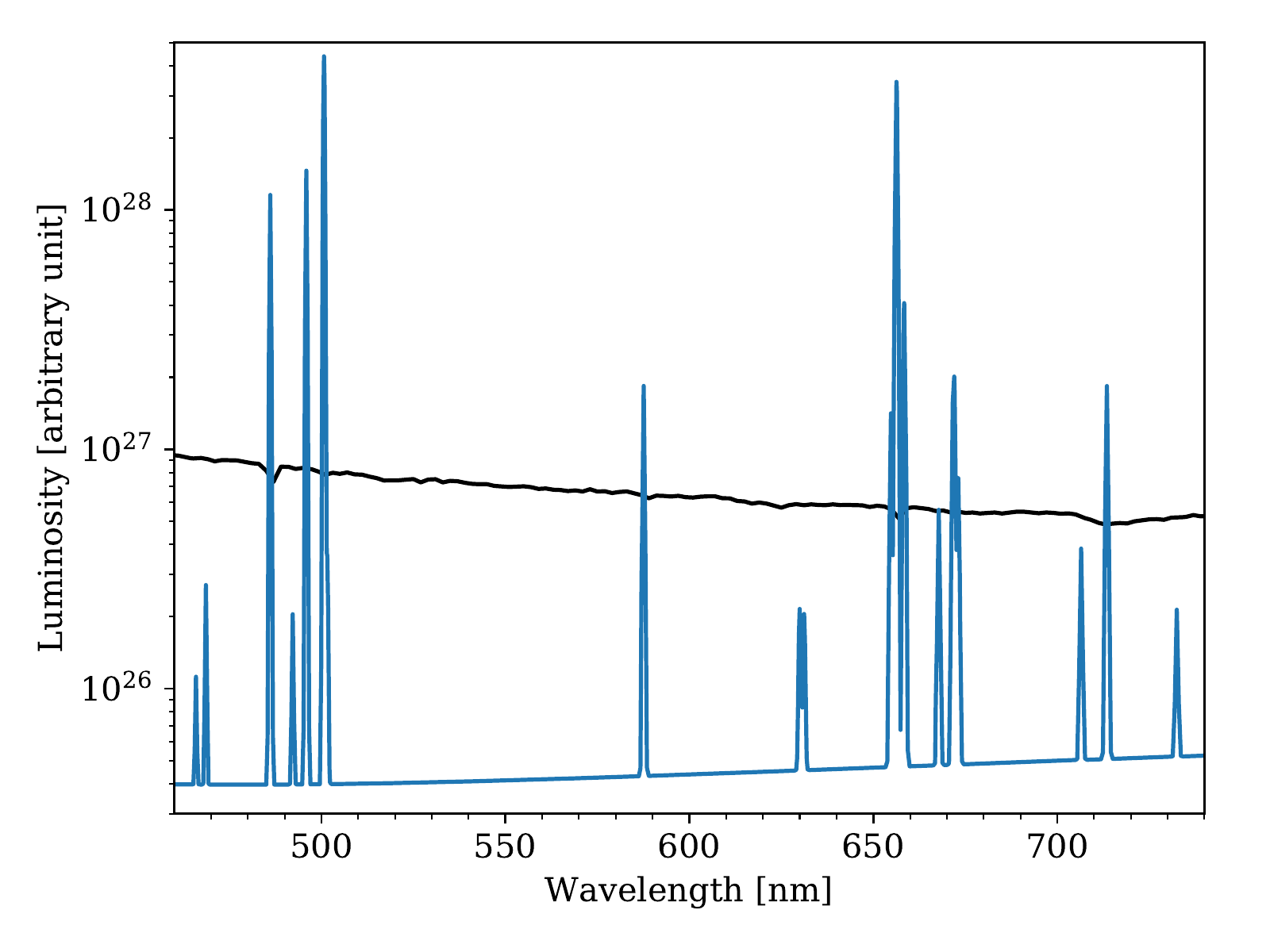}
 \caption{\textit{Left}: Nebular (blue) and stellar (black) FUV to NIR spectra. In total 124 lines from H\,\textsc{ii} regions are taken into account. The nebular continuum takes into account free--free, free--bound, and two--photon processes. The lines are modelled from an improved version of the \texttt{CLOUDY} templates computed in \cite{inoue2011a}. Here the density is set to 100~cm$^{-3}$, the metallicity $Z=0.02$, and the ionisation parameter $\log U=-2$. Both $f_\text{esc}$ and $f_\text{dust}$ are set to 0. The stellar emission is computed with the \texttt{sfh2exp} ($t_1=13000$~Myr, $\tau_1=7000$~Myr, $t_2=25$~Myr, $\tau_2=50$~Myr, and $f=0.1$) and \texttt{bc03} (\cite{salpeter1955a} IMF and $Z=0.02$) modules. \textit{Right}: Zoom in the 460~nm to 740~nm range. At this scale we can see the line width due to gas motions. Here the line width is set to 300~km~s$^{-1}$. \label{fig:nebular}}
\end{figure*}

We present a model of an SED including nebular emission in Fig.~\ref{fig:nebular}.

\subsection{Attenuation laws\label{ssec:modules-attenuation}}

Galaxies contain dust, and this dust is very efficient at absorbing short-wavelength radiation. The energy absorbed from the UV to the NIR is then re--emitted in the mid-- and far-IR. This energy balance principle lies at the core of \texttt{CIGALE}. It is therefore important that the attenuation is properly modelled.

First, we have to distinguish between an extinction curve, which is only dependent on the dust grain mix (composition, size distribution, etc.), and an attenuation curve, which also depends on the geometry. Except for a handful of very nearby objects such as the Magellanic Clouds, observers only see the effect of attenuation laws in galaxies. Because attenuation laws change with redshift and from galaxy to galaxy, \texttt{CIGALE} needs to be able to cover a broad range of such laws both in terms of shape and in terms of normalisation.

The most direct approach would be to consider a mix of dust grains and a geometry. However as we see below, we use different sets of templates to model dust emission. Each template would therefore need to have a specific extinction curve corresponding to the assumed mix. This would be difficult for empirical templates that do not assume specific grain properties. The assumption of a geometry is also a problem as \texttt{CIGALE} can be used on vastly different objects, from small regions in galaxies (down to sub--kpc scale) to large galaxies at all redshifts. In any case, observations of the Milky Way show that the relative distribution of dust and stars can be much more complex than the simple geometries that are often assumed (slab, sandwich, shell, etc.) and would therefore require a full radiative transfer with a realistic geometry \citep[e.g.][]{delooze2014a}.

An indirect but suitably much faster approach is to assume attenuation laws. Numerous studies have focussed on determining attenuation laws in galaxies, finding a remarkable diversity \citep[e.g.][Buat et al. (in press), Decleir et al. (submitted), and many others]{wild2011a,reddy2015a,reddy2016a,lofaro2017a,salim2018a}. This means that these attenuation laws must be flexible so that they can adapt to the broad diversity of observed curves. In this endeavor, we have pursued two ways of modelling attenuation curves in galaxies: the implementation of the \cite{charlot2000a} model, and flexible laws inspired from the starburst curve \citep{calzetti2000a}.

\subsubsection{The \texttt{dustatt\_modified\_CF00} module}

As a first approach to addressing this problem, we implemented the \cite{charlot2000a} model through the \texttt{dustatt\_modified\_CF00} module. The key idea behind this model is the realisation that not only young stars still embedded in their birth cloud suffer from additional attenuation compared to stars that have broken out and escaped into the ISM, but also that the attenuation curves association to the birth cloud and the ISM must be different. In practice, this is modelled by assuming two different power--law attenuation curves of the form $\text{A}\left(\lambda\right)\propto\lambda^\delta$: one for the birth cloud with a default slope of $\delta_{BC}=-1.3$, and one for the ISM with a default slope of $\delta_{ISM}=-0.7$. Because radiation from young stars has to travel through both the birth cloud and the ISM to escape the galaxy, the spectrum of stars younger than 10~Myr are attenuated by both the birth cloud and ISM curves. Stars older than 10~Myr are only attenuated by the ISM curve. Following Sect.~\ref{ssec:stellar-pops}, this age can be configured freely through the stellar populations module. In each case the nebular emission is attenuated following the same law as the stars giving rise to it. The V--band attenuation of both curves are linked through the relation:
\begin{equation}
\mu=\frac{\mathrm{A_V^{ISM}}}{\mathrm{A_V^{BC}+A_V^{ISM}}},
\end{equation}
or, in other words, the ratio of the total attenuation undergone by stars older than 10~Myr to that undergone by stars younger than 10~Myr. This module is flexible beyond a strict implementation of the \cite{charlot2000a} model in the sense that $\mathrm{A_V^{ISM}}$, $\mu$, $\delta_{BC}$, and $\delta_{ISM}$ are all input parameters.

It should be noted that \texttt{CIGALE} also provides the module \texttt{dustatt\_powerlaw}, which should not be confused with this module as it departs in several ways from the \cite{charlot2000a} model, having a single power law for both young and old stars, only with a different absolute attenuation, and the attenuation is set as the total attenuation for each component.

\subsubsection{The \texttt{dustatt\_modified\_starburst} module}

\begin{figure}[!htbp]
 \includegraphics[width=\columnwidth]{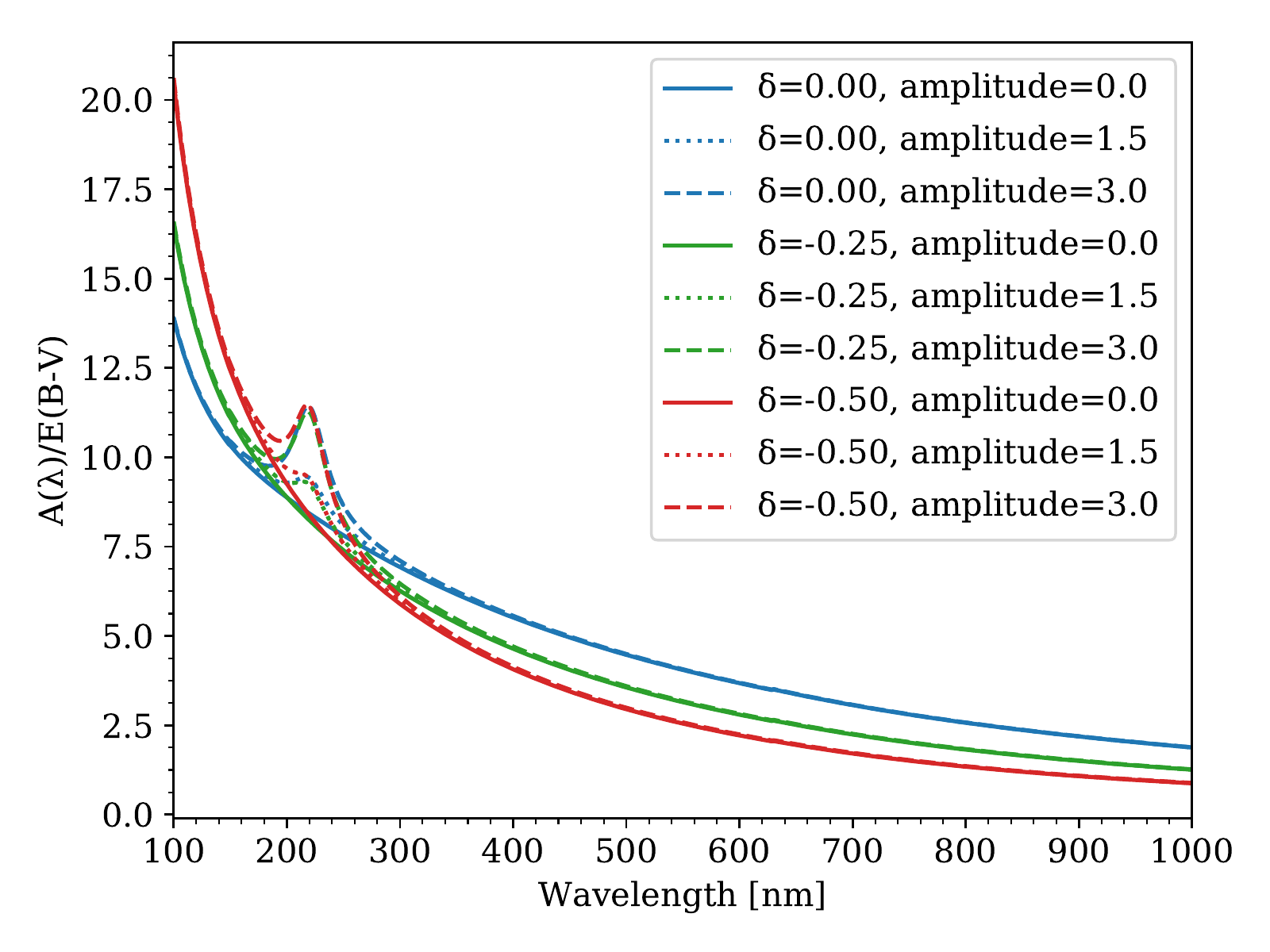}
 \caption{Three stellar attenuation curves generated by the \texttt{dustatt\_modified\_starburst} module of \texttt{CIGALE}. Based on the \cite{calzetti2000a} attenuation law, the red, green, and blue correspond to a power--law modification with indices $\delta$ of $0.00$, $-0.25$, and $-0.50$. In addition, a 220 nm bump has been added with three different amplitudes: 0 (dotted), 1.5 (solid), and 3 (dashed). The difference in the normalisation comes from the fact that $E(B-V)$ is kept constant after multiplying the starburst law (case $\delta=0$) by a power law.\label{fig:attenuation-curves}}
\end{figure}

A more empirical approach is to use a well-known curve as a baseline. Subsequently, this curve can be parametrised to make it more generic and allow for some flexibility, for example in terms of slope or to account for the presence of a bump around 220~nm. We have also adopted this solution with the \texttt{dustatt\_modified\_staburst} module. It is based on the \cite{calzetti2000a} starburst attenuation curve, extended with the \cite{leitherer2002a} curve between the Lyman break and 150~nm. Its slope can be modified by multiplying it by a power law function of slope $\delta$ similar to the one described above and a UV bump can be added. This bump is modelled as a Lorentzian--like Drude profile which is described by three parameters: its central wavelength, its full width at half maximum (FWHM), and its amplitude \citep[see Eq.~3 from][]{noll2009a}. The overall attenuation can be expressed as:

\begin{equation}
k_{\lambda} = \left(k_\lambda^{starburst}\times\left(\lambda/550~\textrm{nm}\right)^\delta+D_\lambda\right)\times\frac{\mathrm{E(B-V)_{\delta=0}}}{\mathrm{E(B-V)_{\delta}}},
\end{equation}
with $D_\lambda$ the Drude profile, and the last term renormalising the curve so that E(B-V) remains equal to the input E(B-V) when $\delta\neq0$. We show a set of stellar attenuation curves representative of the flexibility of our approach in Fig.~\ref{fig:attenuation-curves}.

Formally, the starburst law is defined for the continuum only, the emission lines being dimmed with a Milky Way extinction law. Here we have adopted a slightly more flexible approach, adopting the Milky Way curve of \cite{cardelli1989a} with the \cite{odonnell1994a} update, as well as the Small and Large Magellanic Cloud extinction curves of \cite{pei1992a}. The value of $R_V$ can be modified. The overall normalisation of the curves affecting the stars and the lines is determined according to the reddening of the emission lines $\mathrm{E(B-V)_{lines}}$, with a simple reduction factor $f$ between the two curves:
\begin{equation}
  f=\frac{\mathrm{E(B-V)_{continuum}}}{\mathrm{E(B-V)_{lines}}},
\end{equation}
with lines being more dimmed by dust than stars. 

\subsection{Dust emission\label{ssec:dust-emission}}

As dust absorbs stellar photons from the UV to the NIR, this energy is re--emitted at longer wavelengths, essentially in the mid-- and far--IR domains. In general, dust emission can be split into three broad components. In the mid--IR, around 8~$\mu$m, the emission is dominated by polycyclic aromatic hydrocarbon (PAH) bands. At longer wavelengths, the emission is progressively taken over by  very small, warm grains that tend to be stochastically heated for weak and moderate radiation field intensities but progressively becomes dominated by equilibrium emission at higher intensities. Beyond $\sim100~\mu$m, the emission is increasingly due to  big, relatively cold grains. The different heating mechanisms of these different dust species, their composition, the metallicity, and the intensity and shape of the incident radiation field, all have an impact on the dust SED.

\begin{figure*}[!htbp]
 \includegraphics[width=\columnwidth]{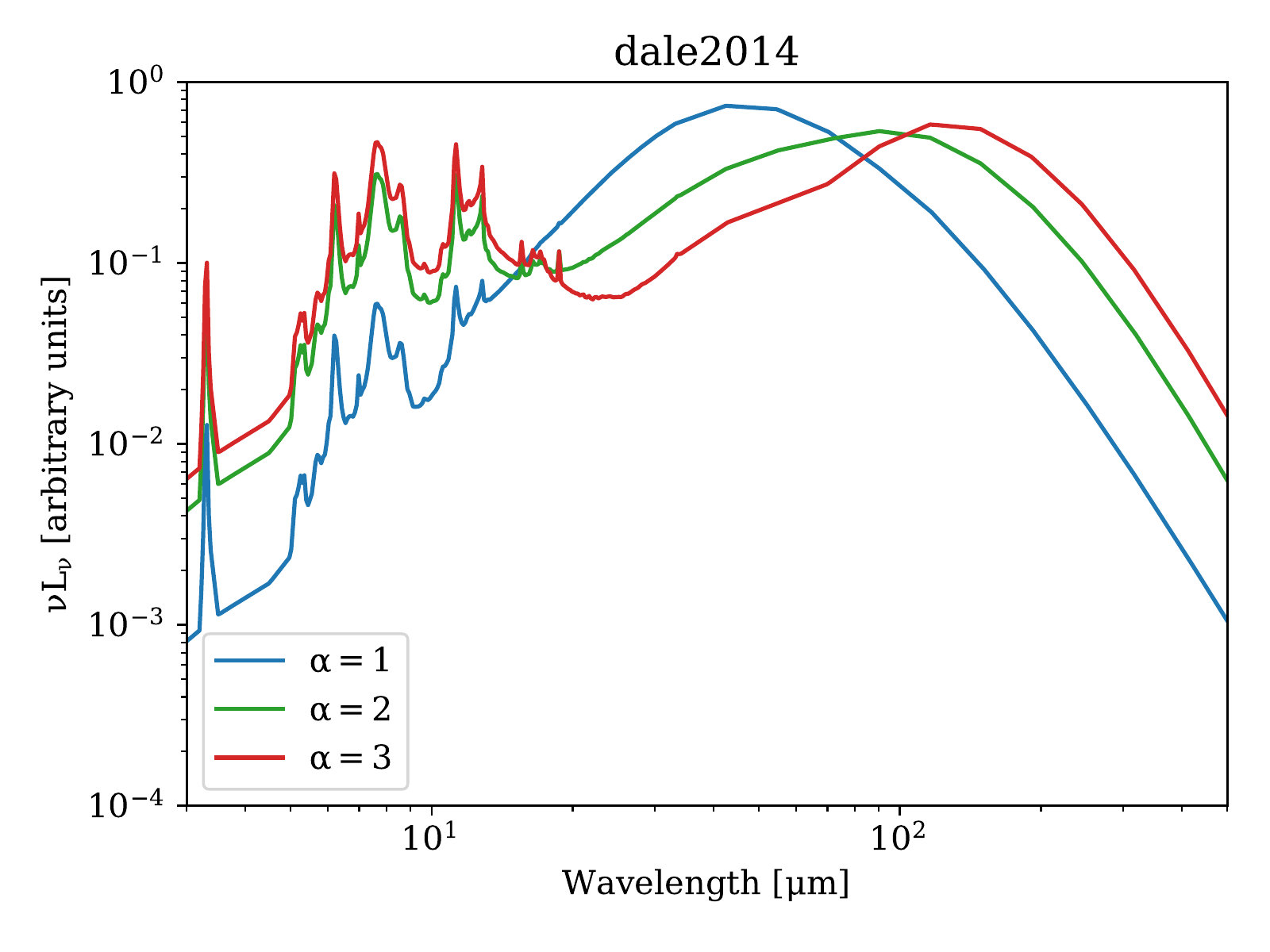}
 \includegraphics[width=\columnwidth]{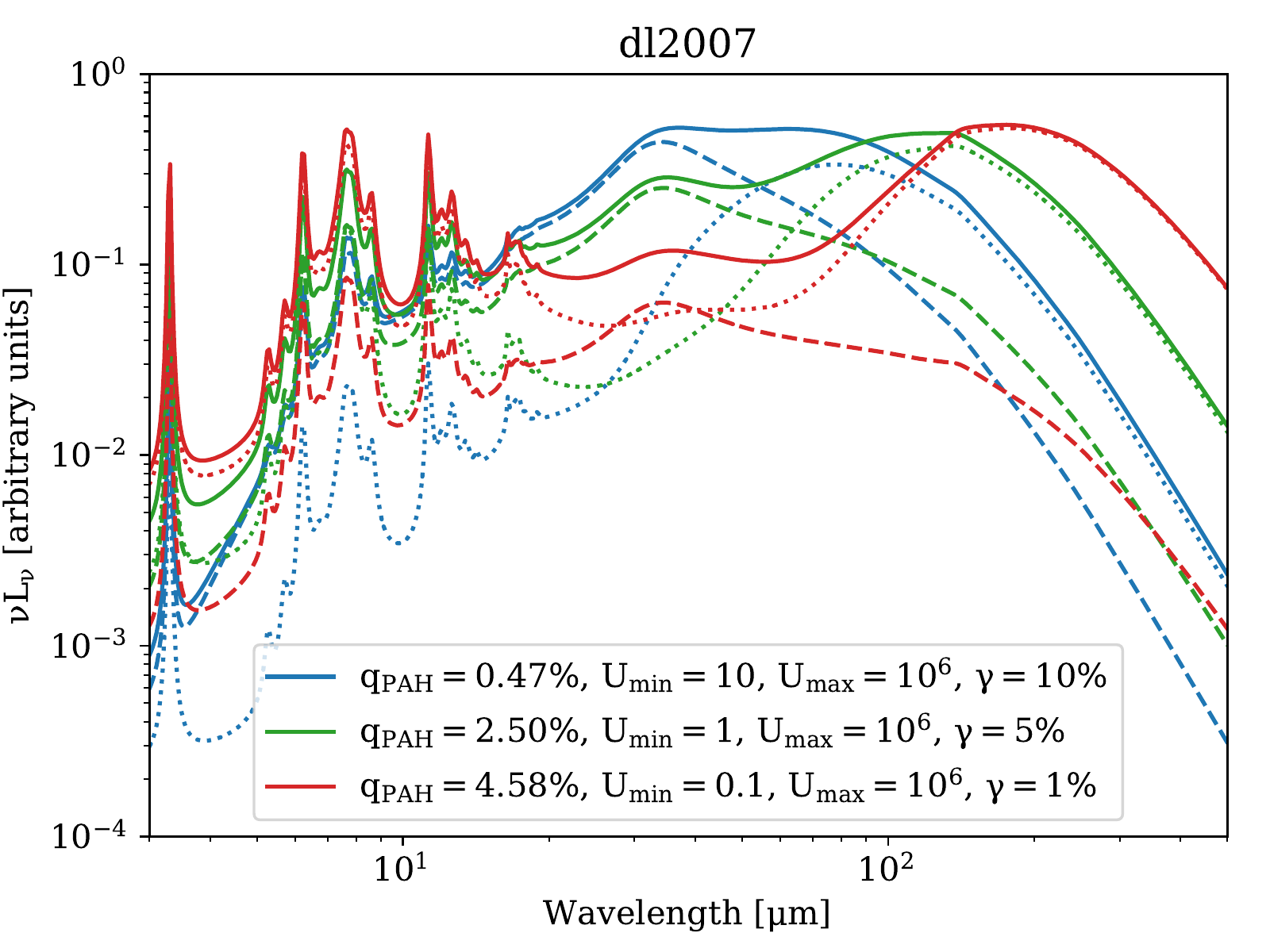}\\
 \includegraphics[width=\columnwidth]{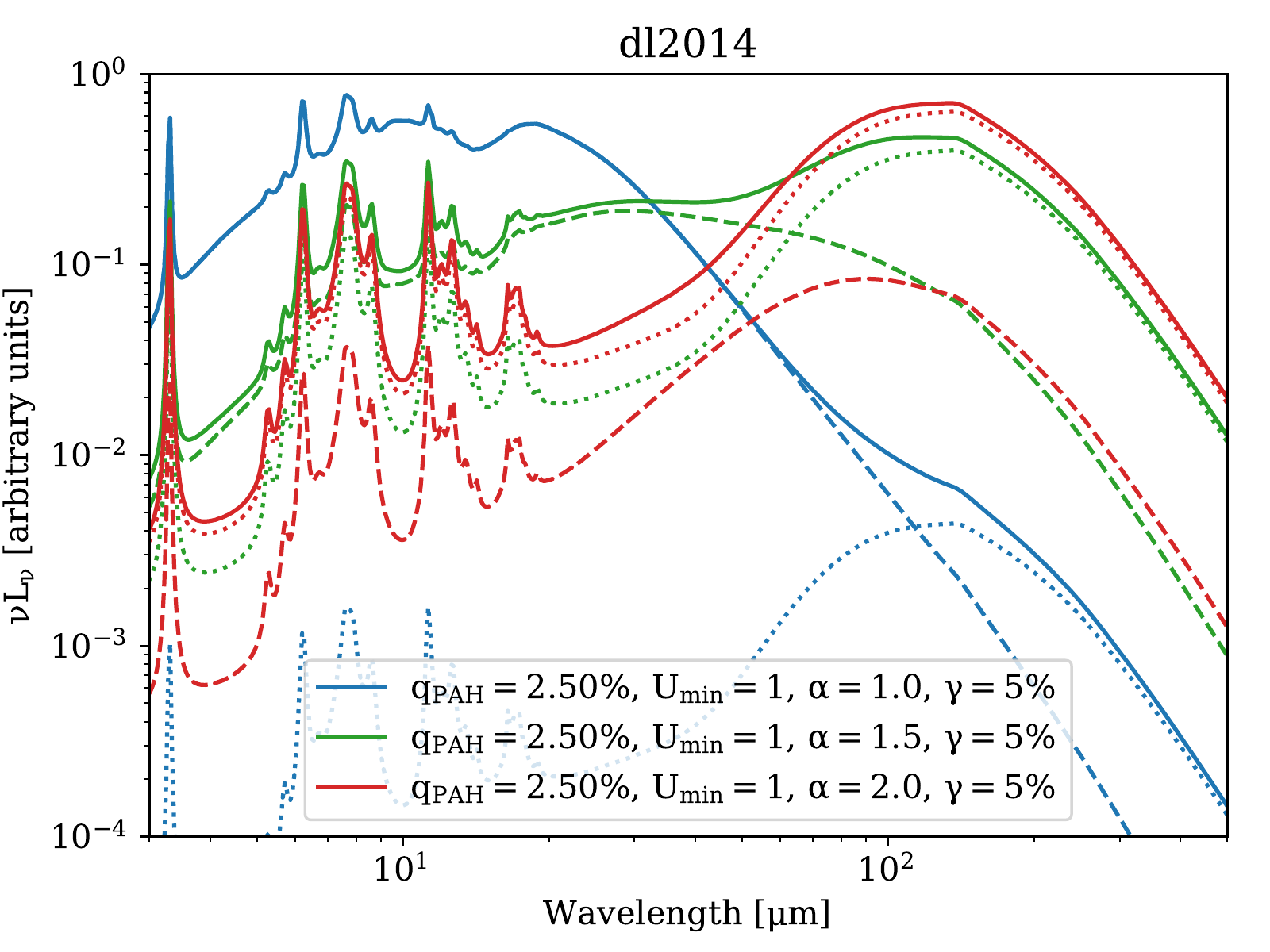}
 \includegraphics[width=\columnwidth]{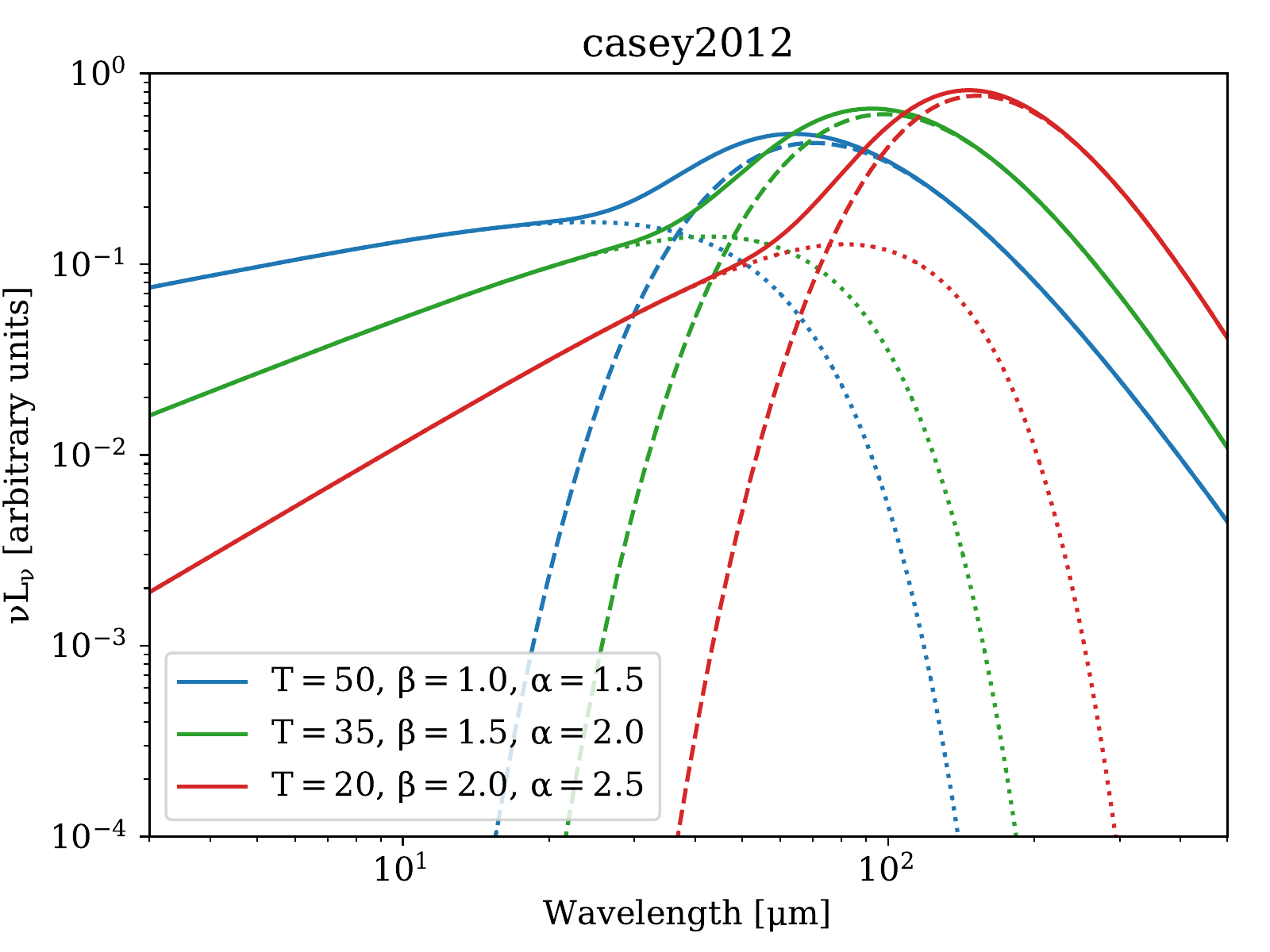}
 \caption{Examples of the SED produced by the four dust modules of \texttt{CIGALE}: \texttt{dale2014} (top--left), \texttt{dl2007} (top--right), \texttt{dl2014} (bottom--left), and \texttt{casey2012} (bottom--right). Each colour indicates a different set of parameters shown in the bottom--right corner. The solid lines represent the total SED, summing up all components. For the \texttt{dl2007} and \texttt{dl2014} modules, the dashed line corresponds to the star--forming regions and the dotted line to the diffuse emission. Finally, for the \texttt{casey2012} module, the dashed line corresponds to the modified black body whereas the dotted line corresponds to the power law.\label{fig:dust}}
\end{figure*}

The modelling of dust emission is a very active domain of research, building on several generations of increasingly more powerful IR instruments, from IRAS to \textit{Herschel}. For \texttt{CIGALE}, we consider three different sets of models: the \cite{dale2014a} empirical templates, the \cite{draine2007a} models \cite[including the updates of][]{draine2014a}, and the \cite{casey2012a} analytic model. The modules are described hereafter and some examples of dust SED are shown in Fig.~\ref{fig:dust}.

\subsubsection{\texttt{dale2014} module}

The dust templates of \cite{dale2014a} are based on a sample of nearby star--forming galaxies originally presented in \cite{dale2002a}. The latest update refines the PAH emission and also adds an optional AGN component as seen in Sect.~\ref{ssec:agn}. Aiming at simplicity, the star--forming component is parametrised by a single parameter $\alpha$ defined as: $dM_d\left(U\right)\propto U^{-\alpha}dU$, with $M_d$ being the dust mass, and $U$ the radiation field intensity. The $\alpha$ parameter is itself tightly linked to the 60--to--100 $\mu$m colour. The main strength of this model is its simplicity, with only one easy-to-interpret parameter. However, the PAH emission relative to the total infrared shows a limited variation with respect to $\alpha$. This can be an issue in particular in metal--poor galaxies which are known to have only little PAH emission \citep[e.g.][]{engelbracht2005a}.

\subsubsection{\texttt{dl2007} and \texttt{dl2014} modules}

Presented in \cite{draine2007a}, these models are based on a dust mixture of amorphous silicate and graphite grains, and PAH. One of the key features of the \cite{draine2007a} templates is the separation of dust emission into two components. The first one models the diffuse dust emission heated by the general stellar population. In this context, the dust is illuminated with a single radiation field $U_{min}$. The second one models dust tightly linked to star-forming regions. In that case the dust is illuminated with a variable radiation field ranging from $U_{min}$ to $U_{max}$ following a power--law index $\alpha$ \citep[see Eq. 23 of][]{draine2007a}. By default it is set to a fixed value of $\alpha=-2$. The dust mass fraction of dust linked to the star--forming regions (respectively diffuse emission) is $\gamma$ (respectively $1-\gamma$). The last parameter of these models is $q_{PAH}$, the mass fraction of the PAH, which is common for the two components. These components are kept separate in \texttt{CIGALE} to allow for their individual inspection.

In recent years, this model has been refined further. Because the parameters are not identical, a different module is available to use the updated models: \texttt{dl2014}. Among the main differences to note, these new models have led to the following: 1) an expansion on the range of radiation field intensities and PAH mass fractions; 2) the power law index $\alpha$ is now a free parameter; 3) $U_{max}$ has been set to $10^7$; 4) a change in the treatment of graphite ; 5) the dust masses have been renormalised \citep{draine2014a}.

One of the main strengths of these models is that they are very flexible. They can account for very different physical conditions with a variety of radiation fields and a variable PAH emission. However, this flexibility comes at the cost of a much larger parameter space to explore compared to the \cite{dale2014a} templates and is therefore more expensive in terms of processing power and memory.

\subsubsection{\texttt{casey2012} module}

The \texttt{casey2012} module implements the analytic model of \cite{casey2012a}. Dust emission is modelled with two components: a single temperature modified black body in the FIR ``representing the reprocessed starburst emission in the whole galaxy'' and a power law in the mid--IR ``which approximates hot--dust emission from AGN heating or clumpy, hot starbursting regions'' \citep{casey2012a}. In practice, the module depends on three parameters: the temperature of the dust, the emissivity index of the dust, and the mid-IR power law index. To distinguish both components and easily assess their relative contributions, \texttt{CIGALE} stores them separately in the SED.

While less physically motivated than the \cite{draine2007a} models and not based on observations as the \cite{dale2014a} templates, the \cite{casey2012a} models are very flexible and can be easily used for local and high-redshift galaxies. The main limitations of this module, however, are that it includes no PAH emission and that the IR is computed from an energy balance and thus AGN heating is in effect not included.

\subsection{Synchrotron radio emission}

With the advent of the Square Kilometre Array, an avalanche of data in the centimetre regime is upon us. At such wavelengths, the emission is split between thermal processes related to the ionisation of the gas by massive stars and non--thermal processes related to the interaction of relativistic electrons from supernovae with the local magnetic field.

While the \texttt{nebular} module naturally models the thermal radio continuum, it lacks synchrotron emission. The exact shape and intensity of the synchrotron spectrum depends on various parameters such as the strength of the magnetic field, the energy of the relativistic electrons propagating through it, and so on. Rather than attempting to model the synchrotron in such detail, the \texttt{synchrotron} module relies on the radio-IR correlation $q_{IR}$ of \cite{helou1985a}, a free power--law spectral slope $\alpha$, and on the assumption that at 21~cm the spectrum is largely dominated by non--thermal emission. In effect, knowing the IR emission, $q_{IR}$ directly provides the luminosity density at 21~cm. It is then a simple matter of computing a spectrum with the requested $\alpha$ and scaling it so that it matches the estimated luminosity. On the other hand, radio data can help to estimate the IR emission if no other data are available in this range.

\subsection{Active galactic nuclei\label{ssec:agn}}

Along with star formation, AGNs are thought to have a dramatic impact on galaxy evolution. Yet, properly disentangling the emission of AGNs from star formation is not necessarily an easy task as they can both strongly emit in the UV, and a large fraction of this emission can be reprocessed by dust and re--emitted at longer wavelengths.

\begin{figure}[!htbp]
 \includegraphics[width=\columnwidth]{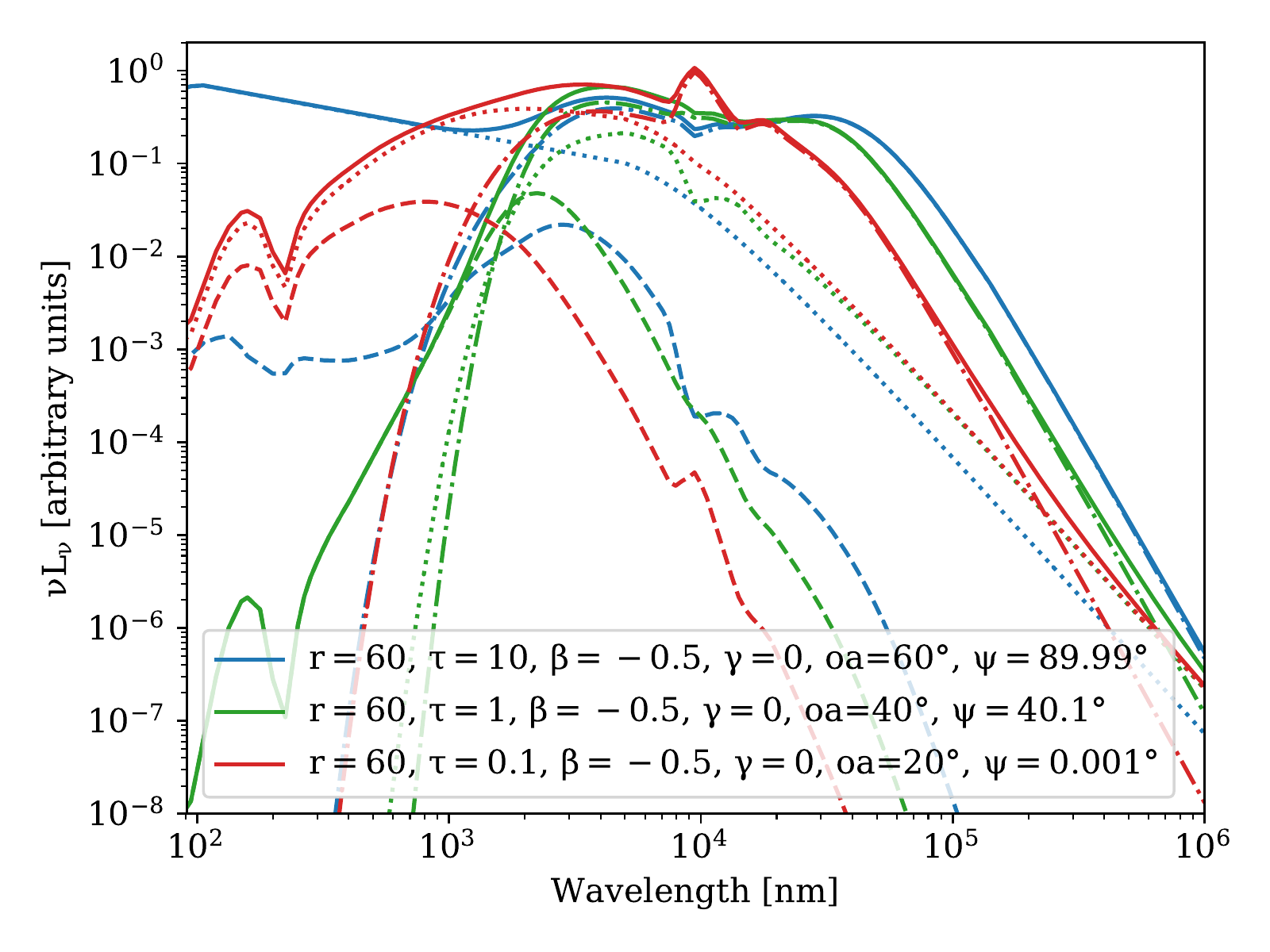}
 \caption{Examples of the SED produced with the \texttt{fritz2006} module. Each colour indicates a different set of parameters shown in the bottom--left corner. The solid lines represent the total emission. The dotted lines represent the AGN accretion disk, the dashed lines the scattered component, and the dash--dot line the thermal emission.\label{fig:agn}}
\end{figure}

Several options are available in \texttt{CIGALE} to model the presence of an AGN from coarse but rapid methods to more detailed but slower methods. If only the IR is fitted, the \texttt{casey2012} module can be used, with the AGN being simply parametrised by the slope of the power law $\alpha$. If the AGN is a quasar, the \texttt{dale2014} module provides a simple template from the UV to the IR. The AGN is simply parametrised with the AGN fraction defined as the ratio of the AGN luminosity to the sum of the AGN and dust luminosities. While these methods are rapid and easy to use, they do not necessarily offer all the flexibility one may want to take into account the variety of AGN SED. \texttt{CIGALE} also provides the detailed AGN models of \cite{fritz2006a}. It explicitly takes into account three components through a radiative transfer model: the primary source located in the torus, the scattered emission by dust, and the thermal dust emission. These modules are determined through a set of seven parameters: $r$ the ratio of the maximum to minimum radii of the dust torus, $\tau$ the optical depth at 9.7~$\mu$m, $\beta$ and $\gamma$ describing the dust density distribution ($\propto r^\beta e^{-\gamma\left|\cos\theta\right|}$) with $r$ the radius and $\theta$ the opening angle of the dust torus, $\psi$ the angle between the AGN axis and the line of sight, and the AGN fraction. We show some examples of the SED generated by the \texttt{fritz2006} module in Fig.~\ref{fig:agn}.

\subsection{Measuring physical properties from the rest--frame spectrum}

The previous modules have allowed us to build a full FUV--to--radio rest--frame spectrum. If we have the physical properties (SFR, attenuation, stellar and dust mass, etc.) associated with each of the components, some observed quantities can only be accurately computed from the full spectrum. The aim of this module is to compute such quantities before redshifting the spectrum. The following quantities are measured:
\begin{itemize}
  \item The UV slope $\beta$ is computed by fitting a straight line to the $F_\lambda$ spectrum in log--log space over the wavelength ranges defined in Table 2 of \cite{calzetti1994a}.
  \item The $D_n4000$ break index is computed from the ratio of the mean fluxes of the $F_\lambda$ spectrum from the rest--frame 400 to 410~nm on the red side and from 385 to 395~nm on the blue side \citep{balogh1999a}.
  \item IRX is computed as the log of the ratio of the dust to rest--frame far--UV (GALEX band) luminosities.
  \item Rest--frame equivalent widths are computed as the ratio of the integral of the spectrum over a user--defined wavelength range, including and excluding nebular lines.
  \item Rest-frame luminosity densities and colours are computed integrating the spectrum over any filter or pair of filters that are present in the filters database.
\end{itemize}

\subsection{Redshifting}

\begin{figure}[!htbp]
 \includegraphics[width=\columnwidth]{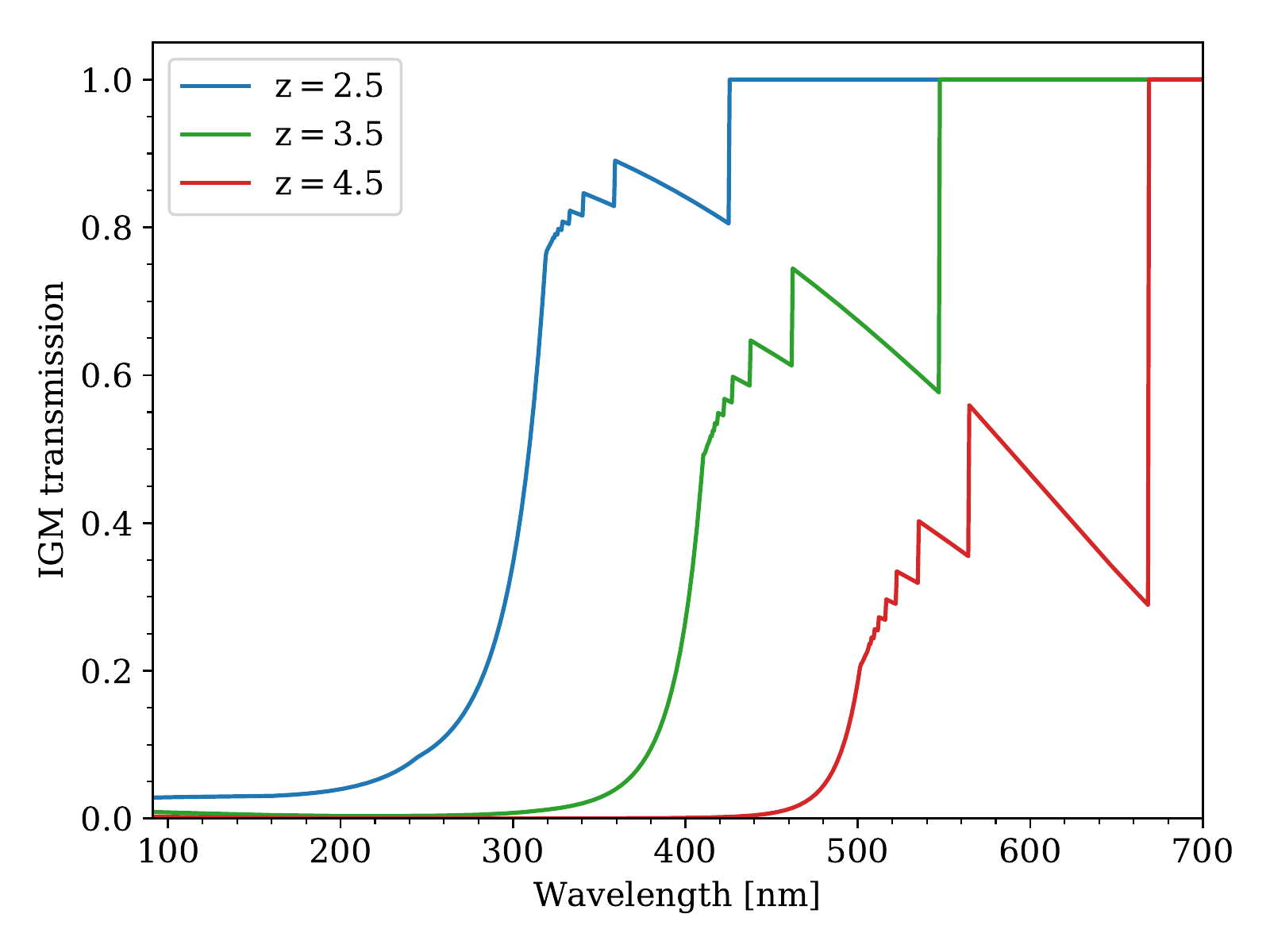}
 \caption{IGM transmission prescription of \cite{meiksin2006a} in the observed frame. The three redshifts are indicated in the top--left corner.\label{fig:igm}}
\end{figure}

Finally, the last module called to build the SED is the \texttt{redshifting} module. It has two effects. First, it redshifts the spectrum and dims it, multiplying the wavelengths by $1+z$ and dividing the spectrum by $1+z$. The second effect of this module is to take into account the absorption of short wavelength radiation by the IGM. To do so the \texttt{CIGALE} applies the prescription of \cite{meiksin2006a}, which is shown in Fig.~\ref{fig:igm}.

We note that this module takes only one parameter, the redshift. However if the redshifts to apply are not indicated in the configuration file, \texttt{CIGALE} then automatically builds a list of redshifts from the input flux file, rounding them to two decimal places by default in order to avoid computing models with many close redshifts. Nevertheless, the physical quantities are computed for the exact input redshift at full precision.

\section{Analysis modules\label{sec:analysis-modules}}

The modules presented in Sect.~\ref{sec:SED-creation} are the building blocks to compute individual models. But these building blocks alone do not provide us with the desired set of SEDs nor do they provide us with estimates of the physical properties of the objects under consideration. Such tasks lie upon the so--called analysis modules. Two analysis modules are available with \texttt{CIGALE}: \texttt{savefluxes} to generate a grid of models and save the outputs (fluxes, SFH, physical properties, etc.) and \texttt{pdf\_analysis} that not only generates a grid of models but also fits these models to observations to estimate various physical properties and save the outputs.

\subsection{Computing physical quantities}

The computation of the physical quantities in \texttt{CIGALE} depends on their nature. Intrinsic intensive and extensive physical properties are computed in the different physical modules presented in Sect.~\ref{sec:SED-creation} where it makes most sense. Fluxes however depend on the observer and are measured after the computation of a given model. The technique differs whether we compute fluxes in bandpasses or whether they are extracted from low- or high-resolution spectra.

\subsubsection{Bandpasses}

The fluxes in bands are computed by integrating the model spectrum through the corresponding filter. The basic method is standard:

\begin{equation}
  f_\lambda = \frac{\int_{\lambda_{low}}^{\lambda_{high}}F_\lambda\left(\lambda\right) \times T\left(\lambda\right) d\lambda}{\int_{\lambda_{low}}^{\lambda_{high}}T\left(\lambda\right) d\lambda},\label{eqn:flux-bandpass}
\end{equation}
with $f_\lambda$ being the flux per wavelength through a filter of transmission $T$ in units of energy\footnote{If the filter is provided in units of photons, it is converted and stored in units of energy in the database of \texttt{CIGALE}.} defined between wavelengths $\lambda_{low}$ and $\lambda_{high}$, and $F_\lambda$ the flux per wavelength of the model spectrum. To preserve the best sampling and not miss features, both the spectrum and the transmission filter are interpolated on one another's wavelengths.

We have to note that the denominator does not depend on the model spectrum and is a constant. To avoid its computation, the transmission curves are normalised in such a way that $\int_{\lambda_{low}}^{\lambda_{high}}T\left(\lambda\right) d\lambda=1$.

The observed fluxes are however provided in units of frequency. The flux per wavelength from the integral can easily be converted to fluxes per frequency through the use of the pivot wavelength $\lambda_{pivot}$ that is independent of the source \citep{koornneef1986a}:
\begin{equation}
f_\nu = \frac{\lambda_{pivot}^2}{c}f_\lambda,
\end{equation}
with $c$ the speed of light in the vacuum and $\lambda_{pivot}$ defined as:
\begin{equation}
  \lambda_{pivot} = \sqrt{\frac{\int_{\lambda_{low}}^{\lambda_{high}}Td\lambda}{\int_{\lambda_{low}}^{\lambda_{high}}T/\lambda^2d\lambda}}.
\end{equation}
In effect, as $\lambda_{pivot}/c$ is a constant, we also integrate it to the normalisation of $T$ and we finally rescale the filter so that the integration of a spectrum in flux per wavelength provides a flux density in mJy.

\subsubsection{Emission lines}

Photometric observations in bandpasses are fairly straightforward to deal with as can be seen above. Measuring emission lines is however more difficult. This is due in no small part to the fact that while bandpasses encompass all of the emission within their range, the continuum has to be subtracted from emission lines. The presence of sometimes strong absorption lines under emission lines makes this a difficult problem and the spectral resolution of the data leads to different strategies.

\paragraph{Low resolution data}

Low resolution spectra and narrow--band observations do not allow to measure the underlying absorption lines. A commonly used technique is to measure the level of the continuum around the emission line and therefore only take into account the flux above the inferred continuum under the line. This has a natural downside of not taking into account the loss of flux due to the underlying absorption line. To model this we have implemented special filters that naturally subtract the continuum. Each filter is made of a positive part with the transmission set to 1 on the emission part. Off the line it is set to a negative value such that $\int_{\lambda_{low}}^{\lambda_{high}}T\left(\lambda\right) d\lambda=0$. The flux $f$ is then directly obtained through $f=\int_{\lambda_{low}}^{\lambda_{high}}F_\lambda\left(\lambda\right)T\left(\lambda\right) d\lambda$ without further normalisation of the filter as we compute a flux rather than a flux density. In effect, the integration of the spectrum on the negative part of the filter will evaluate the flux provided by the continuum and will allow us to subtract it from the flux computed by integrating the positive value. Assuming that the continuum is evaluated well enough, the remainder will be the flux from the line.

For more flexibility, \texttt{CIGALE} provides these filters at different spectral resolutions for the main rest--frame optical lines. To compute the line fluxes from the spectra at any redshift the filters are stretched in wavelength by a factor $1+z$. This stretching is necessary to ensure that each filter remains centred on the line with the same resolution as in the rest frame.

\paragraph{High-resolution data}

If the data are at high resolution or if the lines have been corrected for absorption lines, the previous technique would not provide reliable results. In such a case \texttt{CIGALE} computes the emission line fluxes directly from their theoretical emission based on the nebular emission templates presented above. After normalisation to the number of Lyman continuum photons and extinction by dust, they provide the luminosity in any of the following lines that can be listed in the \texttt{pcigale.ini} file: \texttt{Ly-alpha}, \texttt{CII-133.5}, \texttt{SiIV-139.7}, \texttt{CIV-154.9}, \texttt{HeII-164.0}, \texttt{OIII-166.5}, \texttt{CIII-190.9}, \texttt{CII-232.6}, \texttt{MgII-279.8}, \texttt{OII-372.7}, \texttt{H-10}, \texttt{H-9}, \texttt{NeIII-386.9}, \texttt{HeI-388.9}, \texttt{H-epsilon}, \texttt{SII-407.0}, \texttt{H-delta}, \texttt{H-gamma}, \texttt{H-beta}, \texttt{OIII-495.9}, \texttt{OIII-500.7}, \texttt{OI-630.0}, \texttt{NII-654.8}, \texttt{H-alpha}, \texttt{NII-658.4}, \texttt{SII-671.6}, and \texttt{SII-673.1}.

\subsection{\texttt{savefluxes} module\label{ssec:savefluxes}}

Our ability to compare observations to theoretical models is key to constraining models of galaxy evolution. This can take various forms, from simple colour--colour plots to the computation of the SED of galaxies in numerical simulations and semi--analytic models. The \texttt{savefluxes} module has been designed for this kind of application: it aims at computing and saving the spectra and the properties of arbitrary theoretical galaxies. In practice the steps taken are the following.
\begin{enumerate}
\item From the list of parameters of each SED creation module given in the configuration file (see e.g. Appendix \ref{sec:pcigale.ini} ), \texttt{savefluxes} determines the complete list of parameters for each model to be computed. This essentially consists in finding all the possible combinations of parameters, creating the equivalent of an $n$--dimensional grid, with each dimension corresponding to an individual parameter. Alternatively, the parameters for each model can be explicitly provided in a file (one line per SED and one column per parameter). The former approach is useful to compute a systematic grid of theoretical models whereas the latter is more adapted for galaxies from simulations whose properties do not follow a grid.
\item For each model, the spectrum is computed, and its physical properties (both input and derived) and fluxes in passbands (which can be narrow as well as broad) are stored in memory. Optionally, the full spectrum along with the individual components (stellar populations, nebular emission, dust emission, etc.) and the SFH are saved to disk as FITS tables.
\item The integrated fluxes and the physical properties of all the models are saved to disk both as \texttt{ASCII} and \texttt{FITS} tables.
\end{enumerate}

\subsection{\texttt{pdf\_analysis} module\label{ssec:analysis}}

The SEDs of galaxies contain a treasure trove of information on their physical properties, which we need to access to understand how they form and evolve. To do so, the simplest and probably most common method consists in fitting the observed SED of a galaxy with a set of models and selecting the best-fitting one. Unfortunately, this method suffers from severe drawbacks. First of all, it ignores the degeneracies one can encounter. Models with almost equally good fits can have very different properties. As such the properties corresponding to the best fit are not necessarily representative of the true properties of the object. A related issue revolves around the estimation of the uncertainties on the physical properties. The best fit in itself does not provide information on the uncertainties. Methods such as bootstrap can be applied at the cost of repeating the fitting procedure numerous times.

A technique that has become increasingly popular over the past decade to address these issues is to rely on the goodness of fit of all the models rather than just the best--fitting model. This is generally done through the likelihood. In this case each model in the grid of models (the priors) will have an associated likelihood taken as $\exp\left(-\chi^2/2\right)$. These likelihoods can then be used as weights to estimate both the physical parameters (the likelihood--weighted means of the physical parameters) and the related uncertainties (the likelihood--weighted standard deviations of the physical parameters). This method works well when the probability distribution function is well behaved (e.g. a single peak). For more difficult cases, the marginalised probability distribution function (PDF) provides the full information to estimate the physical properties.

With either method, an important difference between various fitting codes is the algorithm to sample the priors. Two main strategies are commonly considered. Some codes use a Monte--Carlo Markov Chain (MCMC) method (or some variant of it). This is especially the case when the dimension of the problem is very large, for instance when considering non--parametric SFH (the SFH does not follow any given functional form but rather the SFR is free at every single time step) to fit spectra. While the evident upside is that it allows a large volume of priors to be  explored, the sampling can be sparse, with the risk of missing some high-likelihood regions. In addition, the SEDs have to be recomputed (or at least reinterpolated on a grid of precomputed priors) at each step of the exploration of the parameter space and this for each object, which may require particularly long computing times for large catalogues.

An alternative method that we adopt in \texttt{CIGALE} is to rely on a fixed grid of models. The main advantage is that the models need to be computed only once for all the objects. Because of this, numerous optimisations can be applied to compute the grid of models and to fit them to the data. The main downside is that it can be somewhat memory intensive. To get good results, the grid of models needs to be reasonably well sampled. At the same time, for the process to be computationally efficient, the grid along with the associated physical properties to estimate need to remain in memory. The amount of memory that is required primarily depends on 1) the number of models to compute, 2) the number of bands and physical properties to fit, and 3) the number of physical properties to estimate.

In the \texttt{pdf\_analysis} module we implemented the estimation of the physical properties from likelihood--weighted parameters on a fixed grid of models. The computation of the grid of SEDs and associated physical properties follows the same steps as for the \texttt{savefluxes} module described in Sect.~\ref{ssec:savefluxes}, except for minor differences of no consequence here. Once the grid of models has been computed, the high level algorithm to estimate the physical properties is the following.
\begin{enumerate}
 \item From the complete set $S_0$ of models, selection of the subset $S_1$ of models closest to the rounded redshift of the analysed object. By default, the rounding is to two decimal places, but this can be user--defined.
 \item Computation of the multiplicative factors (Eq.~\ref{eq:scaling}) to scale the $S_1$ models to the observations.
 \item Computation of the $\chi^2$ (Eq.~\ref{eq:chi2}) between all the $S_1$ models and the observations.
 \item Computation of the likelihood $\exp\left(-\chi^2/2\right)$ for the $S_1$ models.
 \item Estimation of each physical property along with the associated uncertainty as the likelihood--weighted mean and standard deviation of the $S_1$ models.
 \item Save the estimates and the uncertainties on the physical properties along with the fluxes and the physical properties of the best fitting model.
 \item Optionally, save the spectrum of the best fit with the individual components (stellar populations, nebular emission, dust emission, etc.), the $\chi^2$ distribution, and the marginalised PDF for each analysed variable.
\end{enumerate}

Several of the key steps here require further explanation. First of all, as mentioned earlier the SFH is normalised so that its integral is 1~M$_\odot$ (when stellar populations are available, or to 1~W otherwise for the dust emission). This means that in order to obtain the extensive physical properties such as masses or luminosities, we need to rescale the models by a factor $\alpha$ before computing the $\chi^2$. This can be done analytically:
\begin{equation}
 \alpha=\frac{\sum_{i} f_i\times m_i/\sigma_i^2}{\sum_{i} m_i^2/\sigma_i^2} + \frac{\sum_{j} f_j\times m_j/\sigma_j^2}{\sum_{j} m_j^2/\sigma_j^2},\label{eq:scaling}
\end{equation}
with $f_i$ and $m_i$ being the observed and model fluxes, $f_j$ and $m_j$ the observed and model extensive physical properties, and $\sigma$ being the corresponding observational uncertainties. Then the computation of the $\chi^2$ is straightforward:
\begin{equation}
 \chi^2=\sum_i\left(\frac{f_i-\alpha\times m_i}{\sigma_i}\right)^2+\sum_j\left(\frac{f_j-\alpha\times m_j}{\sigma_j}\right)^2+\sum_k\left(\frac{f_k-m_k}{\sigma_k}\right)^2,\label{eq:chi2}
\end{equation}
with $f_k$ and $m_k$ being the observed and model intensive physical properties. This means that the stellar mass (or the dust luminosity when there is no stellar population included in the model) is not a free parameter even though it is technically possible to treat it as such. This would greatly expand the size of the parameter space by adding an extra dimension, while slowing the computation of the grid and the analysis by a similar amount, and degrade the accuracy of the estimation of the physical properties.

Optionally, \texttt{CIGALE} can also handle fluxes for which only upper limits have been determined. We adopt the method presented in Appendix A2 of \cite{sawicki2012a}. This affects the aforementioned computing steps in several ways. First, the computation of the $\chi^2$ is divided between regular quantities (first three terms corresponding to Eq.~\ref{eq:chi2}) and those with only an upper limit (last three terms):

\begin{equation}
\begin{aligned}
 \chi^2&=\sum_i\left(\frac{f_i-\alpha\times m_i}{\sigma_i}\right)^2
        +\sum_j\left(\frac{f_j-\alpha\times m_j}{\sigma_j}\right)^2
        +\sum_k\left(\frac{f_k-             m_k}{\sigma_k}\right)^2\\
       & -2\sum_i\ln\left(\frac{1}{2}\left[1+\textrm{erf}\left(\frac{f_{ul,i}-\alpha\times m_i}{\sqrt{2}\sigma_i}\right)\right]\right)\\
       & -2\sum_j\ln\left(\frac{1}{2}\left[1+\textrm{erf}\left(\frac{f_{ul,j}-\alpha\times m_j}{\sqrt{2}\sigma_j}\right)\right]\right)\\
       & -2\sum_k\ln\left(\frac{1}{2}\left[1+\textrm{erf}\left(\frac{f_{ul,k}-             m_k}{\sqrt{2}\sigma_k}\right)\right]\right),\label{eq:chi2-ul}
\end{aligned}
\end{equation}
with `erf' being the error function, $f_{ul}$ the flux upper limit, and the indices $i$, $j$, and $k$ indicating respectively the fluxes, extensive properties, and intensive properties. See equations A6 to A10 of \cite{sawicki2012a} for a full derivation\footnote{We should note however that equation A10 from \cite{sawicki2012a}, corresponding to Eq.~\ref{eq:chi2-ul} here, contains a mistake, which we have corrected for.}. The main difficulty is to determine $\alpha$. For this we have to numerically solve $\partial\chi^2/\partial\alpha=0$ for every model, which is equivalent to solving Eq.~A11 from \cite{sawicki2012a}:
\begin{equation}
\begin{aligned}
\sum_i\left(\frac{f_i-\alpha\times m_i}{\sigma_i}\right)\times\left(\frac{m_i}{\sigma_i}\right)
+\sum_j\left(\frac{f_j-\alpha\times m_j}{\sigma_j}\right)\times\left(\frac{m_j}{\sigma_j}\right)&\\
-\sqrt{\frac{2}{\pi}}\sum_i\frac{m_i\times\exp\left(-\left[\left(f_{ul,i}-\alpha\times m_i\right)/\sqrt{2}\sigma_i\right]^2\right)}{\sigma_i\left[1+\textrm{erf}\left(\left(f_{ul,i}-\alpha\times m_i\right)/\sqrt{2}\sigma_i\right)\right]}&\\
-\sqrt{\frac{2}{\pi}}\sum_j\frac{m_j\times\exp\left(-\left[\left(f_{ul,j}-\alpha\times m_j\right)/\sqrt{2}\sigma_j\right]^2\right)}{\sigma_j\left[1+\textrm{erf}\left(\left(f_{ul,j}-\alpha\times m_j\right)/\sqrt{2}\sigma_j\right)\right]}&=0.
\end{aligned}
\end{equation}
We do so using the \texttt{scipy.optimize.root} root-finding method. Once the $\chi^2$ are computed, the subsequent steps no longer depend on the presence or absence of upper limits.

We have to note that objects in a catalogue do not all need to be fitted with the same set of data. \texttt{CIGALE} will automatically only consider the available data for a given object. The lack of certain data for some targets has a direct impact on some of the aforementioned computation steps. For steps 2 and 3, we simply do not include the data that have been marked as invalid in the input file (value lower than 0 or set to ``nan'', by convention) in the computation of $\alpha$ and $\chi^2$. For step 4, we take into account the number of bands to compute the probability that a model will reproduce the observations.

Another feature of this module is the possibility to assess whether or not physical properties can actually be estimated in a reliable way through the analysis of a mock catalogue. The idea is to compare the physical properties of the mock catalogue, which are known exactly, to the estimates from the analysis of the likelihood distribution. To build the mock catalogue we consider the best fit for each object. We then modify each quantity by adding a value taken from a Gaussian distribution with the same standard deviation as the observation. This mock catalogue is then analysed in the exact same way as the original observations. Physical properties for which the exact and estimated values are similar can be estimated reliably.

Applications of the \texttt{pdf\_analysis} module to a sample of representative galaxies is presented in Sect.~\ref{ssec:application-fit}.

\section{Examples of \texttt{CIGALE} use cases\label{sec:versatility}}

As seen above, \texttt{CIGALE} has been designed to be flexible and versatile so that it may have various applications: estimation of the physical properties of an object from the observed SED, generation of theoretical SEDs from analytical considerations or numerical simulations, library to build new tools, or even to serve as a basis for simulating observations. Here we briefly present examples of the former three applications.

\subsection{Example of \texttt{CIGALE} as an SED generation tool}

The automated generation of SEDs for specific parameters or over different sets of models can be useful for multiple applications: quickly generate artificial observations of galaxies whose physical properties are known (e.g. from numerical simulations), compare observations with grids of models without having to resort to full-scale SED modelling of large samples of galaxies, derive theoretical relations depending on one or more parameters, and so on.

\begin{figure*}[!htbp]
 \includegraphics[width=0.333\textwidth]{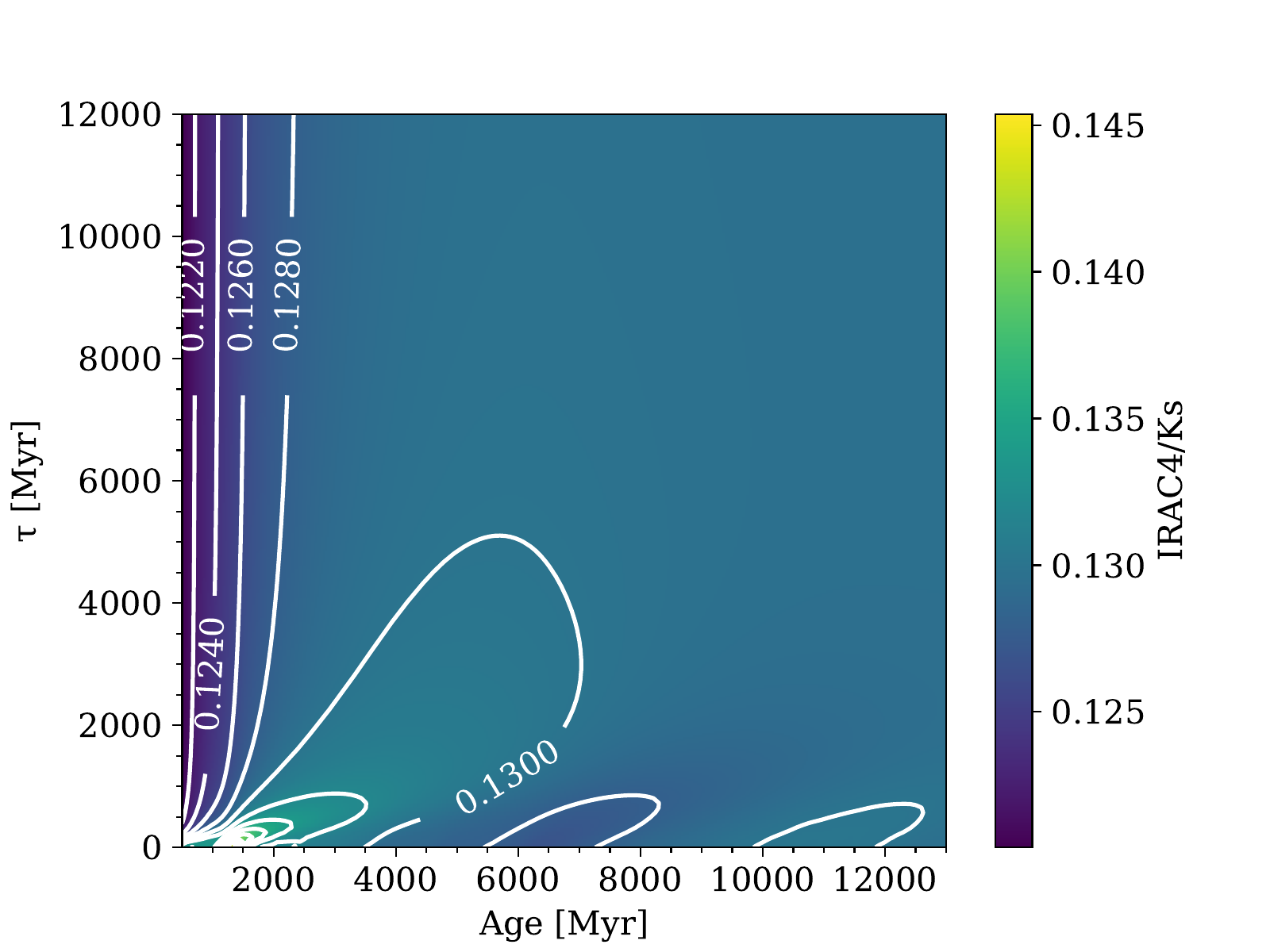}
 \includegraphics[width=0.333\textwidth]{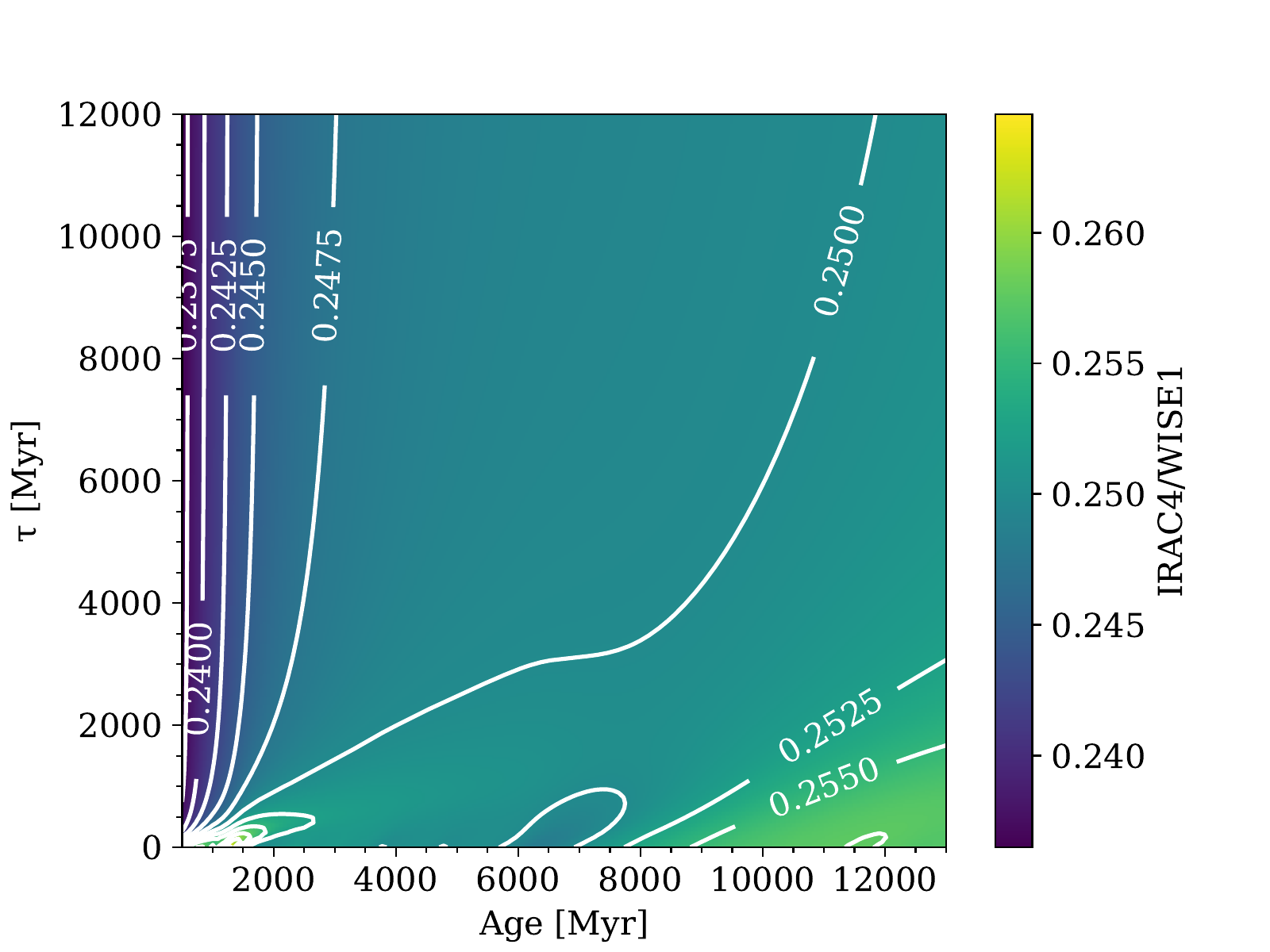}
 \includegraphics[width=0.333\textwidth]{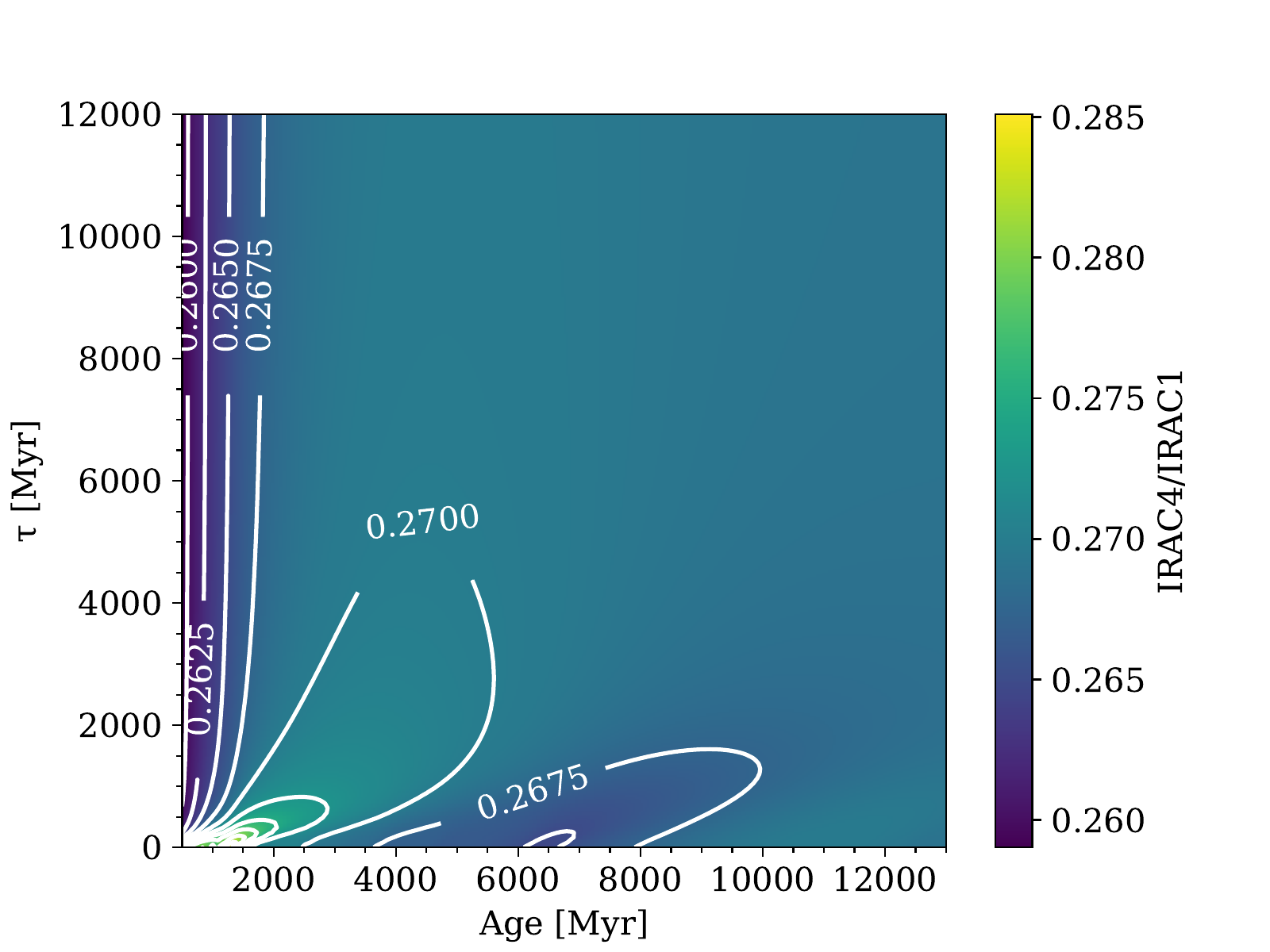}

 \includegraphics[width=0.333\textwidth]{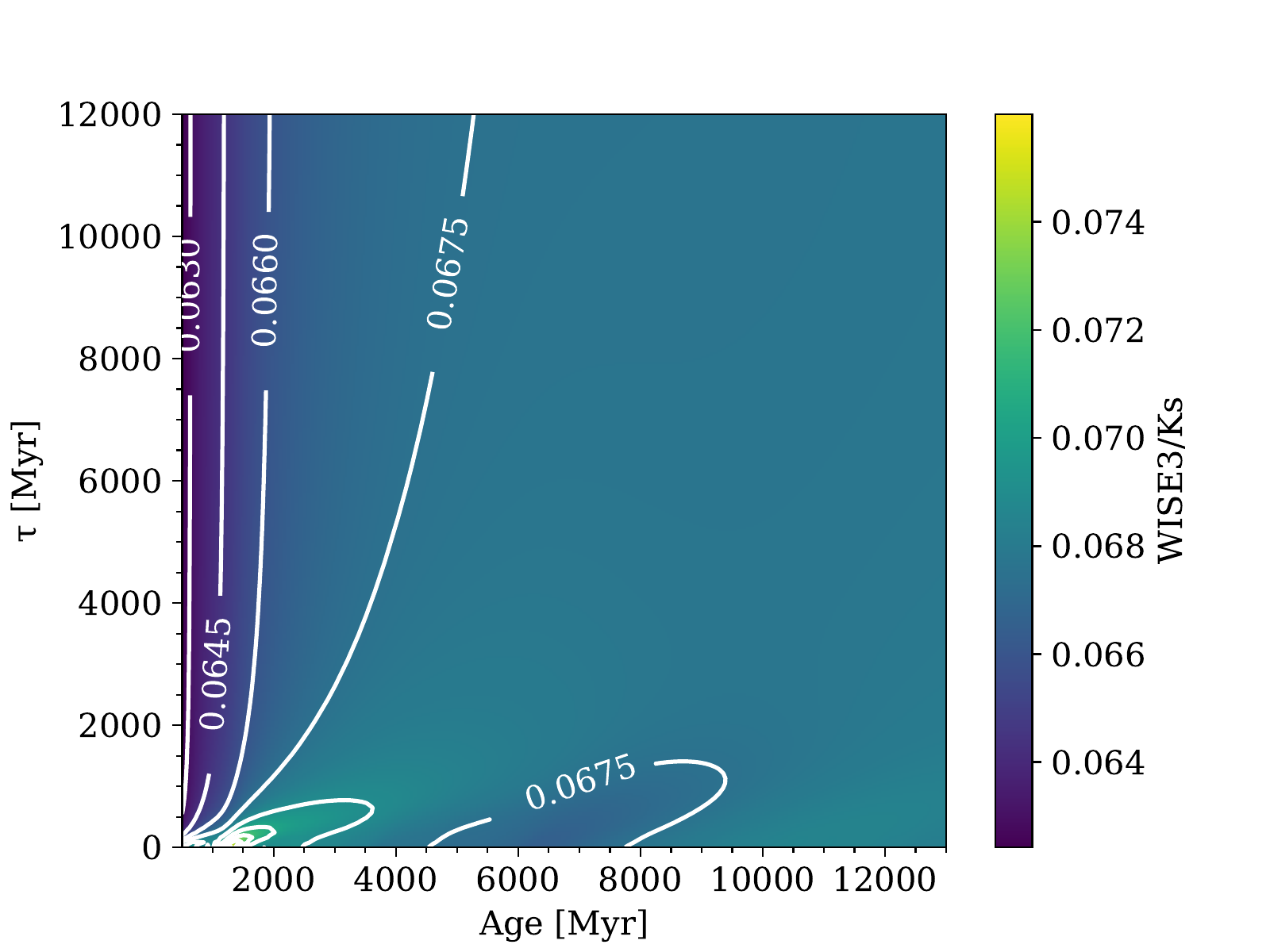}
 \includegraphics[width=0.333\textwidth]{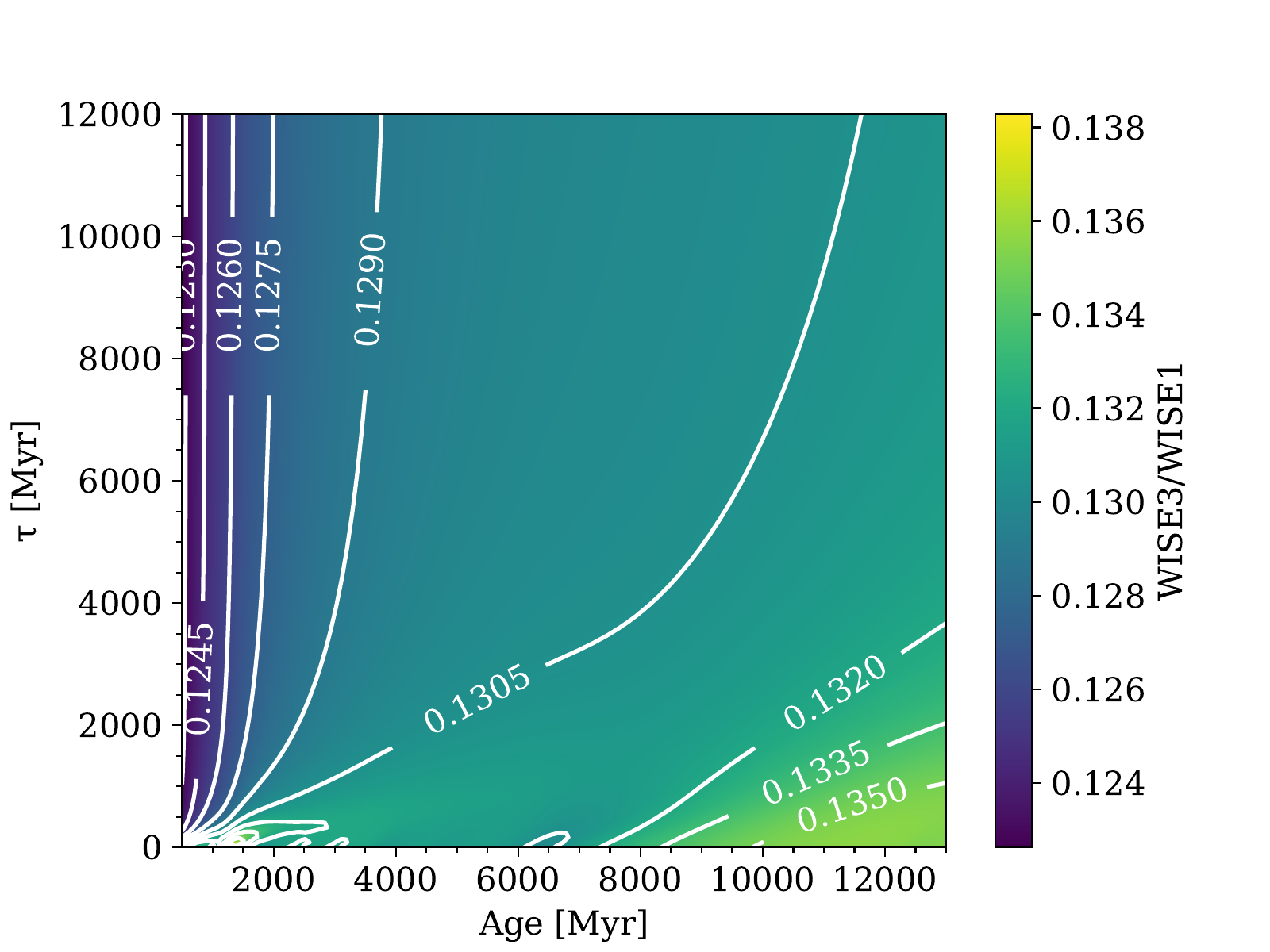}
 \includegraphics[width=0.333\textwidth]{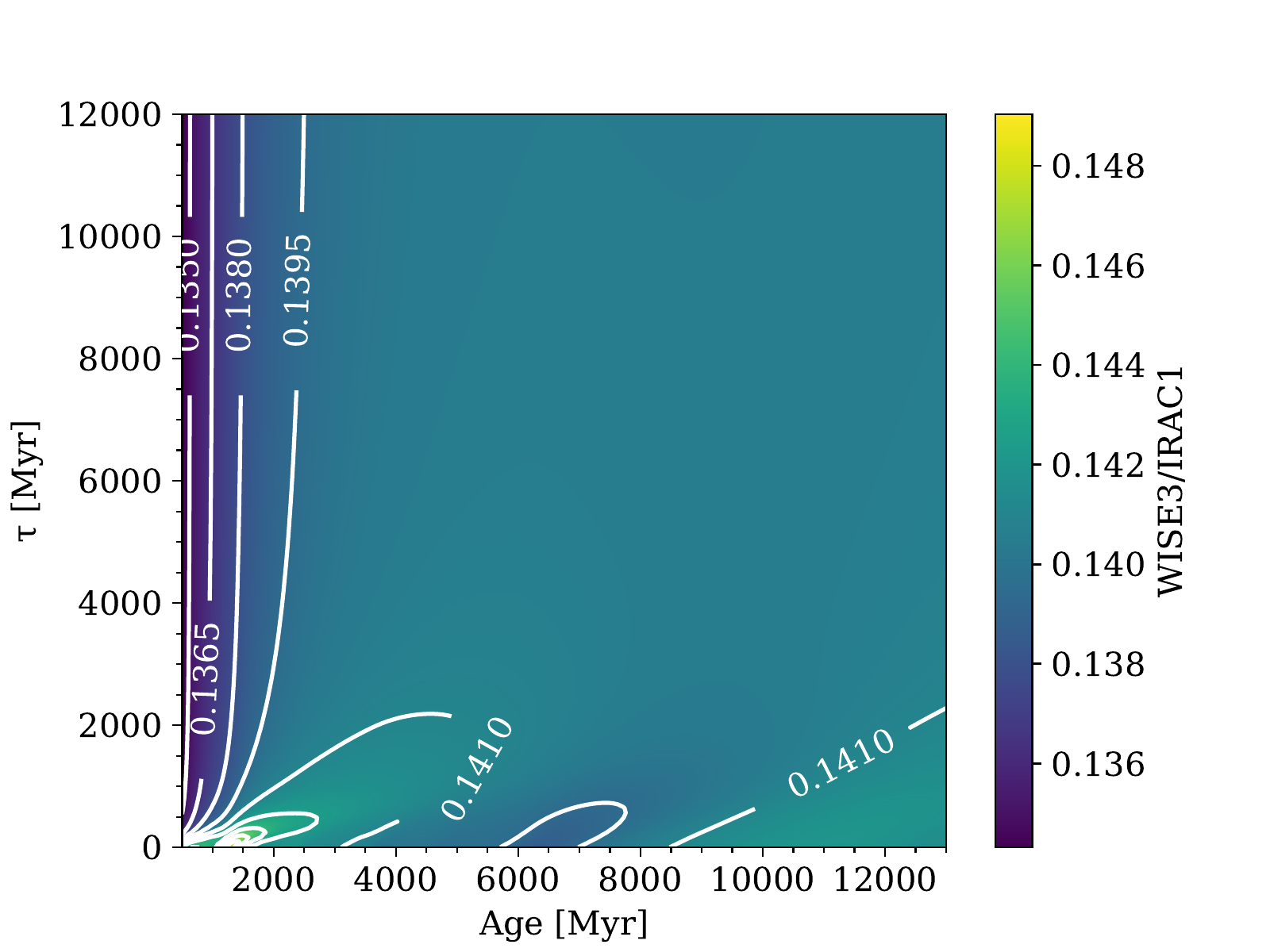}

 \includegraphics[width=0.333\textwidth]{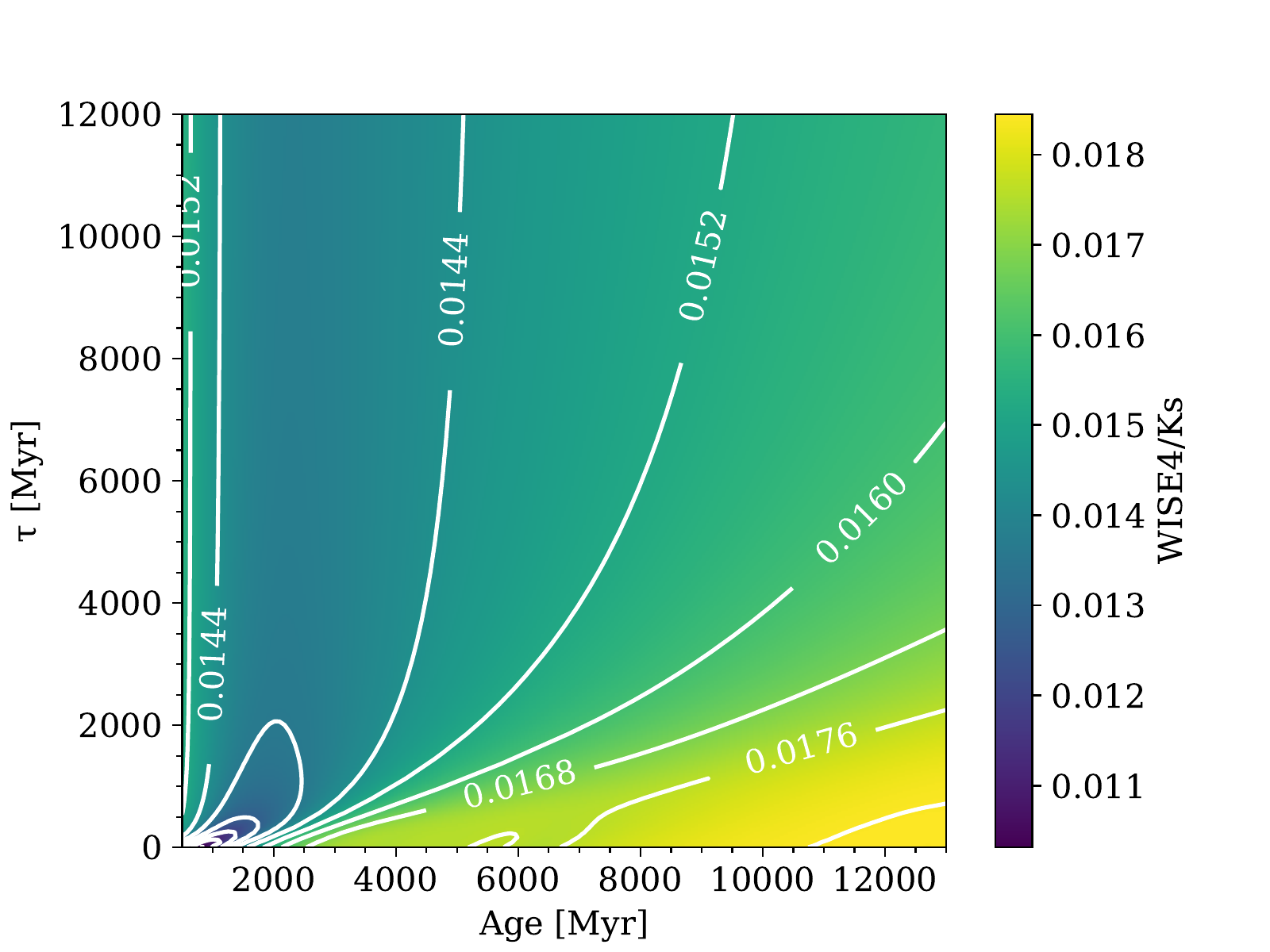}
 \includegraphics[width=0.333\textwidth]{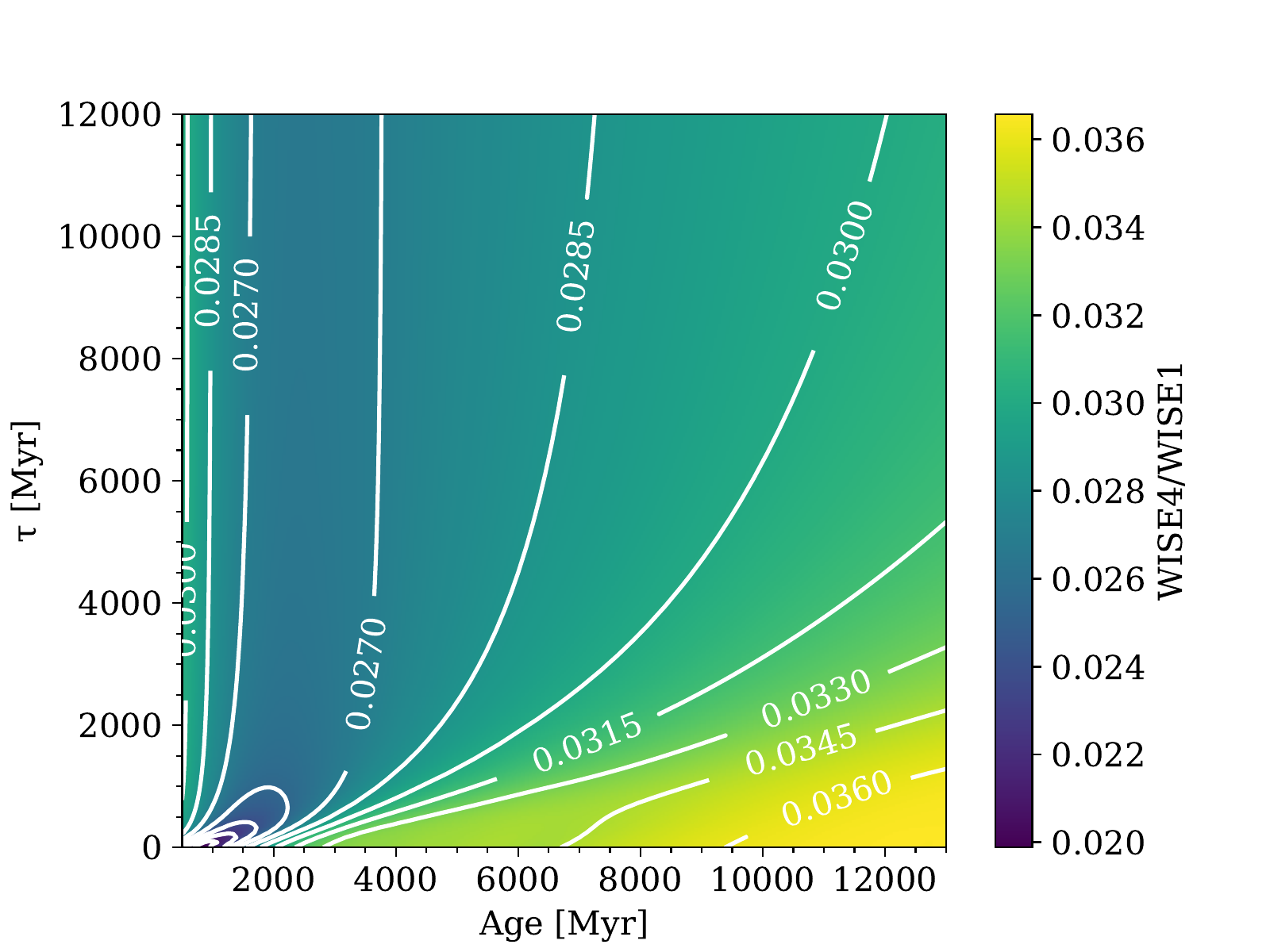}
 \includegraphics[width=0.333\textwidth]{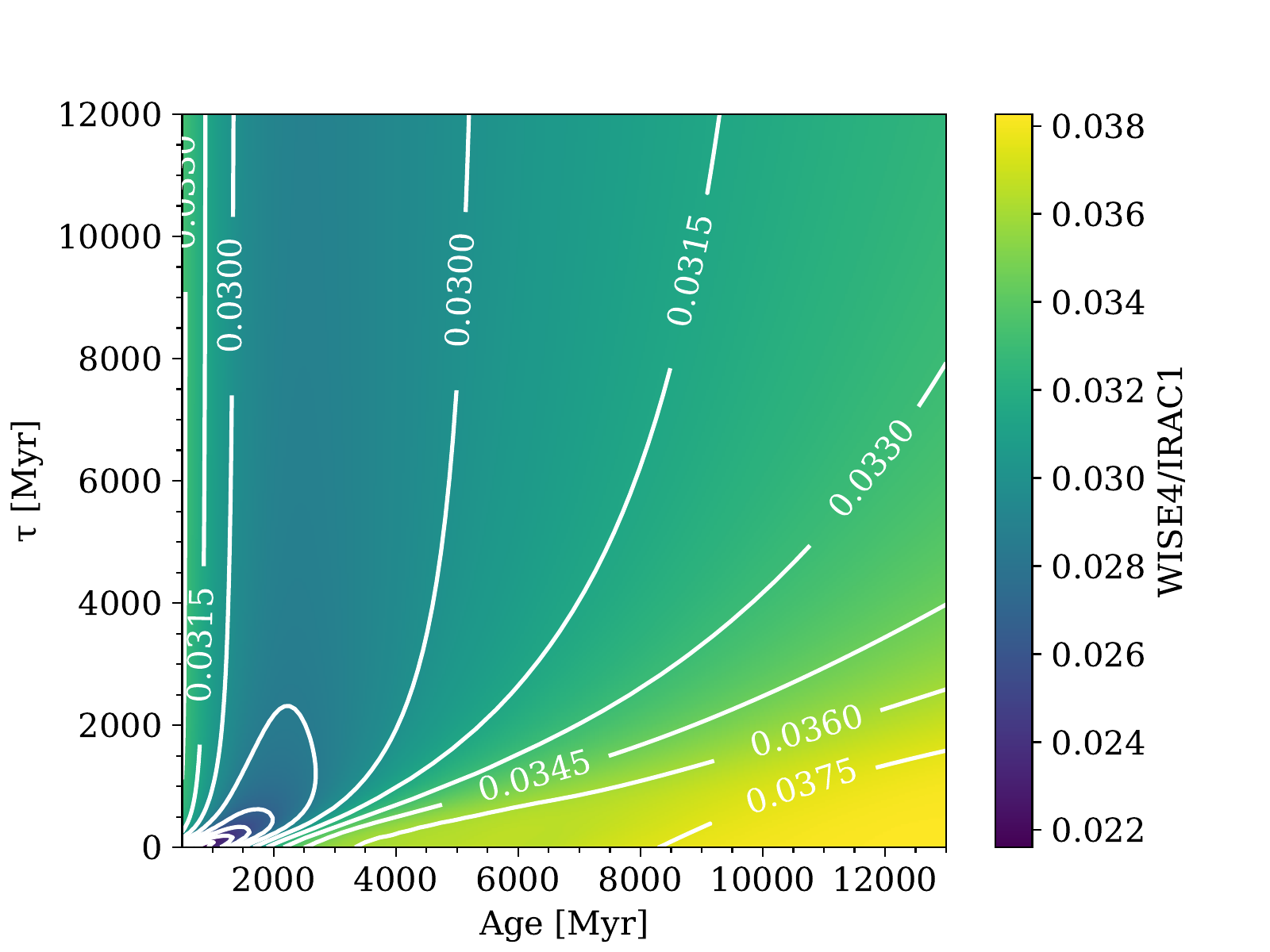}

 \includegraphics[width=0.333\textwidth]{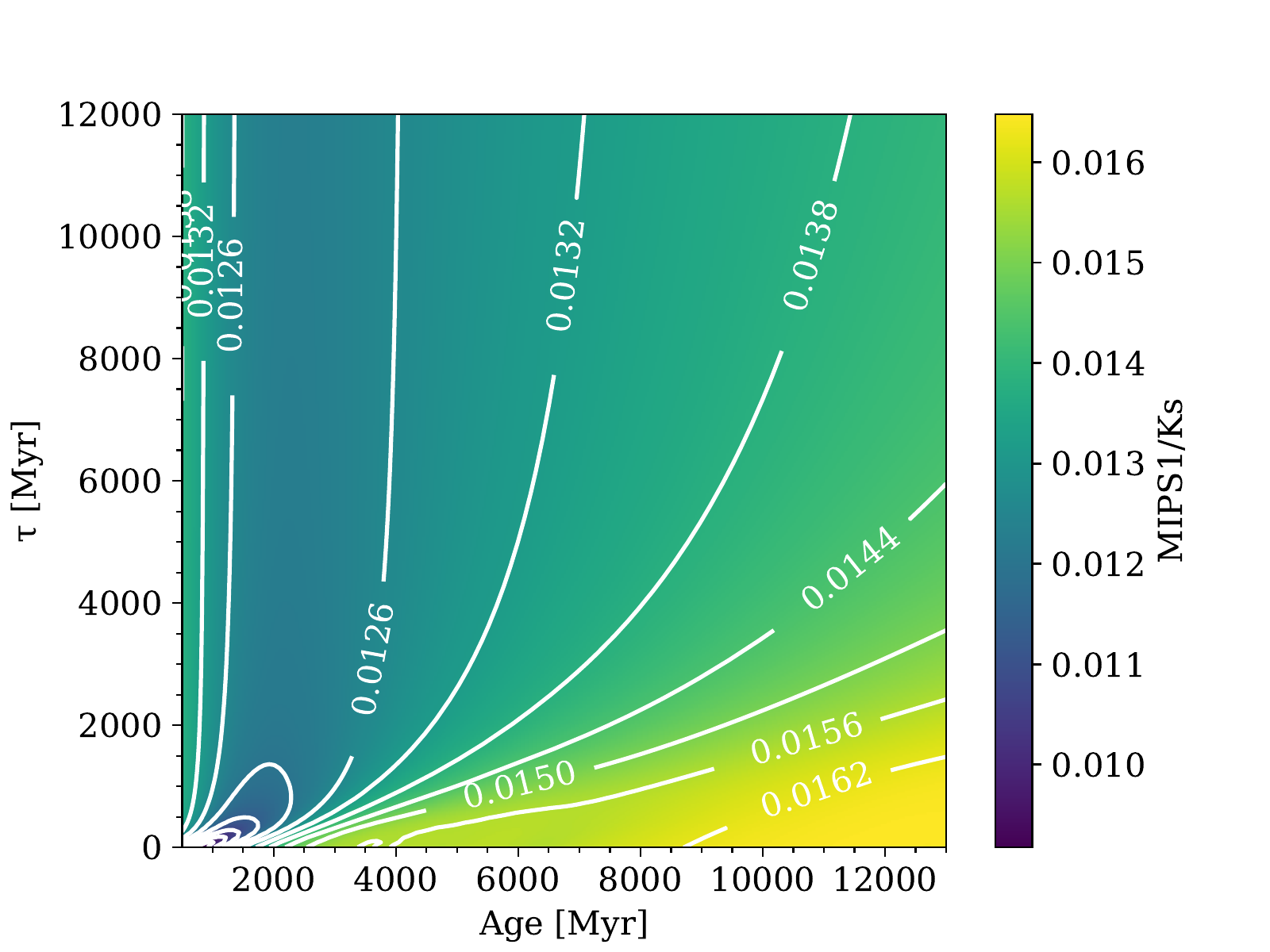}
 \includegraphics[width=0.333\textwidth]{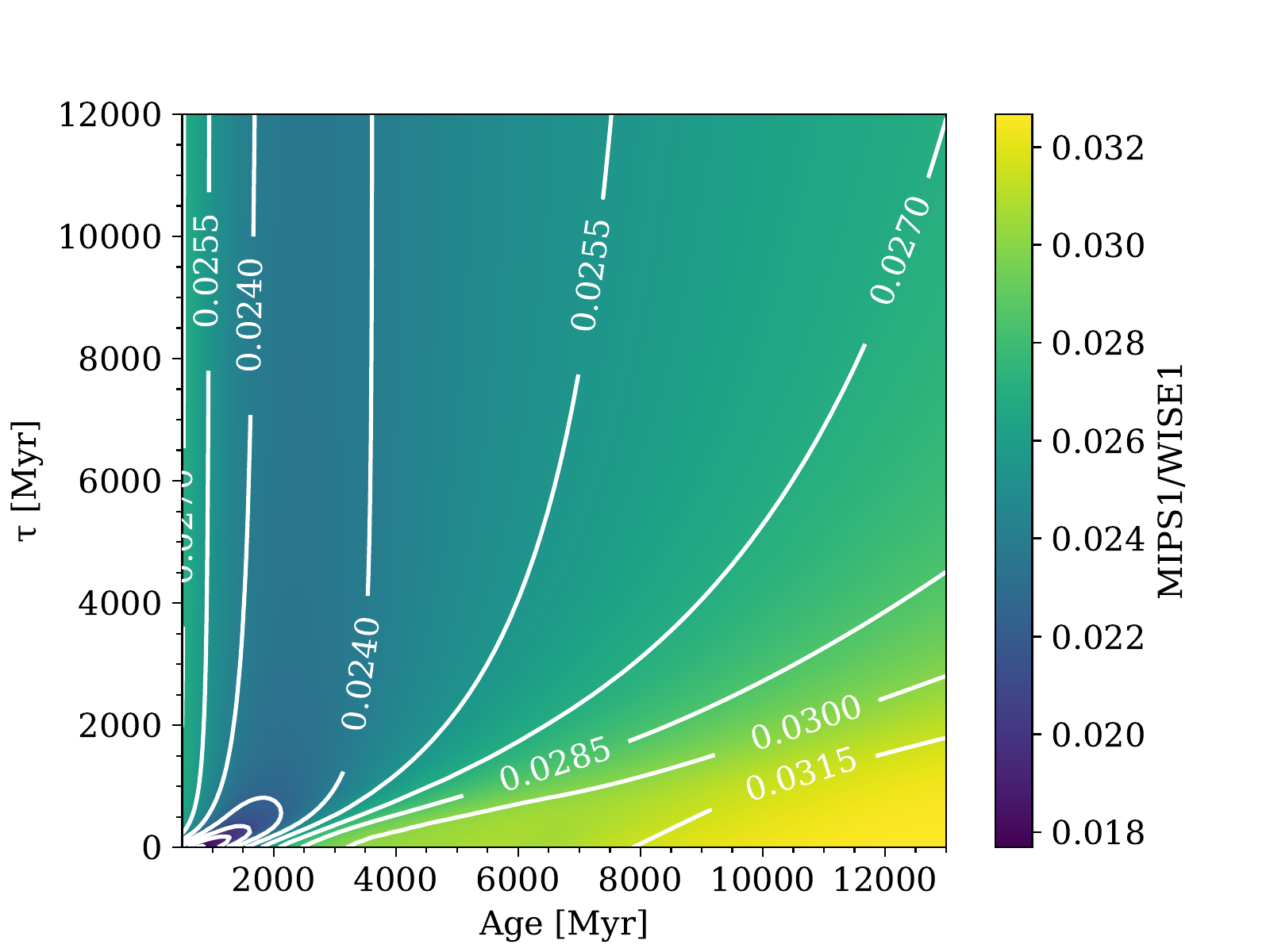}
 \includegraphics[width=0.333\textwidth]{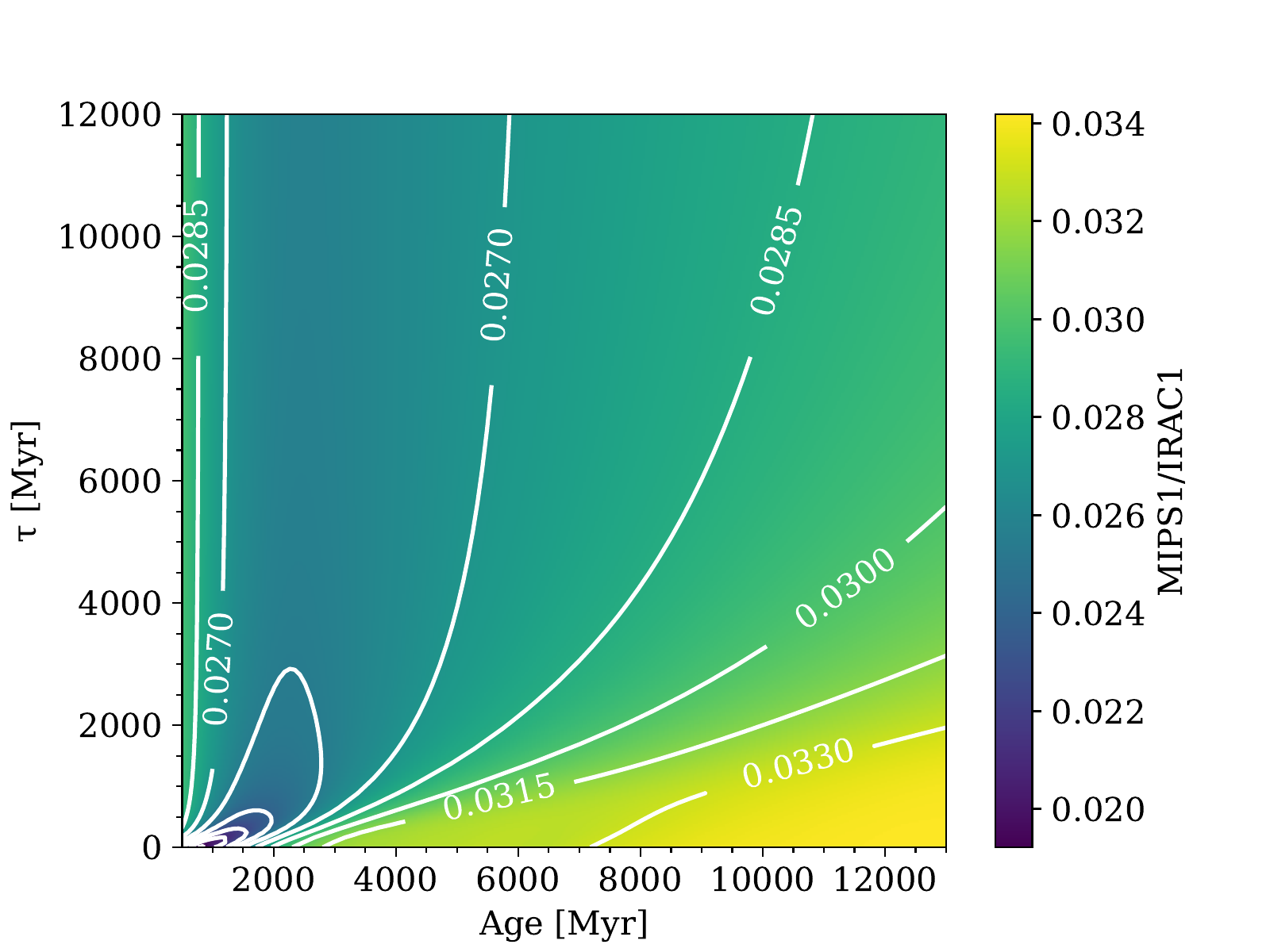}

 \caption{Comparison of the relative fluxes between the stellar fluxes in mid-IR dust--dominated bands (from top to bottom: IRAC 8~$\mu$m, WISE 12~$\mu$m, WISE 22~$\mu$m, and MIPS 24~$\mu$m) and NIR dust--free bands (from left to right: Ks 2.2~$\mu$m, WISE 3.4~$\mu$m, and IRAC 3.6~$\mu$m) bands depending on the age of the galaxy ($x$--axis) and the $\tau$ constant of a ``delayed'' SFH ($y$--axis). The colour indicates the value of the ratio of the fluxes according to the colour bar to the right of each plot, with white lines indicating isocontours. A total of 22500 models were computed representing the combination of 150 parameters on each axis. The \cite{bruzual2003a} models were adopted, assuming a \cite{salpeter1955a} IMF and a metallicity $Z=0.02$.\label{fig:comp-stellar-contrib}}
\end{figure*}

An example of the former use case with \texttt{CIGALE} can be seen in \cite{boquien2014a}. They used a set of SFHs from idealised galaxy simulations at $1\le z\le2$ to simulate observations in star-formation-tracing bands and examine the impact of the SFH on the measure of the SFR. Another example can be found in \cite{alvarez2016a}, where they defined Lyman--break galaxy selection criteria in the redshift range $2.5<z<3.5$ using \texttt{CIGALE}. Here, for the purpose of illustrating the generation of theoretical models, we focus on the latter case. We examine the question of the stellar contribution in the mid-IR. Indeed, even though mid-IR emission in galaxies is often used as a tracer of star--formation \citep[e.g.,][]{calzetti2007a}, the stellar contamination can be important. To correct for it the standard method consists in extrapolating the NIR flux to the wavelength of interest \citep[e.g.][]{helou2004a,ciesla2014a}, exploiting the Rayleigh--Jeans regime of the emission. In practice such methods are calibrated by computing the ratio between near-- and mid--IR fluxes, so that with one NIR band that is free of dust, one can estimate the stellar emission in the mid-IR. In Fig.~\ref{fig:comp-stellar-contrib} we show how mid--to--near-IR ratios vary depending on the age and timescale of a ``delayed'' SFH using the \cite{bruzual2003a} models assuming a \cite{salpeter1955a} IMF and a metallicity of $Z=0.02$.
The simple configuration file to run this example is shown in Appendix~\ref{sec:pcigale.ini}. It allows to generate a total of 22500 models: the combination of 150 values for the age and 150 values for the timescale $\tau$.

The plots show interesting variations depending on the SFH and the selected NIR and mid--IR wavelengths. Unsurprisingly, when the wavelength difference is small, the variation in the mid--to--near-IR flux ratio is limited to typically less than 20\% in the worst cases. In other words, as was noted by \cite{helou2004a}, there is only a weak dependence on the SFH. The variation is larger when considering longer wavelengths, such as WISE 22~$\mu$m or MIPS 24~$\mu$m. In this case there can be variations of up to a factor two with a strong dependence on the SFH. The most important difference depends on the age of early--type galaxies (small values for $\tau$), with $\sim1$~Gyr-old galaxies having much bluer colours. When star formation is still on--going (larger values for $\tau$), the colours show a much weaker dependence on star formation and a constant colour can easily be adopted for late--type galaxies.

This simple example shows how easy it is with \texttt{CIGALE} to explore theoretical grids of models to understand the effects of specific assumptions. We showed here the influence of the age and the timescale of a ``delayed'' SFH, but similar studies can be done for different parametrisations of the SFH and also with different parameters (IMF, stellar population models, metallicity, presence of dust, presence of an active nucleus, etc.). Beyond the creation of grids of models of a broad range of purposes, the generation of models can also be used in connection to numerical simulations. For instance \cite{boquien2014a} used \texttt{CIGALE} to compute the emission of galaxies in star formation tracing bands with respect to time from 23 high--resolution numerical simulations. This allowed them to investigate the effect of the SFH on the determination of the SFR from standard methods and provide new SFR estimators on different timescales. In a similar way, \cite{ciesla2015a} coupled \texttt{CIGALE} with semi--analytic models to investigate the ability of SED modelling to disentangle the emission of active nuclei and measure their properties.

\subsection{Example of \texttt{CIGALE} as a library}

\begin{figure}[!htbp]
 \includegraphics[width=\columnwidth]{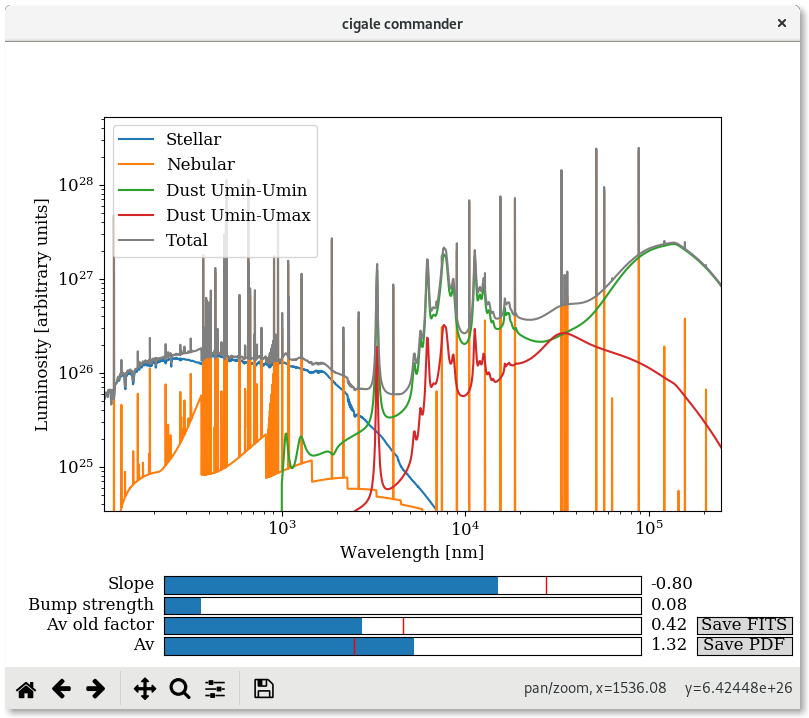}
 \caption{Screenshot of a simple application built using \texttt{CIGALE} as a library. It is designed to interactively explore  the effect of dust attenuation on the FUV--to-far-IR spectrum of a star--forming galaxy. Here the key parameters for the attenuation can easily be changed with sliders, allowing for a rapid examination of the impact of each parameter. The script is approximately only 150 lines long. It transparently uses \texttt{CIGALE} modules to build the SED from user--provided parameters and \texttt{matplotlib} to display the plot and handle the sliders.\label{fig:pcigale-commander}}
\end{figure}

The modular and flexible design of \texttt{CIGALE} enables its use not just as a stand-alone package, but also as a library to build new applications well beyond what it was initially conceived for. As a simple example of such a case, we have created a simple pedagogical tool to interactively explore the effect of dust attenuation on the FUV--to-far-IR spectrum of a star--forming galaxy. We show a screen capture of \texttt{cigale commander} in Fig.~\ref{fig:pcigale-commander}.
While this example is limited to attenuation curves, it could be extended to all parameters of all modules. Not all applications need to be interactive though. It is also possible to use \texttt{CIGALE} as a database to easily access the base models (single stellar populations, dust emission templates, etc.) in a uniform way without having to write additional code to read the original models that come in different specific formats.

\subsection{Example of \texttt{CIGALE} to estimate the physical properties of star--forming galaxies\label{ssec:application-fit}}

Measuring the physical properties of galaxies is probably one of the most common uses of SED-modelling codes such as \texttt{CIGALE} and the literature is rich with examples covering a broad range of questions at all redshifts. Various articles have already covered the subject of the reliability of \texttt{CIGALE} to retrieve a number of the intrinsic physical properties of galaxies \citep[e.g.][]{boquien2012a,boquien2016a,buat2014a,lofaro2017a}. As the outcome of such studies naturally depends on the quality and breadth of the data available along with the sampled priors, we refer to these articles for in--depth analyses. Another interesting question is that of how the estimates of the physical properties from different codes compare. To answer this question, Hunt et al. (submitted) modelled the SED of the galaxies of the KINGFISH sample \citep{kennicutt2011a} comparing results from \texttt{CIGALE}, \texttt{MAGPHYS} \citep{dacunha2008a}, and the latest iteration of \texttt{grasil} \citep{silva1998a}. Examining the average SFR over 100 Myr, stellar mass, the FUV attenuation, and the dust luminosity, \texttt{CIGALE} provides excellent consistency with other codes. In another recent effort, this time more geared towards higher redshift samples, \texttt{CIGALE} equally shows excellent performance (Pacifici et al., in prep.).

\begin{figure}[!htbp]
 \includegraphics[width=\columnwidth]{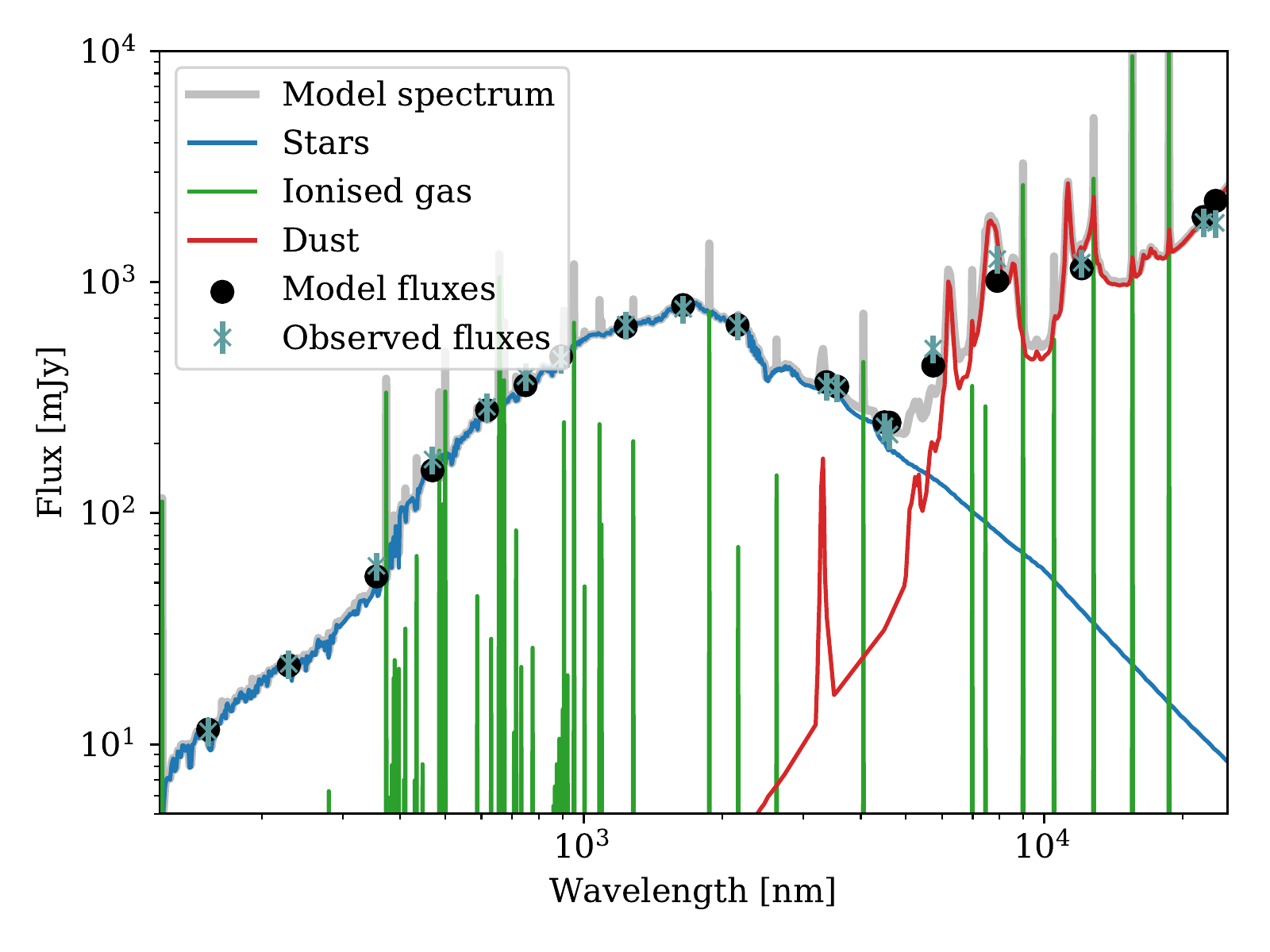}
 \caption{Best-fitting model for the spiral galaxy NGC~4321 (grey) located in the nearby Virgo cluster, showing the total stellar (blue, the dust attenuation has already been accounted for), nebular (green, also including dust attenuation), and dust (red) emission. The model fluxes in passbands computed using Eq.~\ref{eqn:flux-bandpass} are indicated with black circles. These fluxes were then fitted to the observations (turquoise cross with the uncertainties indicated with the vertical lines), yielding a final reduced $\chi^2\simeq 0.5$.\label{fig:4321}}
\end{figure}

As a simple illustration of the capabilities of \texttt{CIGALE}, we present the results of the modelling of the star--forming galaxies from the UV to mid--IR SED atlas of \cite{brown2014a}. This sample is especially interesting as it provides a set of carefully vetted photometric data, which is ideal for SED modelling. We select a subsample of 78 star--forming galaxies which 1) have both FUV and NUV flux, and 2) are classified either as Sa or later-type or as peculiar. We exclude galaxies that are classified as purely AGN by \cite{brown2014a} as this may contaminate in an appreciable way both the UV and/or the mid--IR ends of the spectra, significantly affecting the results\footnote{For simplicity here we do not include AGN models. We refer to \cite{ciesla2015a} for a presentation of the capabilities of \texttt{CIGALE} regarding AGNs.}. We model these galaxies with a representative set of modules as described in Table~\ref{tab:parameters}, leading to a modest grid of 8164800 models.
\begin{table*}
 \centering
 \begin{tabular}{lll}
  \hline\hline
  Module                               &Parameter                          &Value\\\hline
  \texttt{sfhdelayed}                  &\texttt{tau\_main} ($10^6$ years)  &1, 500, 1000, 2000, 3000, 4000, 5000, 6000, 7000, 8000\\
                                       &\texttt{age\_main} ($10^6$ years)  &13000\\
                                       &\texttt{tau\_burst} ($10^6$ years) &$10^9$\\
                                       &\texttt{age\_burst} ($10^6$ years) &5, 10, 25, 50, 100, 200, 350, 500, 750, 1000\\
                                       &\texttt{f\_burst}                  &0, 0.0001, 0.0005, 0.001, 0.005, 0.01, 0.05, 0.1, 0.25\\\hline
  \texttt{bc03}                        &\texttt{imf}                       &1 (Chabrier)\\
                                       &\texttt{metallicity}               &0.02\\\hline
  \texttt{nebular}                     &\texttt{logU}                      &$-3.0$\\
                                       &\texttt{f\_esc}                    &0.0\\
                                       &\texttt{f\_dust}                   &0.0\\
                                       &\texttt{lines\_width} (km s$^{-1}$)&300\\\hline
  \texttt{dustatt\_modified\_starburst}&\texttt{E\_BV\_nebular}       (mag)&0.005, 0.01, 0.025, 0.05, 0.075, 0.10, 0.15, 0.20, 0.25, 0.30, 0.35,\\
                                       &                                   &0.40, 0.45, 0.50, 0.55, 0.60\\
                                       &\texttt{E\_BV\_factor}             &0.25, 0.50, 0.75\\
                                       &\texttt{uv\_bump\_wavelength}  (nm)&217.5\\
                                       &\texttt{uv\_bump\_width}       (nm)&35.0\\
                                       &\texttt{uv\_bump\_amplitude}       &0.0, 1.5, 3.0 (Milky Way)\\
                                       &\texttt{powerlaw\_slope}           &$-0.5$, $-0.4$, $-0.3$, $-0.2$, $-0.1$, 0.0\\
                                       &\texttt{Ext\_law\_emission\_lines} &1 (Milky Way)\\
                                       &\texttt{Rv}                        &3.1\\
                                       &\texttt{filters}                   &FUV, V\_B90\\\hline
  \texttt{dale2014}                    &\texttt{alpha}                     &0.5, 1.0, 1.5, 2.0, 3.5, 3.0, 3.5, 4.0\\\hline
  \texttt{restframe\_parameters}       &\texttt{beta\_calz94}              &True\\
                                       &\texttt{D4000}                     &False\\
                                       &\texttt{IRX}                       &True\\
                                       &\texttt{EW\_lines}                 &500.7/1.0 \& 656.3/1.0\\
                                       &\texttt{luminosity\_filters}       &FUV \& V\_B90\\
                                       &\texttt{colours\_filters}          &FUV-NUV \& NUV-r\_prime\\
                                       \hline
  \texttt{redshifting}                 &\texttt{redshift}                  &0\\\hline
   \end{tabular}
 \caption{Modules and parameter values used to model the sample of \cite{brown2014a}. The grid of models (fluxes and physical properties) is estimated over all the possible combinations of parameters, leading to a total of 8164800 models. The corresponding \texttt{pcigale.ini} file is provided in Appendix \ref{sec:pcigale.ini}.\label{tab:parameters}}
\end{table*}
An example of a typical best fit is shown in Fig.~\ref{fig:4321}.

\begin{figure*}[!htbp]
 \includegraphics[width=0.33\textwidth]{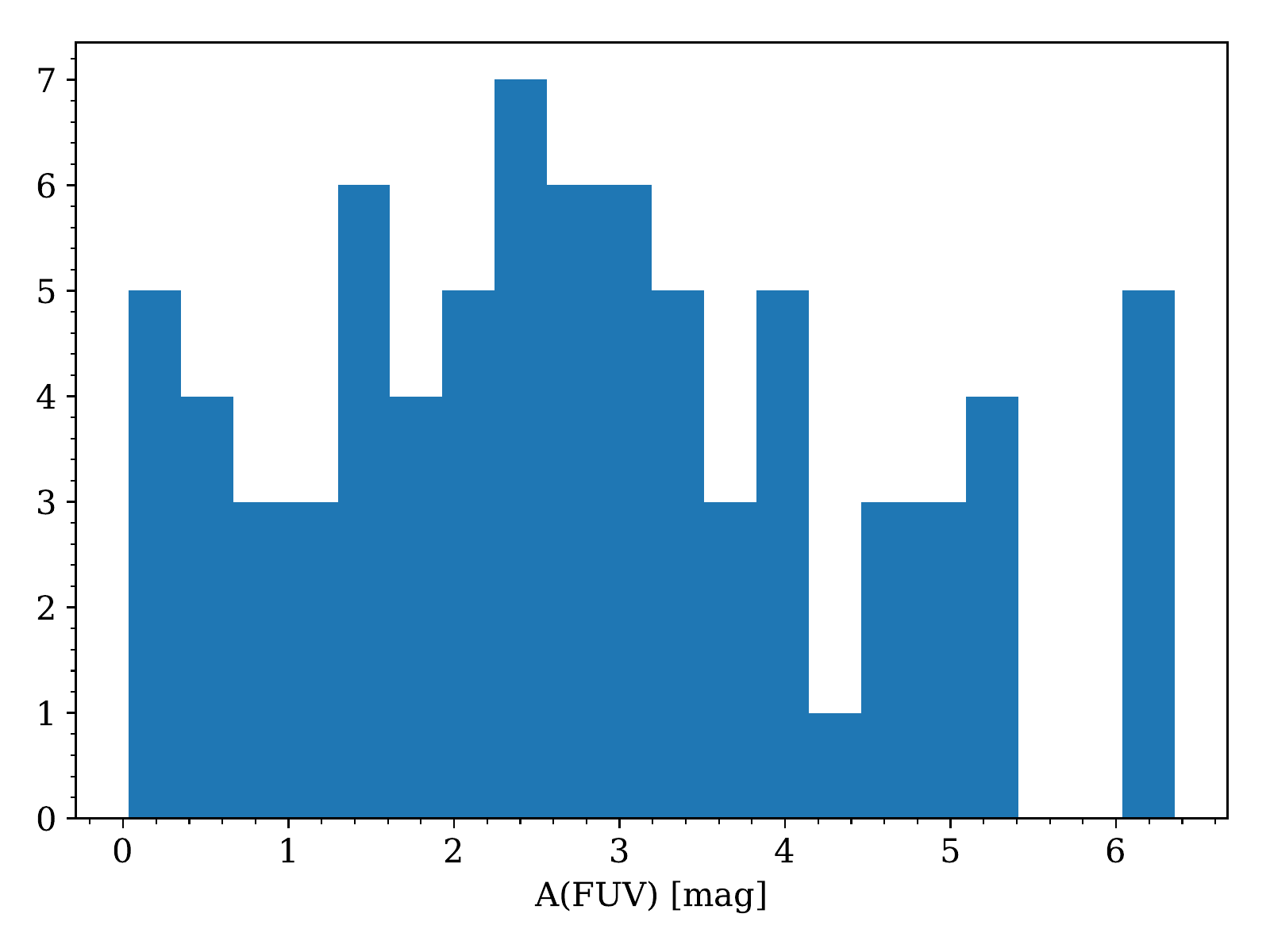}
 \includegraphics[width=0.33\textwidth]{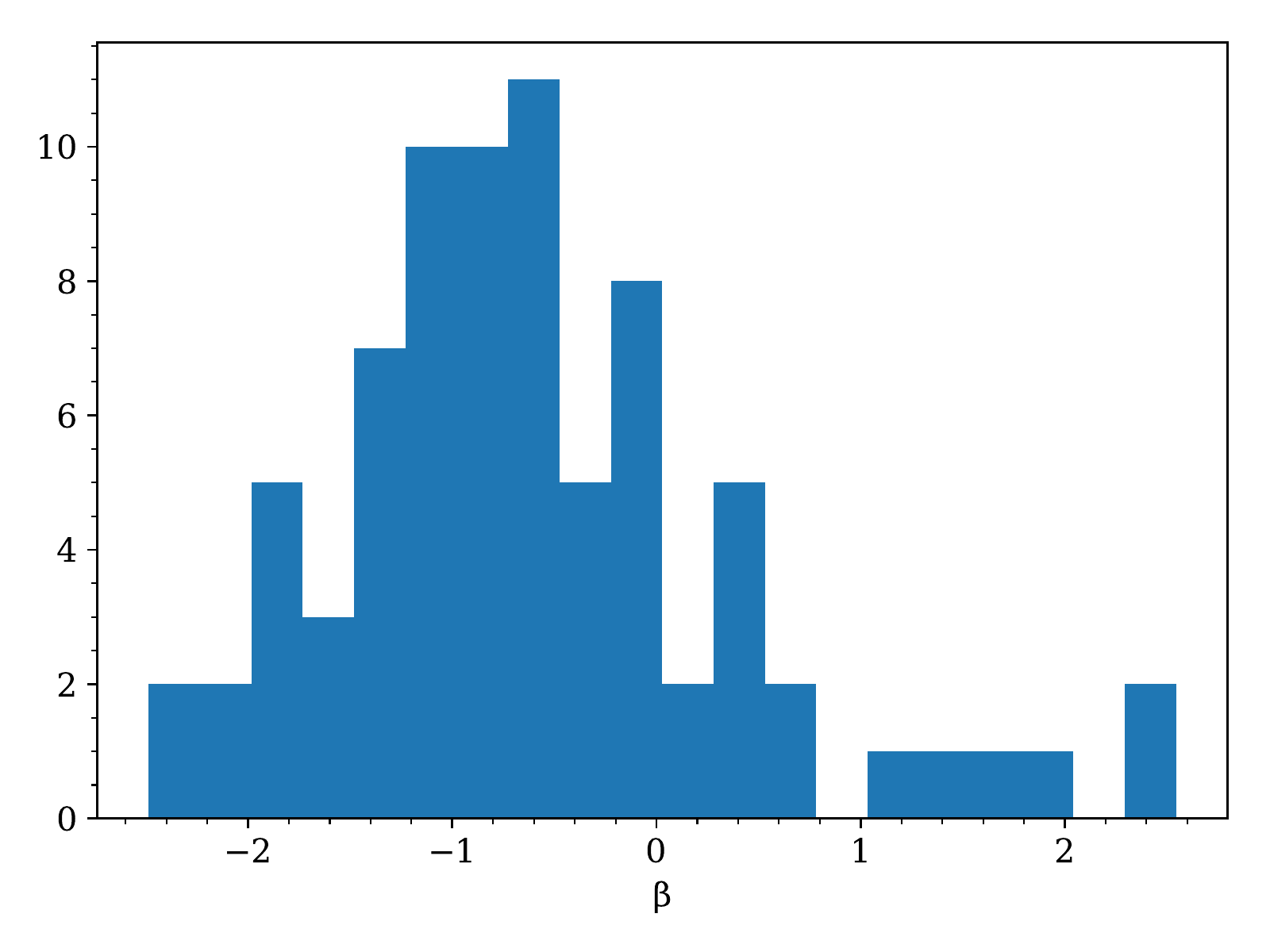}
 \includegraphics[width=0.33\textwidth]{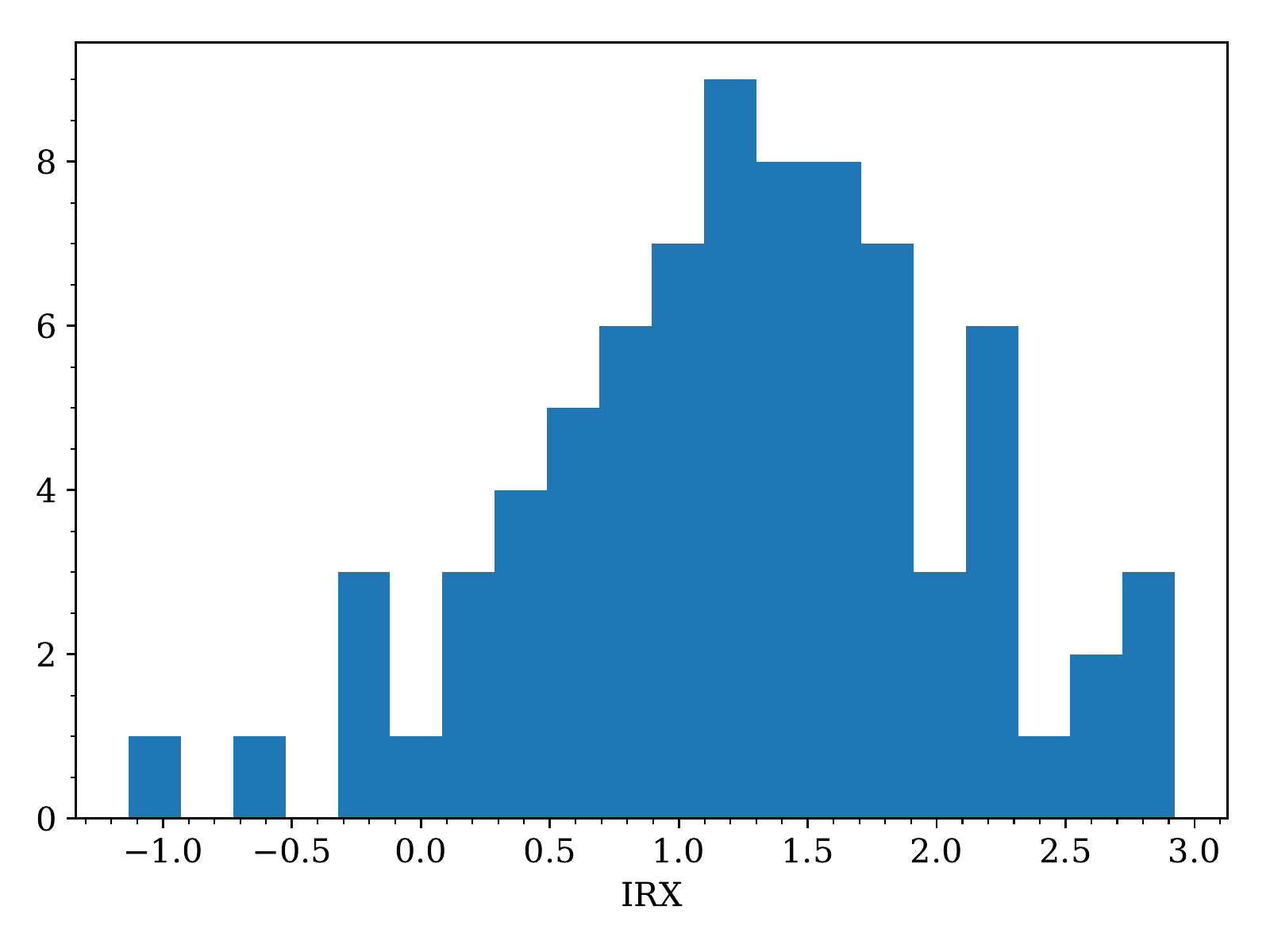}\\
 \includegraphics[width=0.33\textwidth]{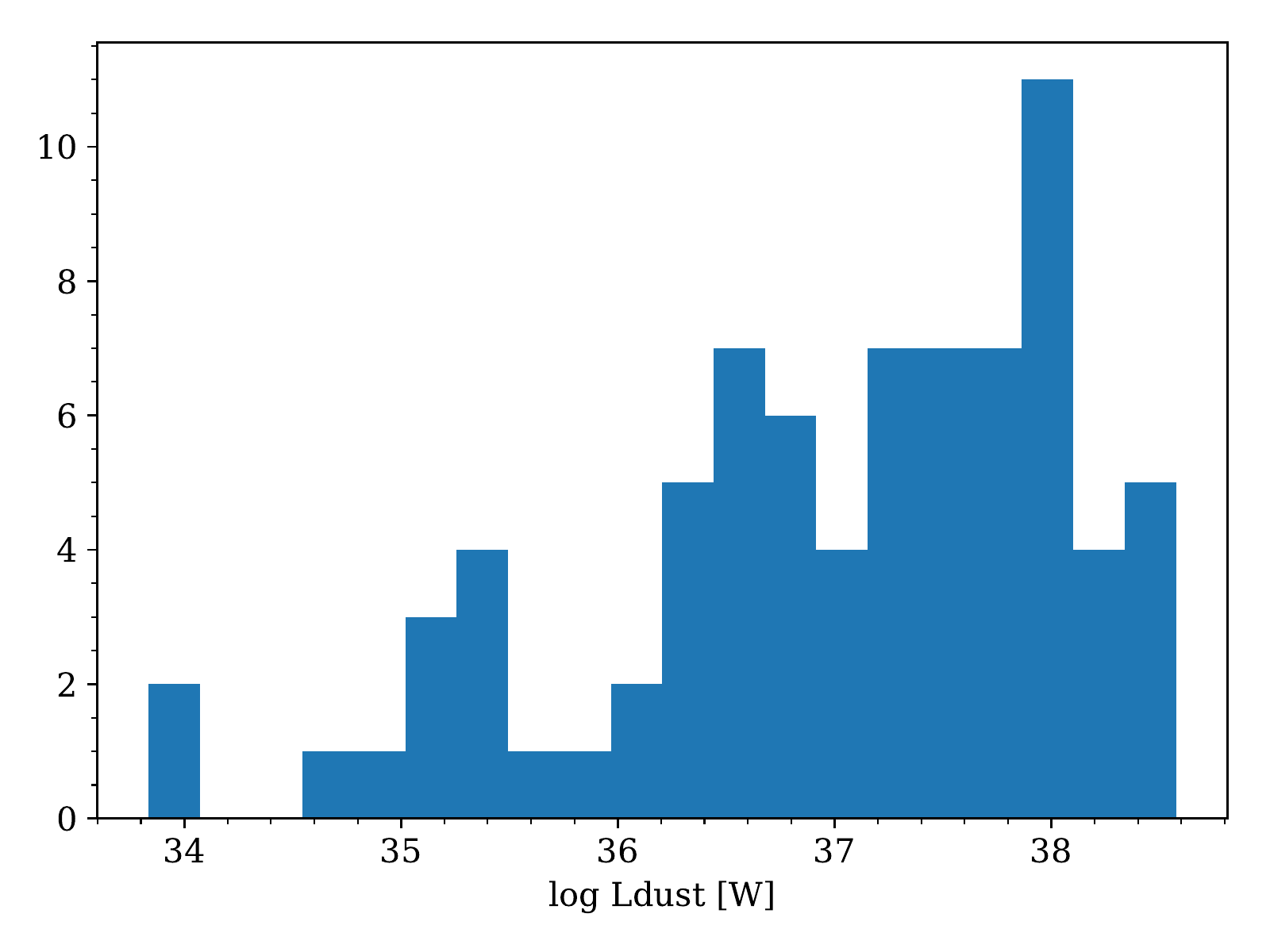}
 \includegraphics[width=0.33\textwidth]{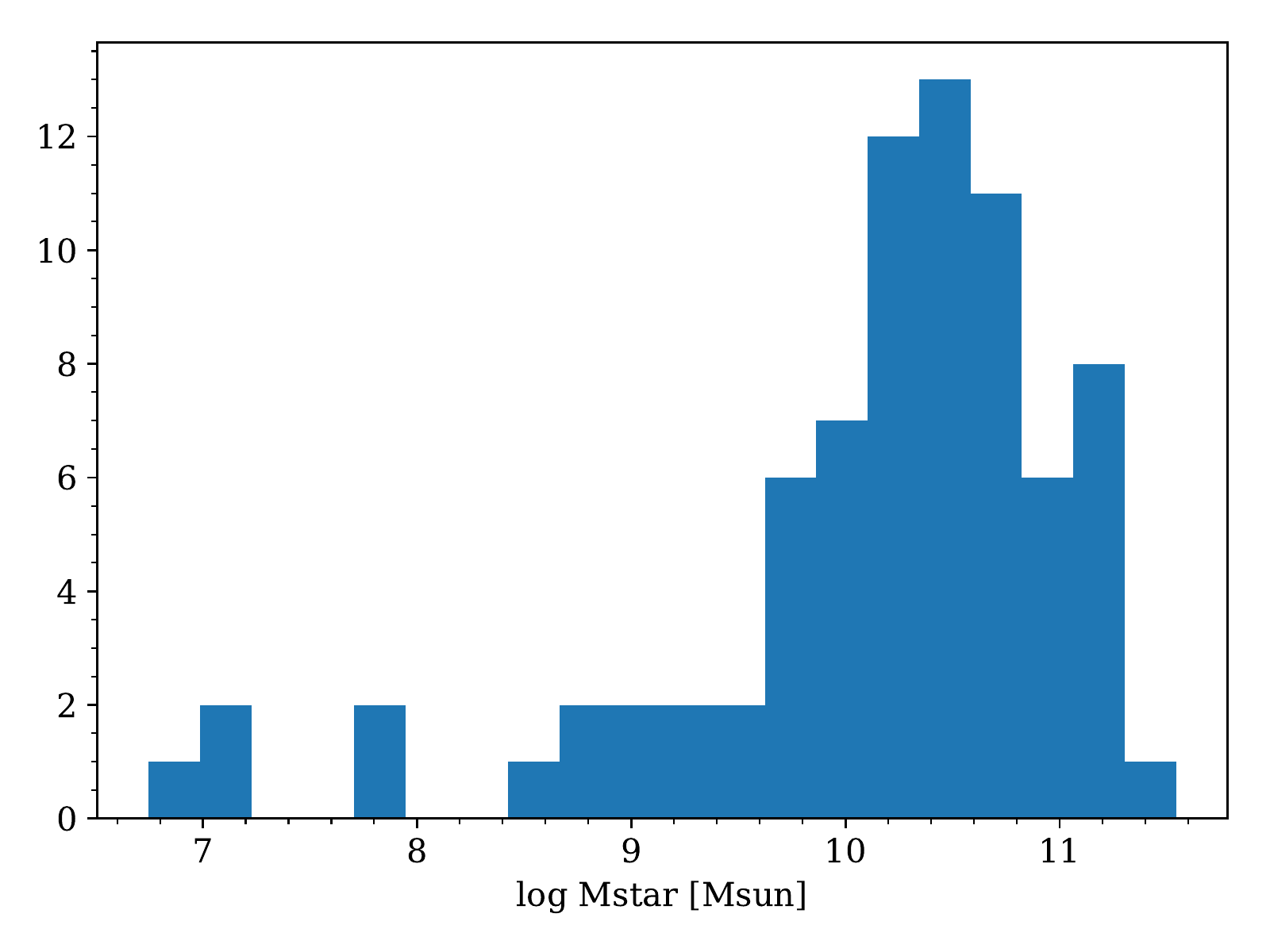}
 \includegraphics[width=0.33\textwidth]{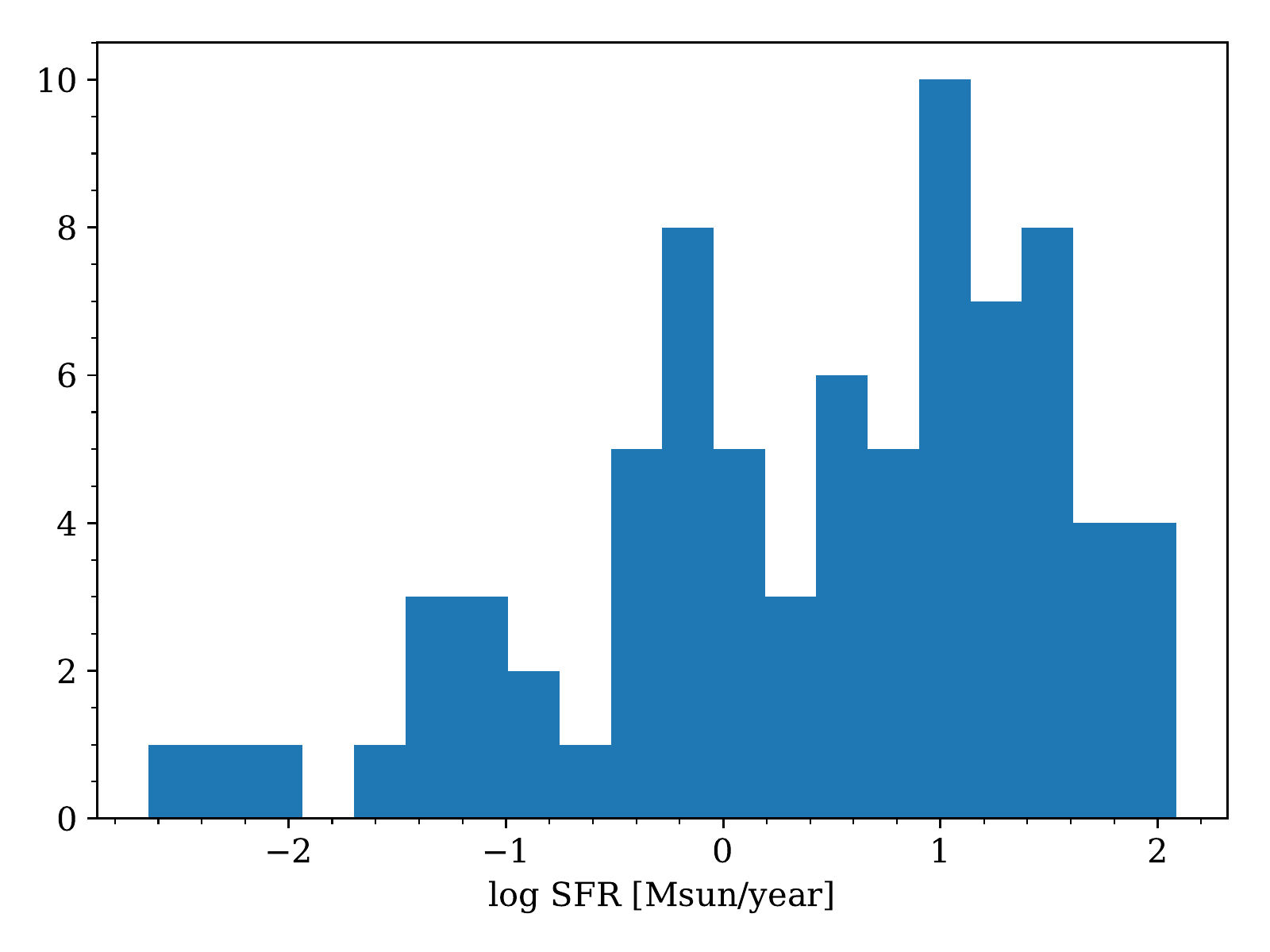}
 \caption{Distribution of the physical properties of the \cite{brown2014a} sample. We see that in a single run \texttt{CIGALE} can model galaxies with a wide range of properties: FUV attenuation, UV slope, IR--to--UV luminosity, dust luminosity, stellar mass, and SFR. This is only a small excerpt of the numerous physical properties that can also be estimated with \texttt{CIGALE} (see Appendix~\ref{sec:properties}).\label{fig:hist}}
\end{figure*}

While \texttt{CIGALE} can provide measurements for numerous physical properties, we chose here to concentrate on a subsample of six quantities that are frequently used to study galaxies: the FUV attenuation ($\mathrm{A_{FUV}}$), the dust luminosity ($\mathrm{L_{dust}}$), the instantaneous SFR, the stellar mass ($\mathrm{M_\star}$), the UV slope \citep[$\beta$,][]{calzetti1994a}, and the decimal logarithm of the ratio between the IR and FUV luminosities (IRX). The distributions of these six physical properties are shown in Fig.~\ref{fig:hist}.

\begin{figure*}[!htbp]
 \includegraphics[width=0.33\textwidth]{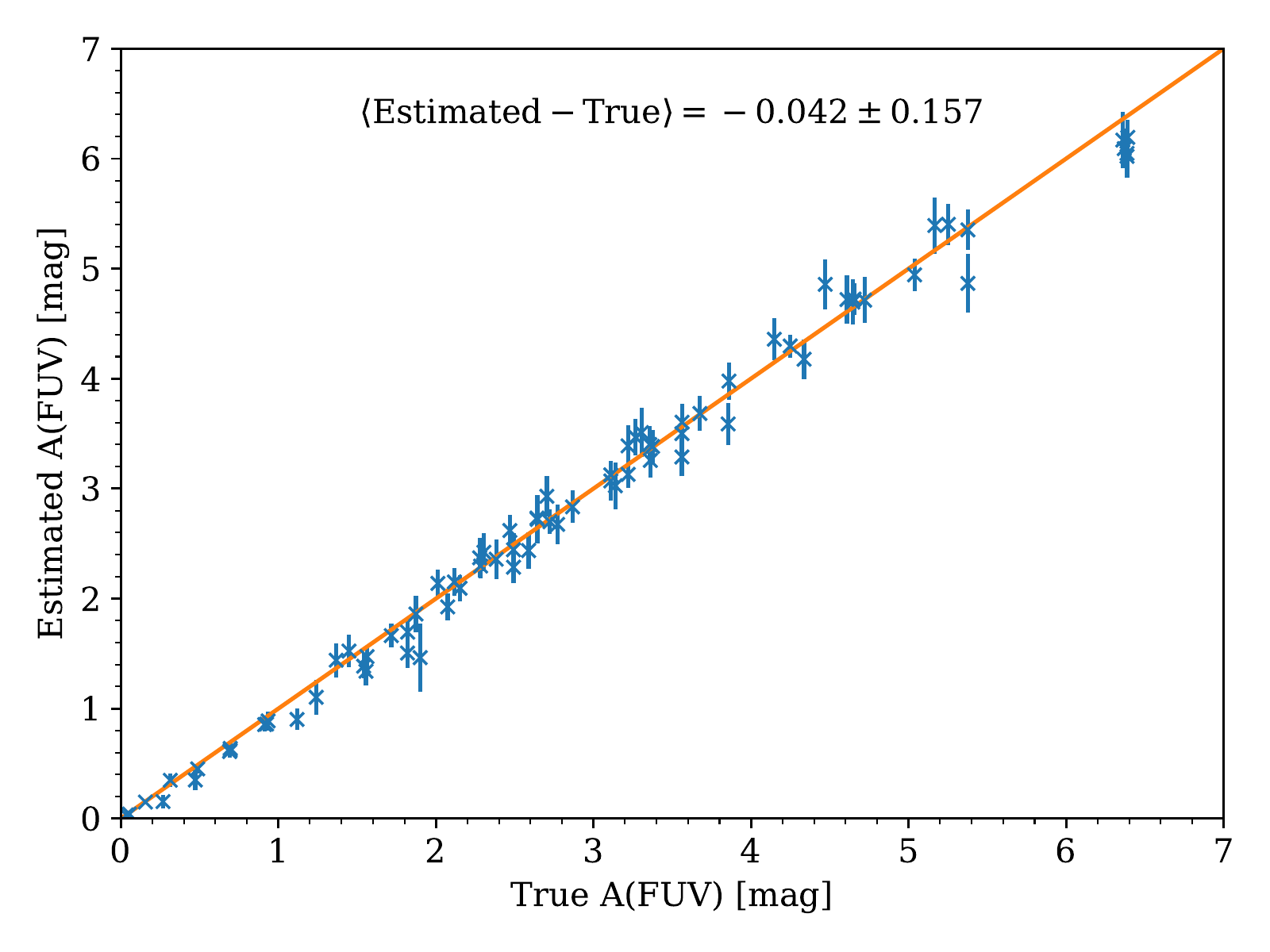}
 \includegraphics[width=0.33\textwidth]{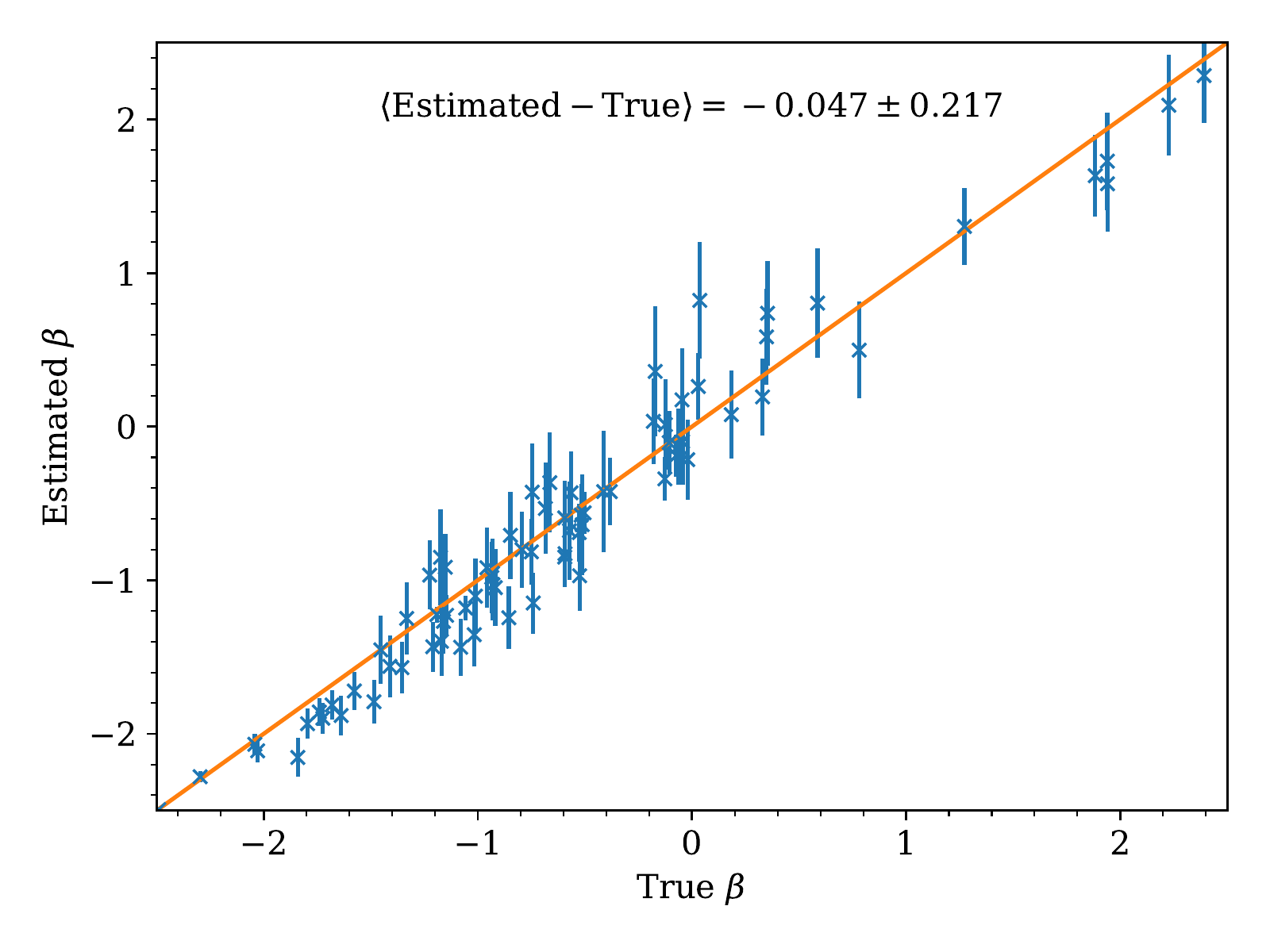}
 \includegraphics[width=0.33\textwidth]{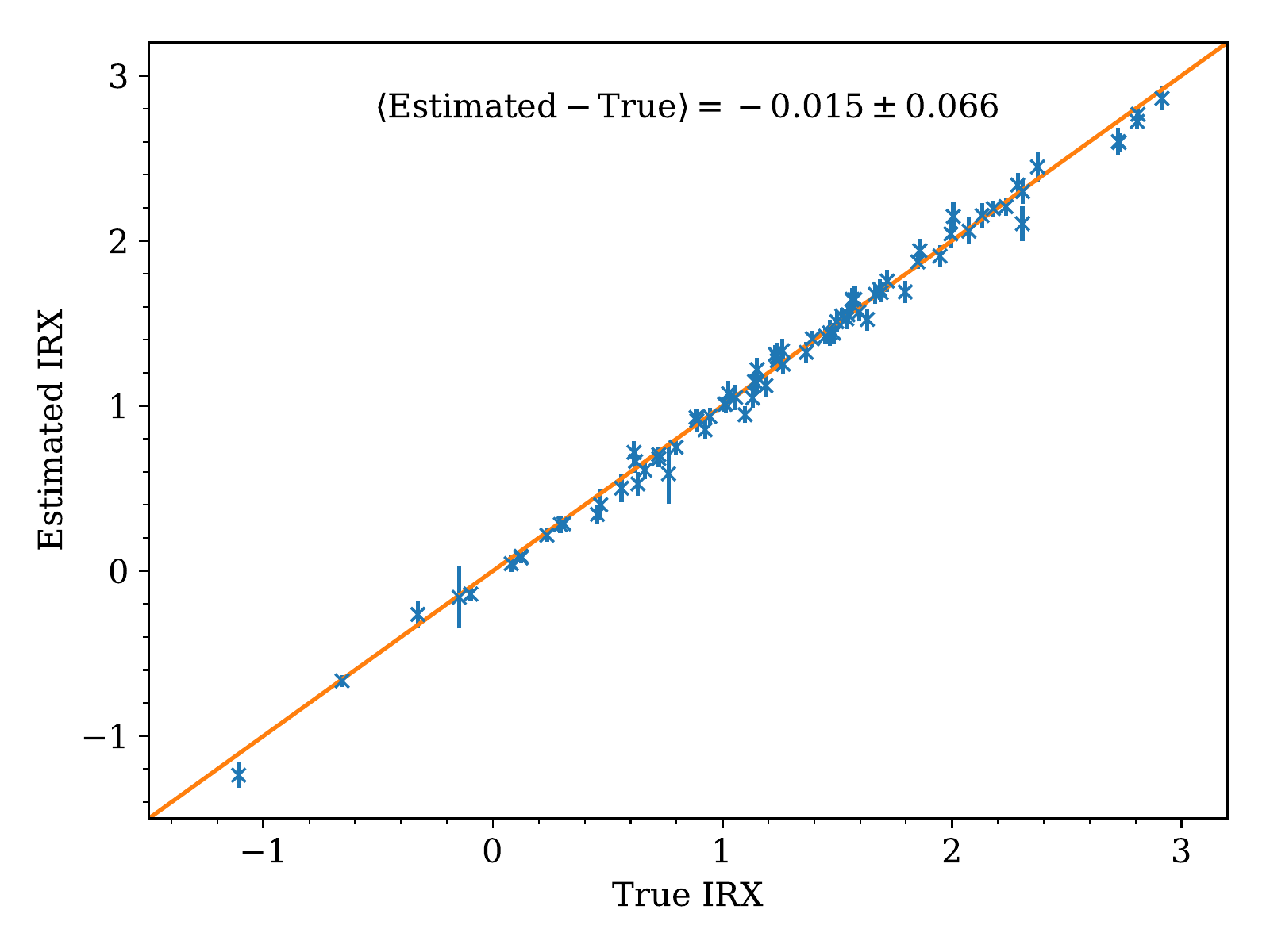}\\
 \includegraphics[width=0.33\textwidth]{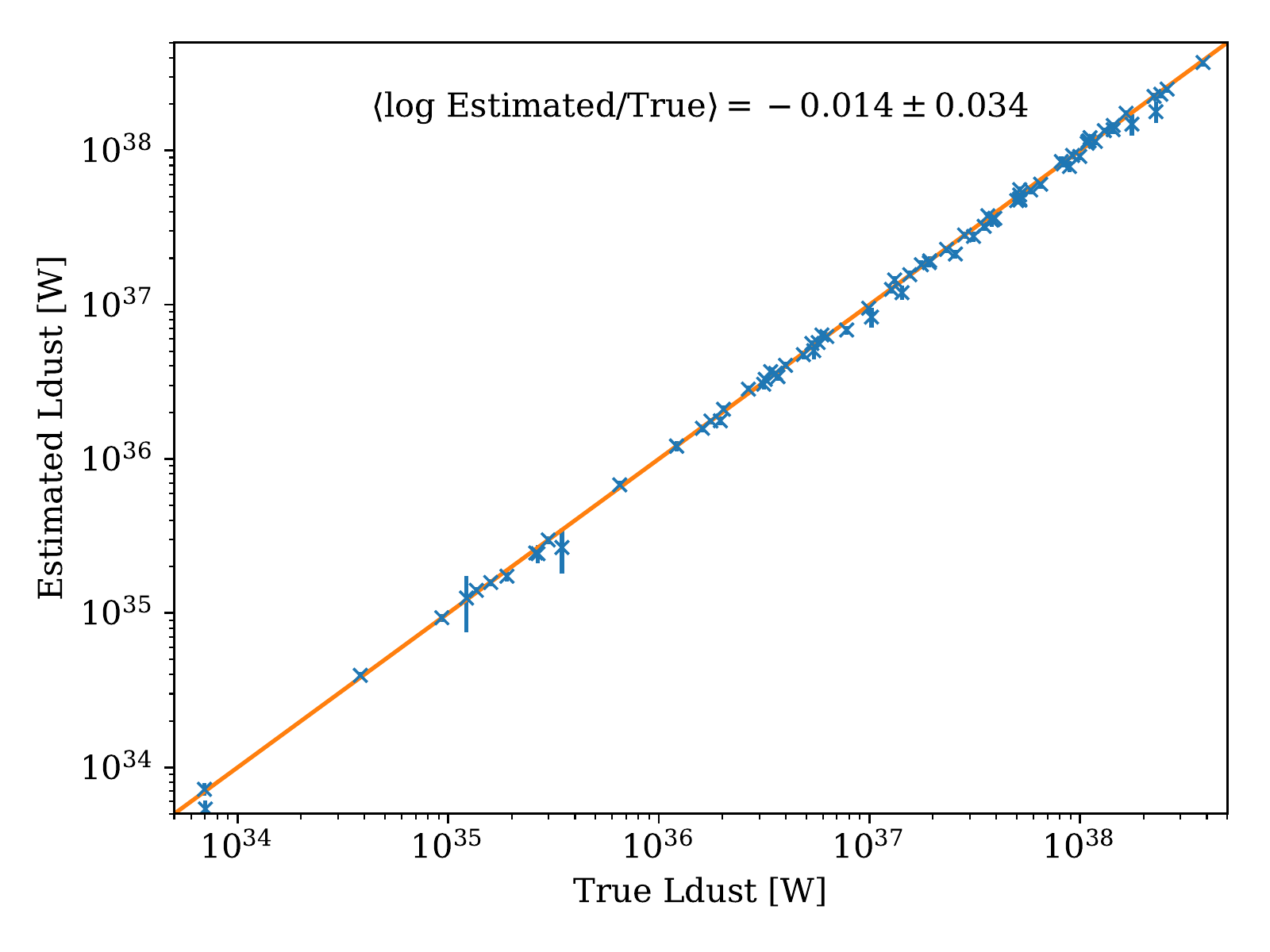}
 \includegraphics[width=0.33\textwidth]{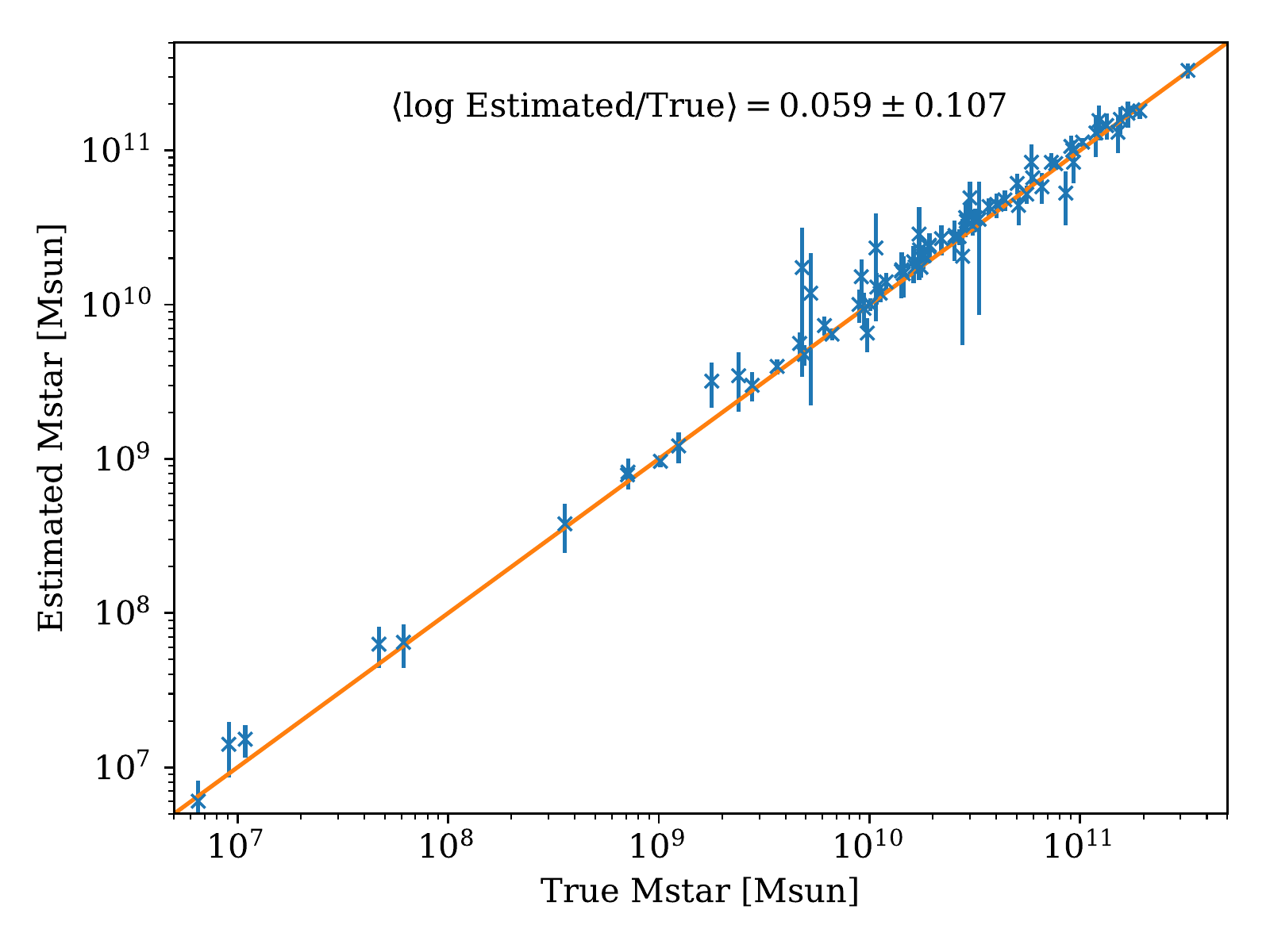}
 \includegraphics[width=0.33\textwidth]{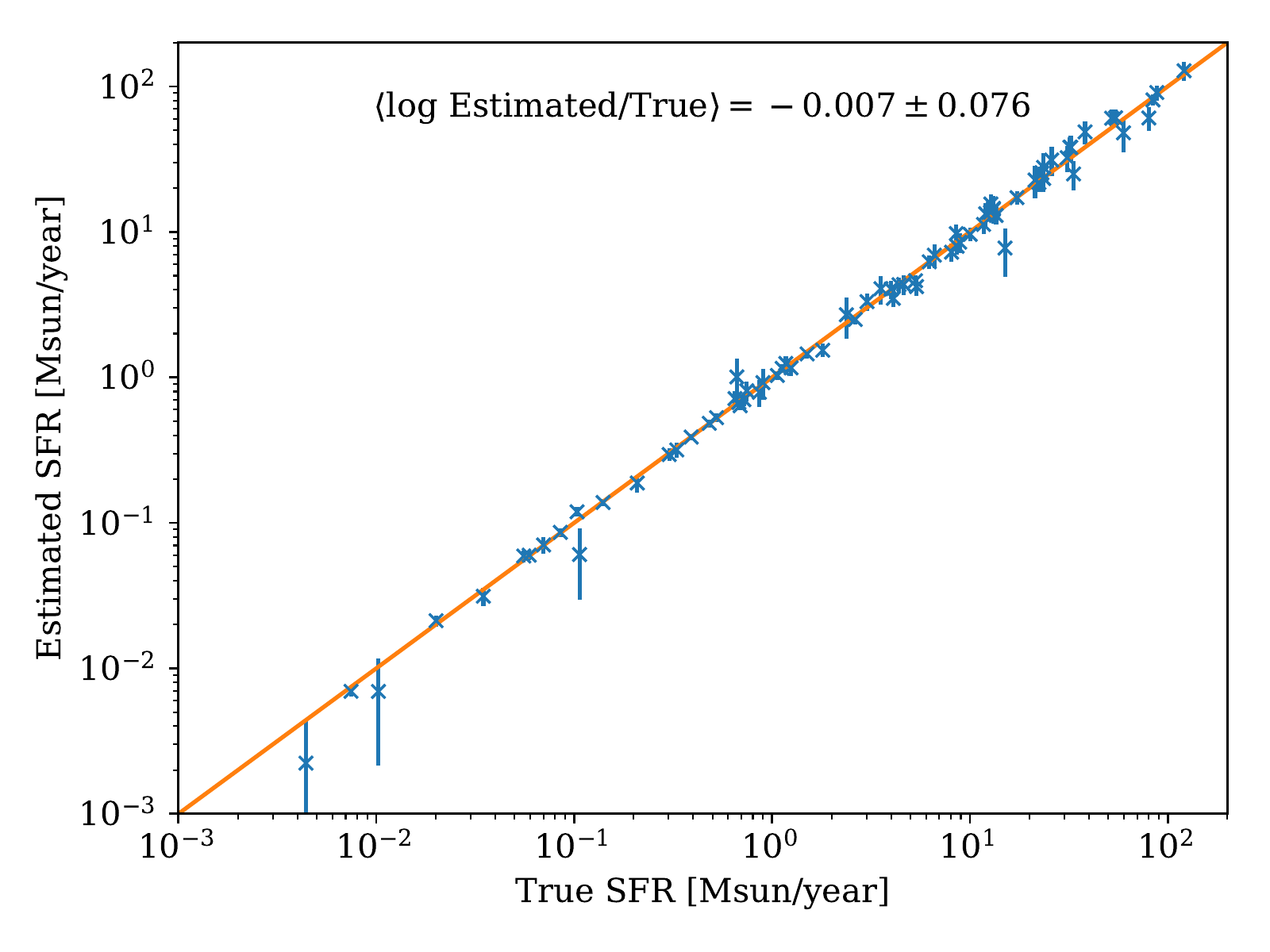}
 \caption{Comparison between the true (x--axis) and estimated (y--axis) values of $\mathrm{A_{FUV}}$), $\mathrm{L_{dust}}$, the SFR, $\mathrm{M_\star}$, and $\beta$ from the upper-left to the lower-right panel. Each galaxy is indicated with a blue cross, with the vertical line giving the 1--$\sigma$ error bar on the estimate. From this analysis it is clear that with this dataset and parameter space  \texttt{CIGALE} can measure these physical properties self--consistently. Not all physical properties are equally well measured however. It is apparent for instance that if the dust luminosity is extremely reliable ($-0.014\pm0.034$~dex), the  performance for $\beta$ for example, even though still excellent, shows somewhat more dispersion ($-0.047\pm0.217$~dex).\label{fig:mocks}}
\end{figure*}

Even though \texttt{CIGALE} will always provide an estimate of the physical properties, this estimate may or may not be reliable depending on the wavelength coverage, the quality of the data, and the explored parameter space, among other factors. A standard way to test whether the physical properties can at least be retrieved in a self--consistent way is through a mock catalogue as described in Sect.~\ref{ssec:analysis}. As a reminder, in a nutshell the idea is to fit the observations and build an artificial catalogue from the best fits. Considering these best fits, we know exactly what the corresponding physical parameters are, so we know the `truth'. Noise is then injected into the fluxes of this new catalogue to simulate new observations. Fitting these artificial observations we can then compare the inferred physical properties from the likelihood distribution to their actual values. We show such an analysis in Fig.~\ref{fig:mocks} for the six aforementioned physical properties.

Overall, all the physical quantities are reliably estimated with the average estimated values  being remarkably close to the true values. Regarding the scatter around this mean value, there is a marked difference in performance. The SFR, $\mathrm{L_{dust}}$, and IRX show very little scatter relative to the true values, lower than 0.1 dex (IRX is already a log--scale quantity). The two quantities with the lower performance appear to be $\beta$ (scatter of 0.217 or less than 5\% of the dynamical range) and $\mathrm{A(FUV)}$ (scatter of 0.157 dex). In either case, this performance remains excellent. This slight difference in performance reflects the fact that not all physical quantities can be determined with the same reliability, partly due to intrinsic degeneracies between the physical processes at play, and partly due to the breadth and quality of the photometric coverage. It is important to note that even though this exercise is useful, it does not take into account the uncertainties due to the reliability of the models themselves, and therefore yields lower limits on the actual uncertainties on the physical properties.

\begin{figure}[!htbp]
 \includegraphics[width=\columnwidth]{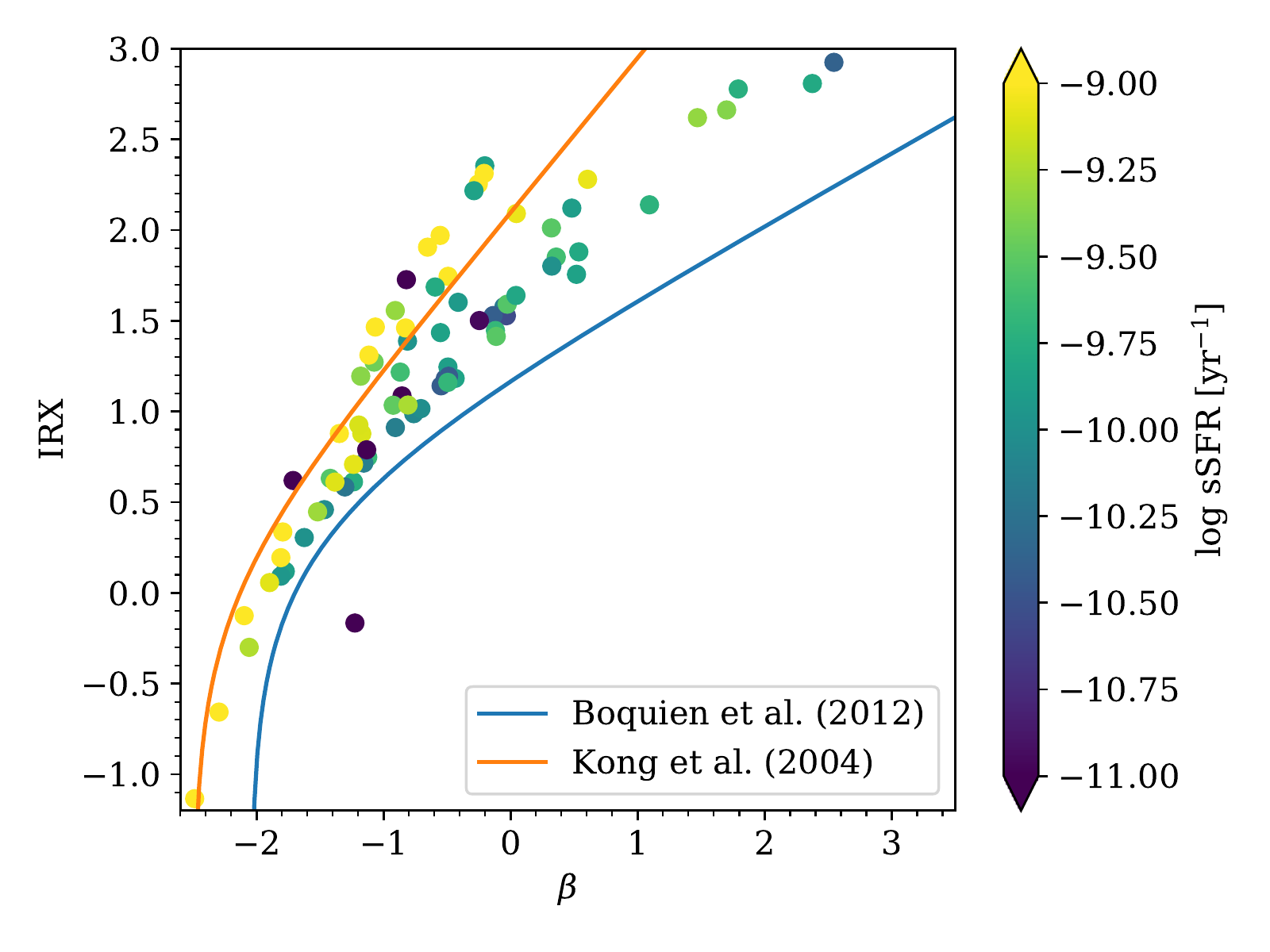}
 \caption{Relation between the observed IR-to-FUV luminosity (IRX) and the UV spectral slope ($\beta$) for the 129 galaxies of the \cite{brown2014a} sample. The colour of each symbol indicates the specific SFR, following the scale given by the colour bar to the right. The locus followed by resolved quiescent star--forming (respectively starburst) galaxies from \cite{boquien2012a} \citep[respectively][]{kong2004a} is indicated by the blue (respectively orange) line.\label{fig:IRX-beta}}
\end{figure}

With this inspection done, well--measured physical quantities can then be used to understand the properties of these galaxies. As a very simple example, we plot here the classical relation between IRX and the
UV spectral slope, $\beta,$ in Fig.~\ref{fig:IRX-beta}.
This relation is important as $\beta$ is often used to estimate the attenuation of distant galaxies where dust observations are missing, and IRX is a simple proxy for the attenuation. Numerous works have found and tried to explain the extensive variations observed in the IRX--$\beta$ relation over the years \citep[e.g. ][and many others]{kong2004a, burgarella2005a, boquien2009a, boquien2012a, overzier2011a, grasha2013a, ye2016a, popping2017a, reddy2018a}. This simple example shows the effect of the SFH to explain why more quiescent galaxies deviate from the relation followed by starburst galaxies, with more active galaxies following the relation of \cite{kong2004a} and more quiescent galaxies progressively moving closer to the relation of \cite{boquien2012a}. It is possible to investigate such questions, and many more, thanks to the flexibility of \texttt{CIGALE} in modelling a broad range of galaxies from intense starburst to elliptical galaxies and estimating their physical properties.

\section{Summary\label{sec:summary}}

In this paper, we present the new generation of the Code Investigating GALaxy Emission, \texttt{CIGALE}. Three principles have guided its development: modularity (it is easy to add new modules, and we encourage and support such developments, or swap modules modelling the same component), clarity (the code is easy to use and to understand), and efficiency (it runs quickly and is parallelised to take advantage of modern processors with multiple cores), both for the developers and for the users.

In practical terms, CIGALE is based on an energy balance principle (the energy absorbed by dust in the UV--to--near--IR domain is re--emitted self--consistently in the mid-- and far--IR). The models from the FUV to the radio are built in a modular way, taking into account flexible SFH and stellar populations \citep{bruzual2003a,maraston2005a}, ionised gas \citep{inoue2011a}, attenuation by dust \citep{calzetti2000a,charlot2000a} and re--emission of the energy at longer wavelengths \citep{draine2007a,casey2012a,dale2014a}, active nuclei \citep{fritz2006a,dale2014a}, and the IGM \citep{meiksin2006a}. The computation of these models is done in a parallel way on grids of models that can reach several hundred million elements. These models can then be simply saved for theoretical studies or can be used to evaluate a wide range of physical properties for observed objects. Such evaluation is based on the likelihood--weighted mean and standard deviation, taking into account the presence of upper limits. Finally, thanks to its versatility, \texttt{CIGALE} can also be used as a library to build new applications.

This article has presented a snapshot of the current state of \texttt{CIGALE}. It is however a constantly evolving code and we will present in upcoming papers new major evolutions to adapt it to the ever changing challenges of panchromatic modelling. The code along with its documentation is publicly available at \url{http://cigale.lam.fr}.

\begin{acknowledgements}
We would like to thank Daniel Dale, Caitlin Casey, Bruce Draine, Jacopo Fritz for their help to implement the latest version of their models in \texttt{CIGALE}. We would like to thank Simone Bianchi, Wouter Dobbels, Nimisha Kumari, Samir Salim, Manal Yassin, and Vivienne Wild for their feedback on early versions that have helped improve the ease of use of the code.

We also thank the anonymous referee who has helped clarify and improve various aspects of this article.

This research has made use of the NASA/IPAC Extragalactic Database (NED), which is operated by the Jet Propulsion Laboratory, California Institute of Technology, under contract with the National Aeronautics and Space Administration. 

MB was supported by the FONDECYT regular project 1170618.
\end{acknowledgements}
\bibliographystyle{aa}
\bibliography{article}

\begin{thebibliography}{101}
\expandafter\ifx\csname natexlab\endcsname\relax\def\natexlab#1{#1}\fi

\bibitem[{{{\'A}lvarez-M{\'a}rquez} {et~al.}(2016){{\'A}lvarez-M{\'a}rquez},
  {Burgarella}, {Heinis}, {Buat}, {Lo Faro}, {B{\'e}thermin},
  {L{\'o}pez-Fort{\'{\i}}n}, {Cooray}, {Farrah}, {Hurley}, {Ibar}, {Ilbert},
  {Koekemoer}, {Lemaux}, {P{\'e}rez-Fournon}, {Rodighiero}, {Salvato}, {Scott},
  {Taniguchi}, {Vieira}, \& {Wang}}]{alvarez2016a}
{{\'A}lvarez-M{\'a}rquez}, J., {Burgarella}, D., {Heinis}, S., {et~al.} 2016,
  \aap, 587, A122

\bibitem[{{Anders} \& {Fritze-v.~Alvensleben}(2003)}]{anders2003a}
{Anders}, P. \& {Fritze-v.~Alvensleben}, U. 2003, \aap, 401, 1063

\bibitem[{{Astropy Collaboration} {et~al.}(2018){Astropy Collaboration},
  {Price-Whelan}, {Sip{\H o}cz}, {G{\"u}nther}, {Lim}, {Crawford}, {Conseil},
  {Shupe}, {Craig}, {Dencheva}, {Ginsburg}, {VanderPlas}, {Bradley},
  {P{\'e}rez-Su{\'a}rez}, {de Val-Borro}, {Aldcroft}, {Cruz}, {Robitaille},
  {Tollerud}, {Ardelean}, {Babej}, {Bach}, {Bachetti}, {Bakanov}, {Bamford},
  {Barentsen}, {Barmby}, {Baumbach}, {Berry}, {Biscani}, {Boquien}, {Bostroem},
  {Bouma}, {Brammer}, {Bray}, {Breytenbach}, {Buddelmeijer}, {Burke},
  {Calderone}, {Cano Rodr{\'{\i}}guez}, {Cara}, {Cardoso}, {Cheedella},
  {Copin}, {Corrales}, {Crichton}, {D'Avella}, {Deil}, {Depagne}, {Dietrich},
  {Donath}, {Droettboom}, {Earl}, {Erben}, {Fabbro}, {Ferreira}, {Finethy},
  {Fox}, {Garrison}, {Gibbons}, {Goldstein}, {Gommers}, {Greco}, {Greenfield},
  {Groener}, {Grollier}, {Hagen}, {Hirst}, {Homeier}, {Horton}, {Hosseinzadeh},
  {Hu}, {Hunkeler}, {Ivezi{\'c}}, {Jain}, {Jenness}, {Kanarek}, {Kendrew},
  {Kern}, {Kerzendorf}, {Khvalko}, {King}, {Kirkby}, {Kulkarni}, {Kumar},
  {Lee}, {Lenz}, {Littlefair}, {Ma}, {Macleod}, {Mastropietro}, {McCully},
  {Montagnac}, {Morris}, {Mueller}, {Mumford}, {Muna}, {Murphy}, {Nelson},
  {Nguyen}, {Ninan}, {N{\"o}the}, {Ogaz}, {Oh}, {Parejko}, {Parley}, {Pascual},
  {Patil}, {Patil}, {Plunkett}, {Prochaska}, {Rastogi}, {Reddy Janga},
  {Sabater}, {Sakurikar}, {Seifert}, {Sherbert}, {Sherwood-Taylor}, {Shih},
  {Sick}, {Silbiger}, {Singanamalla}, {Singer}, {Sladen}, {Sooley},
  {Sornarajah}, {Streicher}, {Teuben}, {Thomas}, {Tremblay}, {Turner},
  {Terr{\'o}n}, {van Kerkwijk}, {de la Vega}, {Watkins}, {Weaver}, {Whitmore},
  {Woillez}, {Zabalza}, \& {Astropy Contributors}}]{astropy2018a}
{Astropy Collaboration}, {Price-Whelan}, A.~M., {Sip{\H o}cz}, B.~M., {et~al.}
  2018, \aj, 156, 123

\bibitem[{{Astropy Collaboration} {et~al.}(2013){Astropy Collaboration},
  {Robitaille}, {Tollerud}, {Greenfield}, {Droettboom}, {Bray}, {Aldcroft},
  {Davis}, {Ginsburg}, {Price-Whelan}, {Kerzendorf}, {Conley}, {Crighton},
  {Barbary}, {Muna}, {Ferguson}, {Grollier}, {Parikh}, {Nair}, {Unther},
  {Deil}, {Woillez}, {Conseil}, {Kramer}, {Turner}, {Singer}, {Fox}, {Weaver},
  {Zabalza}, {Edwards}, {Azalee Bostroem}, {Burke}, {Casey}, {Crawford},
  {Dencheva}, {Ely}, {Jenness}, {Labrie}, {Lim}, {Pierfederici}, {Pontzen},
  {Ptak}, {Refsdal}, {Servillat}, \& {Streicher}}]{astropy2013a}
{Astropy Collaboration}, {Robitaille}, T.~P., {Tollerud}, E.~J., {et~al.} 2013,
  \aap, 558, A33

\bibitem[{{Balogh} {et~al.}(1999){Balogh}, {Morris}, {Yee}, {Carlberg}, \&
  {Ellingson}}]{balogh1999a}
{Balogh}, M.~L., {Morris}, S.~L., {Yee}, H.~K.~C., {Carlberg}, R.~G., \&
  {Ellingson}, E. 1999, \apj, 527, 54

\bibitem[{{Bianchi}(2008)}]{bianchi2008a}
{Bianchi}, S. 2008, \aap, 490, 461

\bibitem[{{Bitsakis} {et~al.}(2016){Bitsakis}, {Dultzin}, {Ciesla},
  {D{\'{\i}}az-Santos}, {Appleton}, {Charmandaris}, {Krongold}, {Guillard},
  {Alatalo}, {Zezas}, {Gonz{\'a}lez}, \& {Lanz}}]{bitsakis2016a}
{Bitsakis}, T., {Dultzin}, D., {Ciesla}, L., {et~al.} 2016, \mnras, 459, 957

\bibitem[{{Boissier} {et~al.}(2003){Boissier}, {Prantzos}, {Boselli}, \&
  {Gavazzi}}]{boissier2003a}
{Boissier}, S., {Prantzos}, N., {Boselli}, A., \& {Gavazzi}, G. 2003, \mnras,
  346, 1215

\bibitem[{{Boquien} {et~al.}(2013){Boquien}, {Boselli}, {Buat}, {Baes},
  {Bendo}, {Boissier}, {Ciesla}, {Cooray}, {Cortese}, {Eales}, {Koda},
  {Lebouteiller}, {de Looze}, {Smith}, {Spinoglio}, \& {Wilson}}]{boquien2013a}
{Boquien}, M., {Boselli}, A., {Buat}, V., {et~al.} 2013, \aap, 554, A14

\bibitem[{{Boquien} {et~al.}(2012){Boquien}, {Buat}, {Boselli}, {Baes},
  {Bendo}, {Ciesla}, {Cooray}, {Cortese}, {Eales}, {Gavazzi}, {Gomez},
  {Lebouteiller}, {Pappalardo}, {Pohlen}, {Smith}, \&
  {Spinoglio}}]{boquien2012a}
{Boquien}, M., {Buat}, V., {Boselli}, A., {et~al.} 2012, \aap, 539, A145

\bibitem[{{Boquien} {et~al.}(2014){Boquien}, {Buat}, \&
  {Perret}}]{boquien2014a}
{Boquien}, M., {Buat}, V., \& {Perret}, V. 2014, \aap, 571, A72

\bibitem[{{Boquien} {et~al.}(2010){Boquien}, {Duc}, {Galliano}, {Braine},
  {Lisenfeld}, {Charmandaris}, \& {Appleton}}]{boquien2010c}
{Boquien}, M., {Duc}, P., {Galliano}, F., {et~al.} 2010, \aj, 140, 2124

\bibitem[{{Boquien} {et~al.}(2009){Boquien}, {Duc}, {Wu}, {Charmandaris},
  {Lisenfeld}, {Braine}, {Brinks}, {Iglesias-P{\'a}ramo}, \&
  {Xu}}]{boquien2009a}
{Boquien}, M., {Duc}, P., {Wu}, Y., {et~al.} 2009, \aj, 137, 4561

\bibitem[{{Boquien} {et~al.}(2016){Boquien}, {Kennicutt}, {Calzetti}, {Dale},
  {Galametz}, {Sauvage}, {Croxall}, {Draine}, {Kirkpatrick}, {Kumari}, {Hunt},
  {De Looze}, {Pellegrini}, {Rela{\~n}o}, {Smith}, \&
  {Tabatabaei}}]{boquien2016a}
{Boquien}, M., {Kennicutt}, R., {Calzetti}, D., {et~al.} 2016, \aap, 591, A6

\bibitem[{{Brown} {et~al.}(2014){Brown}, {Moustakas}, {Smith}, {da Cunha},
  {Jarrett}, {Imanishi}, {Armus}, {Brandl}, \& {Peek}}]{brown2014a}
{Brown}, M.~J.~I., {Moustakas}, J., {Smith}, J.-D.~T., {et~al.} 2014, \apjs,
  212, 18

\bibitem[{{Bruzual} \& {Charlot}(2003)}]{bruzual2003a}
{Bruzual}, G. \& {Charlot}, S. 2003, \mnras, 344, 1000

\bibitem[{{Buat} {et~al.}(2008){Buat}, {Boissier}, {Burgarella}, {Takeuchi},
  {Le Floc'h}, {Marcillac}, {Huang}, {Nagashima}, \& {Enoki}}]{buat2008a}
{Buat}, V., {Boissier}, S., {Burgarella}, D., {et~al.} 2008, \aap, 483, 107

\bibitem[{{Buat} {et~al.}(2011){Buat}, {Giovannoli}, {Heinis}, {Charmandaris},
  {Coia}, {Daddi}, {Dickinson}, {Elbaz}, {Hwang}, {Morrison}, {Dasyra},
  {Aussel}, {Altieri}, {Dannerbauer}, {Kartaltepe}, {Leiton}, {Magdis},
  {Magnelli}, \& {Popesso}}]{buat2011b}
{Buat}, V., {Giovannoli}, E., {Heinis}, S., {et~al.} 2011, \aap, 533, A93+

\bibitem[{{Buat} {et~al.}(2014){Buat}, {Heinis}, {Boquien}, {Burgarella},
  {Charmandaris}, {Boissier}, {Boselli}, {Le Borgne}, \&
  {Morrison}}]{buat2014a}
{Buat}, V., {Heinis}, S., {Boquien}, M., {et~al.} 2014, \aap, 561, A39

\bibitem[{{Buat} {et~al.}(2012){Buat}, {Noll}, {Burgarella}, {Giovannoli},
  {Charmandaris}, {Pannella}, {Hwang}, {Elbaz}, {Dickinson}, {Magdis}, {Reddy},
  \& {Murphy}}]{buat2012a}
{Buat}, V., {Noll}, S., {Burgarella}, D., {et~al.} 2012, \aap, 545, A141

\bibitem[{{Burgarella} {et~al.}(2005){Burgarella}, {Buat}, \&
  {Iglesias-P{\'a}ramo}}]{burgarella2005a}
{Burgarella}, D., {Buat}, V., \& {Iglesias-P{\'a}ramo}, J. 2005, \mnras, 360,
  1413

\bibitem[{{Burgarella} {et~al.}(2011){Burgarella}, {Heinis}, {Magdis}, {Auld},
  {Blain}, {Bock}, {Brisbin}, {Buat}, {Chanial}, {Clements}, {Cooray}, {Eales},
  {Franceschini}, {Giovannoli}, {Glenn}, {Gonz{\'a}lez Solares}, {Griffin},
  {Hwang}, {Ilbert}, {Marchetti}, {Mortier}, {Oliver}, {Page}, {Papageorgiou},
  {Pearson}, {P{\'e}rez-Fournon}, {Pohlen}, {Rawlings}, {Raymond},
  {Rigopoulou}, {Rodighiero}, {Roseboom}, {Rowan-Robinson}, {Scott}, {Seymour},
  {Smith}, {Symeonidis}, {Tugwell}, {Vaccari}, {Vieira}, {Viero}, {Vigroux},
  {Wang}, \& {Wright}}]{burgarella2011a}
{Burgarella}, D., {Heinis}, S., {Magdis}, G., {et~al.} 2011, \apjl, 734, L12

\bibitem[{{Calzetti} {et~al.}(2000){Calzetti}, {Armus}, {Bohlin}, {Kinney},
  {Koornneef}, \& {Storchi-Bergmann}}]{calzetti2000a}
{Calzetti}, D., {Armus}, L., {Bohlin}, R.~C., {et~al.} 2000, \apj, 533, 682

\bibitem[{{Calzetti} {et~al.}(2007){Calzetti}, {Kennicutt}, {Engelbracht},
  {Leitherer}, {Draine}, {Kewley}, {Moustakas}, {Sosey}, {Dale}, {Gordon},
  {Helou}, {Hollenbach}, {Armus}, {Bendo}, {Bot}, {Buckalew}, {Jarrett}, {Li},
  {Meyer}, {Murphy}, {Prescott}, {Regan}, {Rieke}, {Roussel}, {Sheth}, {Smith},
  {Thornley}, \& {Walter}}]{calzetti2007a}
{Calzetti}, D., {Kennicutt}, R.~C., {Engelbracht}, C.~W., {et~al.} 2007, \apj,
  666, 870

\bibitem[{{Calzetti} {et~al.}(1994){Calzetti}, {Kinney}, \&
  {Storchi-Bergmann}}]{calzetti1994a}
{Calzetti}, D., {Kinney}, A.~L., \& {Storchi-Bergmann}, T. 1994, \apj, 429, 582

\bibitem[{{Cardelli} {et~al.}(1989){Cardelli}, {Clayton}, \&
  {Mathis}}]{cardelli1989a}
{Cardelli}, J.~A., {Clayton}, G.~C., \& {Mathis}, J.~S. 1989, \apj, 345, 245

\bibitem[{{Casey}(2012)}]{casey2012a}
{Casey}, C.~M. 2012, \mnras, 425, 3094

\bibitem[{{Chabrier}(2003)}]{chabrier2003b}
{Chabrier}, G. 2003, \pasp, 115, 763

\bibitem[{{Chan} {et~al.}(1979){Chan}, {Golub}, \& {LeVeque}}]{chan1979a}
{Chan}, T.~F., {Golub}, G.~H., \& {LeVeque}, R.~J. 1979, Technical Report
  STAN-CS-79-773, Department of Computer Science, Stanford University

\bibitem[{{Charlot} \& {Fall}(2000)}]{charlot2000a}
{Charlot}, S. \& {Fall}, S.~M. 2000, \apj, 539, 718

\bibitem[{{Ciesla} {et~al.}(2014){Ciesla}, {Boquien}, {Boselli}, {Buat},
  {Cortese}, {Bendo}, {Heinis}, {Galametz}, {Eales}, {Smith}, {Baes},
  {Bianchi}, {de Looze}, {di Serego Alighieri}, {Galliano}, {Hughes}, {Madden},
  {Pierini}, {R{\'e}my-Ruyer}, {Spinoglio}, {Vaccari}, {Viaene}, \&
  {Vlahakis}}]{ciesla2014a}
{Ciesla}, L., {Boquien}, M., {Boselli}, A., {et~al.} 2014, \aap, 565, A128

\bibitem[{{Ciesla} {et~al.}(2016){Ciesla}, {Boselli}, {Elbaz}, {Boissier},
  {Buat}, {Charmandaris}, {Schreiber}, {B{\'e}thermin}, {Baes}, {Boquien}, {De
  Looze}, {Fern{\'a}ndez-Ontiveros}, {Pappalardo}, {Spinoglio}, \&
  {Viaene}}]{ciesla2016a}
{Ciesla}, L., {Boselli}, A., {Elbaz}, D., {et~al.} 2016, \aap, 585, A43

\bibitem[{{Ciesla} {et~al.}(2015){Ciesla}, {Charmandaris}, {Georgakakis},
  {Bernhard}, {Mitchell}, {Buat}, {Elbaz}, {LeFloc'h}, {Lacey}, {Magdis}, \&
  {Xilouris}}]{ciesla2015a}
{Ciesla}, L., {Charmandaris}, V., {Georgakakis}, A., {et~al.} 2015, \aap, 576,
  A10

\bibitem[{{Ciesla} {et~al.}(2017){Ciesla}, {Elbaz}, \& {Fensch}}]{ciesla2017a}
{Ciesla}, L., {Elbaz}, D., \& {Fensch}, J. 2017, \aap, 608, A41

\bibitem[{{Conroy} \& {Gunn}(2010)}]{conroy2010b}
{Conroy}, C. \& {Gunn}, J.~E. 2010, \apj, 712, 833

\bibitem[{{Conroy} {et~al.}(2009){Conroy}, {Gunn}, \& {White}}]{conroy2009a}
{Conroy}, C., {Gunn}, J.~E., \& {White}, M. 2009, \apj, 699, 486

\bibitem[{{da Cunha} {et~al.}(2008){da Cunha}, {Charlot}, \&
  {Elbaz}}]{dacunha2008a}
{da Cunha}, E., {Charlot}, S., \& {Elbaz}, D. 2008, \mnras, 388, 1595

\bibitem[{{Dale} \& {Helou}(2002)}]{dale2002a}
{Dale}, D.~A. \& {Helou}, G. 2002, \apj, 576, 159

\bibitem[{{Dale} {et~al.}(2014){Dale}, {Helou}, {Magdis}, {Armus},
  {D{\'{\i}}az-Santos}, \& {Shi}}]{dale2014a}
{Dale}, D.~A., {Helou}, G., {Magdis}, G.~E., {et~al.} 2014, \apj, 784, 83

\bibitem[{{de Barros} {et~al.}(2014){de Barros}, {Schaerer}, \&
  {Stark}}]{debarros2014a}
{de Barros}, S., {Schaerer}, D., \& {Stark}, D.~P. 2014, \aap, 563, A81

\bibitem[{{de Looze} {et~al.}(2012){de Looze}, {Baes}, {Bendo}, {Ciesla},
  {Cortese}, {de Geyter}, {Groves}, {Boquien}, {Boselli}, {Brondeel}, {Cooray},
  {Eales}, {Fritz}, {Galliano}, {Gentile}, {Gordon}, {Hony}, {Law}, {Madden},
  {Sauvage}, {Smith}, {Spinoglio}, \& {Verstappen}}]{delooze2012b}
{de Looze}, I., {Baes}, M., {Bendo}, G.~J., {et~al.} 2012, \mnras, 427, 2797

\bibitem[{{De Looze} {et~al.}(2014){De Looze}, {Fritz}, {Baes}, {Bendo},
  {Cortese}, {Boquien}, {Boselli}, {Camps}, {Cooray}, {Cormier}, {Davies}, {De
  Geyter}, {Hughes}, {Jones}, {Karczewski}, {Lebouteiller}, {Lu}, {Madden},
  {R{\'e}my-Ruyer}, {Spinoglio}, {Smith}, {Viaene}, \& {Wilson}}]{delooze2014a}
{De Looze}, I., {Fritz}, J., {Baes}, M., {et~al.} 2014, \aap, 571, A69

\bibitem[{{Draine} {et~al.}(2014){Draine}, {Aniano}, {Krause}, {Groves},
  {Sandstrom}, {Braun}, {Leroy}, {Klaas}, {Linz}, {Rix}, {Schinnerer},
  {Schmiedeke}, \& {Walter}}]{draine2014a}
{Draine}, B.~T., {Aniano}, G., {Krause}, O., {et~al.} 2014, \apj, 780, 172

\bibitem[{{Draine} \& {Li}(2007)}]{draine2007a}
{Draine}, B.~T. \& {Li}, A. 2007, \apj, 657, 810

\bibitem[{{Engelbracht} {et~al.}(2005){Engelbracht}, {Gordon}, {Rieke},
  {Werner}, {Dale}, \& {Latter}}]{engelbracht2005a}
{Engelbracht}, C.~W., {Gordon}, K.~D., {Rieke}, G.~H., {et~al.} 2005, \apjl,
  628, L29

\bibitem[{{Ferland}(1980)}]{ferland1980a}
{Ferland}, G.~J. 1980, \pasp, 92, 596

\bibitem[{{Ferland} {et~al.}(1998){Ferland}, {Korista}, {Verner}, {Ferguson},
  {Kingdon}, \& {Verner}}]{ferland1998a}
{Ferland}, G.~J., {Korista}, K.~T., {Verner}, D.~A., {et~al.} 1998, \pasp, 110,
  761

\bibitem[{{Ferland} {et~al.}(2013){Ferland}, {Porter}, {van Hoof}, {Williams},
  {Abel}, {Lykins}, {Shaw}, {Henney}, \& {Stancil}}]{ferland2013a}
{Ferland}, G.~J., {Porter}, R.~L., {van Hoof}, P.~A.~M., {et~al.} 2013, \rmxaa,
  49, 137

\bibitem[{{Fritz} {et~al.}(2006){Fritz}, {Franceschini}, \&
  {Hatziminaoglou}}]{fritz2006a}
{Fritz}, J., {Franceschini}, A., \& {Hatziminaoglou}, E. 2006, \mnras, 366, 767

\bibitem[{{Giovannoli} {et~al.}(2011){Giovannoli}, {Buat}, {Noll},
  {Burgarella}, \& {Magnelli}}]{giovannoli2011a}
{Giovannoli}, E., {Buat}, V., {Noll}, S., {Burgarella}, D., \& {Magnelli}, B.
  2011, \aap, 525, A150+

\bibitem[{{Gordon} {et~al.}(2001){Gordon}, {Misselt}, {Witt}, \&
  {Clayton}}]{gordon2001a}
{Gordon}, K.~D., {Misselt}, K.~A., {Witt}, A.~N., \& {Clayton}, G.~C. 2001,
  \apj, 551, 269

\bibitem[{{Grasha} {et~al.}(2013){Grasha}, {Calzetti}, {Andrews}, {Lee}, \&
  {Dale}}]{grasha2013a}
{Grasha}, K., {Calzetti}, D., {Andrews}, J.~E., {Lee}, J.~C., \& {Dale}, D.~A.
  2013, \apj, 773, 174

\bibitem[{{Hayes} {et~al.}(2011){Hayes}, {Schaerer}, {{\"O}stlin}, {Mas-Hesse},
  {Atek}, \& {Kunth}}]{hayes2011a}
{Hayes}, M., {Schaerer}, D., {{\"O}stlin}, G., {et~al.} 2011, \apj, 730, 8

\bibitem[{{Helou} {et~al.}(2004){Helou}, {Roussel}, {Appleton}, {Frayer},
  {Stolovy}, {Storrie-Lombardi}, {Hurt}, {Lowrance}, {Makovoz}, {Masci},
  {Surace}, {Gordon}, {Alonso-Herrero}, {Engelbracht}, {Misselt}, {Rieke},
  {Rieke}, {Willner}, {Pahre}, {Ashby}, {Fazio}, \& {Smith}}]{helou2004a}
{Helou}, G., {Roussel}, H., {Appleton}, P., {et~al.} 2004, \apjs, 154, 253

\bibitem[{{Helou} {et~al.}(1985){Helou}, {Soifer}, \&
  {Rowan-Robinson}}]{helou1985a}
{Helou}, G., {Soifer}, B.~T., \& {Rowan-Robinson}, M. 1985, \apjl, 298, L7

\bibitem[{{Hirashita} {et~al.}(2017){Hirashita}, {Burgarella}, \&
  {Bouwens}}]{hirashita2017a}
{Hirashita}, H., {Burgarella}, D., \& {Bouwens}, R.~J. 2017, \mnras, 472, 4587

\bibitem[{Hunter(2007)}]{matplotlib}
Hunter, J.~D. 2007, Computing in Science and Engineering, 9, 90

\bibitem[{{Inoue}(2001)}]{inoue2001a}
{Inoue}, A.~K. 2001, \aj, 122, 1788

\bibitem[{{Inoue}(2010)}]{inoue2010a}
{Inoue}, A.~K. 2010, \mnras, 401, 1325

\bibitem[{{Inoue}(2011)}]{inoue2011a}
{Inoue}, A.~K. 2011, \mnras, 415, 2920

\bibitem[{{Inoue} {et~al.}(2001){Inoue}, {Hirashita}, \& {Kamaya}}]{inoue2001b}
{Inoue}, A.~K., {Hirashita}, H., \& {Kamaya}, H. 2001, \apj, 555, 613

\bibitem[{{Inoue} {et~al.}(2006){Inoue}, {Iwata}, \& {Deharveng}}]{inoue2006a}
{Inoue}, A.~K., {Iwata}, I., \& {Deharveng}, J.-M. 2006, \mnras, 371, L1

\bibitem[{{Johnston} {et~al.}(2015){Johnston}, {Vaccari}, {Jarvis}, {Smith},
  {Giovannoli}, {H{\"a}u{\ss}ler}, \& {Prescott}}]{johnston2015a}
{Johnston}, R., {Vaccari}, M., {Jarvis}, M., {et~al.} 2015, \mnras, 453, 2540

\bibitem[{Jones {et~al.}(2001--)Jones, Oliphant, Peterson, {et~al.}}]{scipy}
Jones, E., Oliphant, T., Peterson, P., {et~al.} 2001--, {SciPy}: Open source
  scientific tools for {Python}

\bibitem[{{Kennicutt} {et~al.}(2011){Kennicutt}, {Calzetti}, {Aniano},
  {Appleton}, {Armus}, {Beir{\~a}o}, {Bolatto}, {Brandl}, {Crocker}, {Croxall},
  {Dale}, {Meyer}, {Draine}, {Engelbracht}, {Galametz}, {Gordon}, {Groves},
  {Hao}, {Helou}, {Hinz}, {Hunt}, {Johnson}, {Koda}, {Krause}, {Leroy}, {Li},
  {Meidt}, {Montiel}, {Murphy}, {Rahman}, {Rix}, {Roussel}, {Sandstrom},
  {Sauvage}, {Schinnerer}, {Skibba}, {Smith}, {Srinivasan}, {Vigroux},
  {Walter}, {Wilson}, {Wolfire}, \& {Zibetti}}]{kennicutt2011a}
{Kennicutt}, R.~C., {Calzetti}, D., {Aniano}, G., {et~al.} 2011, \pasp, 123,
  1347

\bibitem[{{Kong} {et~al.}(2004){Kong}, {Charlot}, {Brinchmann}, \&
  {Fall}}]{kong2004a}
{Kong}, X., {Charlot}, S., {Brinchmann}, J., \& {Fall}, S.~M. 2004, \mnras,
  349, 769

\bibitem[{{Koornneef} {et~al.}(1986){Koornneef}, {Bohlin}, {Buser}, {Horne}, \&
  {Turnshek}}]{koornneef1986a}
{Koornneef}, J., {Bohlin}, R., {Buser}, R., {Horne}, K., \& {Turnshek}, D.
  1986, Highlights of Astronomy, 7, 833

\bibitem[{{Kroupa}(2001)}]{kroupa2001a}
{Kroupa}, P. 2001, \mnras, 322, 231

\bibitem[{{Leitherer} {et~al.}(2002){Leitherer}, {Li}, {Calzetti}, \&
  {Heckman}}]{leitherer2002a}
{Leitherer}, C., {Li}, I.-H., {Calzetti}, D., \& {Heckman}, T.~M. 2002, \apjs,
  140, 303

\bibitem[{{Lo Faro} {et~al.}(2017){Lo Faro}, {Buat}, {Roehlly},
  {Alvarez-Marquez}, {Burgarella}, {Silva}, \& {Efstathiou}}]{lofaro2017a}
{Lo Faro}, B., {Buat}, V., {Roehlly}, Y., {et~al.} 2017, \mnras, 472, 1372

\bibitem[{{Ma{\l}ek} {et~al.}(2014){Ma{\l}ek}, {Pollo}, {Takeuchi}, {Buat},
  {Burgarella}, {Malkan}, {Giovannoli}, {Kurek}, \& {Matsuura}}]{malek2014a}
{Ma{\l}ek}, K., {Pollo}, A., {Takeuchi}, T.~T., {et~al.} 2014, \aap, 562, A15

\bibitem[{{Maraston}(2005)}]{maraston2005a}
{Maraston}, C. 2005, \mnras, 362, 799

\bibitem[{{Meiksin}(2006)}]{meiksin2006a}
{Meiksin}, A. 2006, \mnras, 365, 807

\bibitem[{{Nagao} {et~al.}(2011){Nagao}, {Maiolino}, {Marconi}, \&
  {Matsuhara}}]{nagao2011a}
{Nagao}, T., {Maiolino}, R., {Marconi}, A., \& {Matsuhara}, H. 2011, \aap, 526,
  A149

\bibitem[{{Noll} {et~al.}(2009){Noll}, {Burgarella}, {Giovannoli}, {Buat},
  {Marcillac}, \& {Mu{\~n}oz-Mateos}}]{noll2009a}
{Noll}, S., {Burgarella}, D., {Giovannoli}, E., {et~al.} 2009, \aap, 507, 1793

\bibitem[{{O'Donnell}(1994)}]{odonnell1994a}
{O'Donnell}, J.~E. 1994, \apj, 422, 158

\bibitem[{Oliphant(2007)}]{numpy}
Oliphant, T.~E. 2007, Computing in Science and Engineering, 9, 10

\bibitem[{{Overzier} {et~al.}(2011){Overzier}, {Heckman}, {Wang}, {Armus},
  {Buat}, {Howell}, {Meurer}, {Seibert}, {Siana}, {Basu-Zych}, {Charlot},
  {Gon{\c c}alves}, {Martin}, {Neill}, {Rich}, {Salim}, \&
  {Schiminovich}}]{overzier2011a}
{Overzier}, R.~A., {Heckman}, T.~M., {Wang}, J., {et~al.} 2011, \apjl, 726, L7+

\bibitem[{{Pacifici} {et~al.}(2012){Pacifici}, {Charlot}, {Blaizot}, \&
  {Brinchmann}}]{pacifici2012a}
{Pacifici}, C., {Charlot}, S., {Blaizot}, J., \& {Brinchmann}, J. 2012, \mnras,
  421, 2002

\bibitem[{{Papovich} {et~al.}(2001){Papovich}, {Dickinson}, \&
  {Ferguson}}]{papovich2001a}
{Papovich}, C., {Dickinson}, M., \& {Ferguson}, H.~C. 2001, \apj, 559, 620

\bibitem[{{Pappalardo} {et~al.}(2016){Pappalardo}, {Bizzocchi}, {Fritz},
  {Boselli}, {Boquien}, {Boissier}, {Baes}, {Ciesla}, {Bianchi}, {Clemens},
  {Viaene}, {Bendo}, {De Looze}, {Smith}, \& {Davies}}]{pappalardo2016a}
{Pappalardo}, C., {Bizzocchi}, L., {Fritz}, J., {et~al.} 2016, \aap, 589, A11

\bibitem[{{Pei}(1992)}]{pei1992a}
{Pei}, Y.~C. 1992, \apj, 395, 130

\bibitem[{{Popescu} {et~al.}(2000){Popescu}, {Misiriotis}, {Kylafis}, {Tuffs},
  \& {Fischera}}]{popescu2000a}
{Popescu}, C.~C., {Misiriotis}, A., {Kylafis}, N.~D., {Tuffs}, R.~J., \&
  {Fischera}, J. 2000, \aap, 362, 138

\bibitem[{{Popping} {et~al.}(2017){Popping}, {Puglisi}, \&
  {Norman}}]{popping2017a}
{Popping}, G., {Puglisi}, A., \& {Norman}, C.~A. 2017, \mnras, 472, 2315

\bibitem[{{Reddy} {et~al.}(2015){Reddy}, {Kriek}, {Shapley}, {Freeman},
  {Siana}, {Coil}, {Mobasher}, {Price}, {Sanders}, \& {Shivaei}}]{reddy2015a}
{Reddy}, N.~A., {Kriek}, M., {Shapley}, A.~E., {et~al.} 2015, \apj, 806, 259

\bibitem[{{Reddy} {et~al.}(2018){Reddy}, {Oesch}, {Bouwens}, {Montes},
  {Illingworth}, {Steidel}, {van Dokkum}, {Atek}, {Carollo}, {Cibinel},
  {Holden}, {Labb{\'e}}, {Magee}, {Morselli}, {Nelson}, \&
  {Wilkins}}]{reddy2018a}
{Reddy}, N.~A., {Oesch}, P.~A., {Bouwens}, R.~J., {et~al.} 2018, \apj, 853, 56

\bibitem[{{Reddy} {et~al.}(2016){Reddy}, {Steidel}, {Pettini}, \&
  {Bogosavljevi{\'c}}}]{reddy2016a}
{Reddy}, N.~A., {Steidel}, C.~C., {Pettini}, M., \& {Bogosavljevi{\'c}}, M.
  2016, \apj, 828, 107

\bibitem[{{Salim} {et~al.}(2018){Salim}, {Boquien}, \& {Lee}}]{salim2018a}
{Salim}, S., {Boquien}, M., \& {Lee}, J.~C. 2018, \apj, 859, 11

\bibitem[{{Salpeter}(1955)}]{salpeter1955a}
{Salpeter}, E.~E. 1955, \apj, 121, 161

\bibitem[{{Sandage}(1986)}]{sandage1986a}
{Sandage}, A. 1986, \aap, 161, 89

\bibitem[{{Sawicki}(2012)}]{sawicki2012a}
{Sawicki}, M. 2012, \pasp, 124, 1208

\bibitem[{{Silva} {et~al.}(1998){Silva}, {Granato}, {Bressan}, \&
  {Danese}}]{silva1998a}
{Silva}, L., {Granato}, G.~L., {Bressan}, A., \& {Danese}, L. 1998, \apj, 509,
  103

\bibitem[{{Stark} {et~al.}(2013){Stark}, {Schenker}, {Ellis}, {Robertson},
  {McLure}, \& {Dunlop}}]{stark2013a}
{Stark}, D.~P., {Schenker}, M.~A., {Ellis}, R., {et~al.} 2013, \apj, 763, 129

\bibitem[{{Trayford} {et~al.}(2017){Trayford}, {Camps}, {Theuns}, {Baes},
  {Bower}, {Crain}, {Gunawardhana}, {Schaller}, {Schaye}, \&
  {Frenk}}]{trayford2017a}
{Trayford}, J.~W., {Camps}, P., {Theuns}, T., {et~al.} 2017, \mnras, 470, 771

\bibitem[{{Tuffs} {et~al.}(2004){Tuffs}, {Popescu}, {V{\"o}lk}, {Kylafis}, \&
  {Dopita}}]{tuffs2004a}
{Tuffs}, R.~J., {Popescu}, C.~C., {V{\"o}lk}, H.~J., {Kylafis}, N.~D., \&
  {Dopita}, M.~A. 2004, \aap, 419, 821

\bibitem[{{Viaene} {et~al.}(2017){Viaene}, {Baes}, {Tamm}, {Tempel}, {Bendo},
  {Blommaert}, {Boquien}, {Boselli}, {Camps}, {Cooray}, {De Looze}, {De Vis},
  {Fern{\'a}ndez-Ontiveros}, {Fritz}, {Galametz}, {Gentile}, {Madden}, {Smith},
  {Spinoglio}, \& {Verstocken}}]{viaene2017a}
{Viaene}, S., {Baes}, M., {Tamm}, A., {et~al.} 2017, \aap, 599, A64

\bibitem[{{Vika} {et~al.}(2017){Vika}, {Ciesla}, {Charmandaris}, {Xilouris}, \&
  {Lebouteiller}}]{vika2017a}
{Vika}, M., {Ciesla}, L., {Charmandaris}, V., {Xilouris}, E.~M., \&
  {Lebouteiller}, V. 2017, \aap, 597, A51

\bibitem[{{Wild} {et~al.}(2011){Wild}, {Charlot}, {Brinchmann}, {Heckman},
  {Vince}, {Pacifici}, \& {Chevallard}}]{wild2011a}
{Wild}, V., {Charlot}, S., {Brinchmann}, J., {et~al.} 2011, \mnras, 417, 1760

\bibitem[{{Worthey}(1994)}]{worthey1994a}
{Worthey}, G. 1994, \apjs, 95, 107

\bibitem[{{Xilouris} {et~al.}(1999){Xilouris}, {Byun}, {Kylafis}, {Paleologou},
  \& {Papamastorakis}}]{xilouris1999a}
{Xilouris}, E.~M., {Byun}, Y.~I., {Kylafis}, N.~D., {Paleologou}, E.~V., \&
  {Papamastorakis}, J. 1999, \aap, 344, 868

\bibitem[{{Ye} {et~al.}(2016){Ye}, {Zou}, {Lin}, {Lian}, {Hu}, \&
  {Kong}}]{ye2016a}
{Ye}, C., {Zou}, H., {Lin}, L., {et~al.} 2016, \apj, 826, 209

\end{thebibliography}

\appendix

\section{General workflow}

A general schematic of \texttt{CIGALE} including these important choices detailed in Appendix~\ref{sec:choices} are shown in Fig.~\ref{fig:workflow}.

\begin{figure*}[!htbp]
 \includegraphics[width=\textwidth]{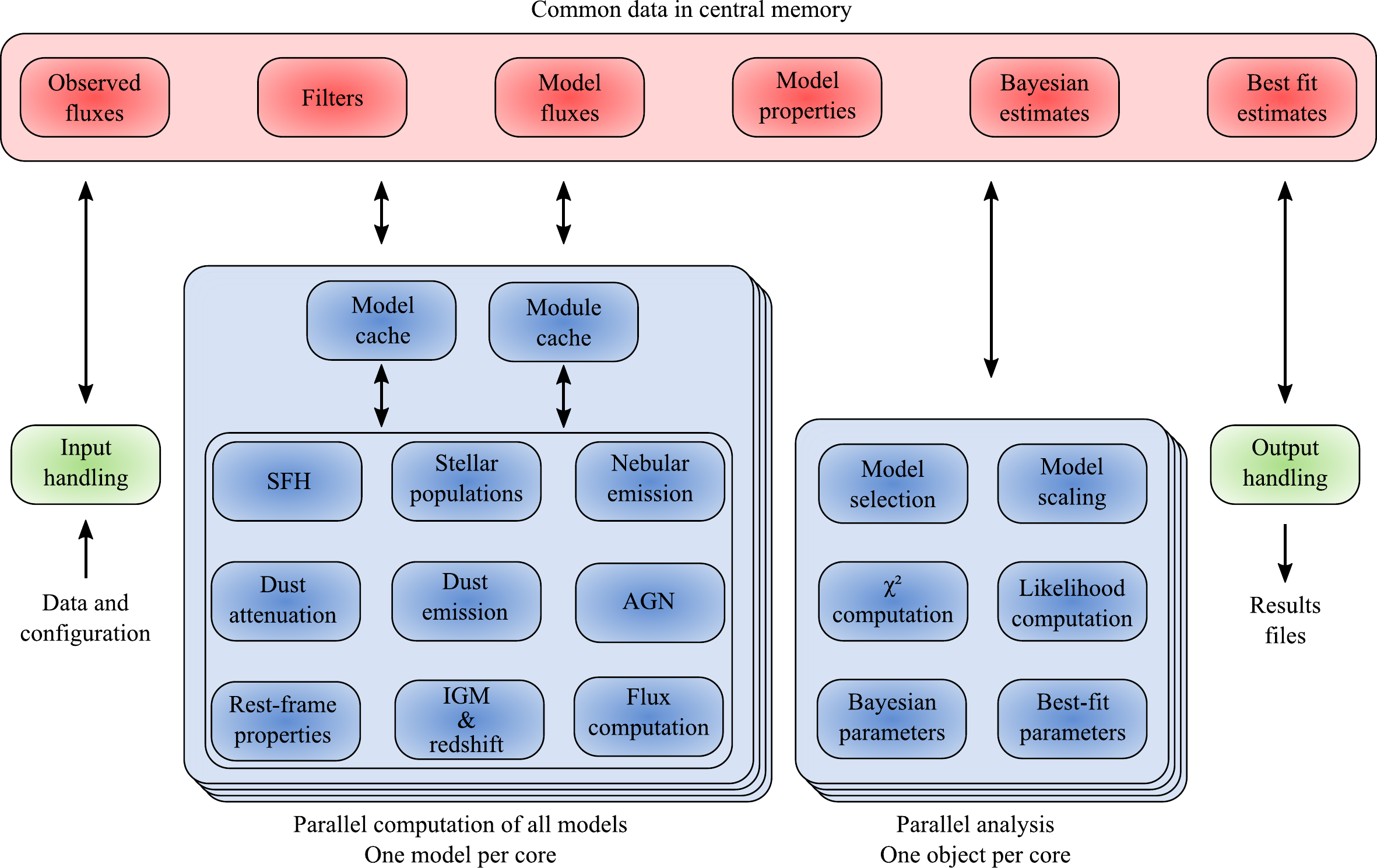}
 \caption{General work flow of \texttt{CIGALE} for SED fitting and parameter estimation. It can be split into 4 broad consecutive steps: 1) input handling in which configuration and data files are processed, 2) model generation in which the model SEDs and the associated physical properties are computed and stored in memory, 3) analysis in which the models are fitted to the observations and physical properties are estimated through the PDF, and finally 4) output handling in which the results are saved to disk. The input and output steps are shown in green, the parallel computation of the models and of the fits are shown in blue, and the main data residing in shared memory are shown in red.\label{fig:workflow}}
\end{figure*}

\section{Important implementation choices\label{sec:choices}}

The algorithms described above are central for computing models and estimating the physical properties of galaxies. However, applied as is, they would lead to extremely poor performance. We describe here two implementation choices we have made in \texttt{CIGALE} to achieve high performance with limited resources: parallelisation, caching, and computation by blocks of models.

\subsection{Parallelisation strategy}

With the increasing number of cores available in off--the--shelf computers, it is important that \texttt{CIGALE} can make full use of them. We have therefore parallelised the model generation and the analysis parts of \texttt{pcigale} as well as the generation of plots with \texttt{pcigale-plots}. The implementation makes use of the \texttt{multiprocessing} module of \texttt{python}, which allows for a process--based parallelism. The strategy is to create an arbitrary number of child processes (in general, but this is user--defined, corresponding to the number of available cores), which each computing one model or analysing one object at a time, on--demand from the main process.

With this strategy, costly communications between the main process and individual sub--processes can prove to be a bottleneck. To limit this issue, the computed models fluxes and physical properties are written into shared arrays accessible from all sub--processes rather than being communicated back to the main process. These shared arrays are also accessed directly by each sub--process carrying out the analysis of the objects and results are written into another set of shared arrays. This optimises both the computation speed and memory usage as no copy of the shared arrays needs to be made for individual sub--processes.

While launching the sub--processes incurs some slight start--up delay, the computation speed increase initially grows with the number of physical cores. However, the communication between the memory and the processor progressively becomes the main bottleneck with increasing numbers of cores, yielding sub--linear gains. Finally, while the gain of this method is substantial, there is however a small memory cost from using sub--processes.

\subsection{Cache system}

Another important aspect of the implementation of \texttt{CIGALE} to minimise computing time is the cache system. Caching is done at two levels. We cache the model creation modules and we cache partially computed SEDs.

The SED of an object is built by adding different physical components, and each of these physical components is added by a specific module, which is fully determined by its input parameters. Rather than instantiating repeatedly the same module each time we need to apply it to an SED, we instantiate it only once and store it in cache. This is especially valuable because the initialisation of a module can be expensive as it loads and computes all the data that do not depend on other modules. With caching, rather than being carried out each time we apply this module to an SED, the computations are done only once for the entire run.

We also store intermediate SED in cache to avoid having to compute each model from scratch. For instance the cache can contain an SED built with steps 1--to--6 (that is the full model but without redshift); steps 7 and 8 can then easily be applied to compute the flux at different redshifts without computing steps 1--to--6 again. Because SED objects can be particularly voluminous, they contain the full spectrum for each of the physical components, intermediate SED that are no longer of use to construct new SEDs are discarded. This is the case for instance for an SED that corresponds to an SFH not used by any of the models still to be computed. The models are computed in an order designed to optimise the efficiency of the cache. In practice at any given time it only contains a small number of intermediate SEDs necessary to ensure the full speedup, maximising cache lookup speed and minimising memory usage.

\subsection{Computation by blocks of models}

As detailed in Sect.~\ref{sec:analysis-modules}, the grid of models is computed in one step. While there are advantages in having all the models in memory (in particular in terms of speed), this can be very demanding for large grids. To allow the computation of grids of models larger than the size of the computer memory, \texttt{CIGALE} can compute models in several blocks. After the computation of each block, the physical properties of the targets are estimated using exclusively the models from this block and the results (likelihood--weighted means, likelihood--weighted standard deviations, best fit, sum of the likelihoods) stored in memory. After the physical properties have been estimated with all block, the results are combined to determine the overall likelihood--weighted means and standard deviations, and the overall best fit. If finding the overall best fit is trivial, as it is simply of lowest $\chi^2$ considering all the blocks, the estimation of the likelihood weighted physical properties is more complex. To do so, we adopt the variance parallel algorithm presented in \cite{chan1979a}, where we replace the number of elements in each block with the sum of the likelihood. Let $w_i$ be the sum of the likelihoods and $m_i$ the likelihood--weighted mean of a physical property for the block of models $j$. The overall likelihood--weighted mean $m$ is:
\begin{equation}
m = \frac{\sum_i m_i\times w_i}{\sum_i w_i}.
\end{equation}

The computation of the overall likelihood--weighted standard deviation $\sigma_i$ from the likelihood--weighted standard deviations for individual blocks $\sigma_i$ is more involved:
\begin{equation}
\sigma^2 = \sum_i \sigma_i^2 \times w_i + \frac{\sum_j \left(m_{i+1}-m_i\right)^2/\sum_i w_i}{\sum_i w_i}
.\end{equation}

\onecolumn

\section{Input and output physical properties\label{sec:properties}}

\subsection{Star formation history}

\FloatBarrier
\subsubsection{\texttt{sfh2exp}}
\begin{table*}[!htbp]
 \centering
 \begin{tabular}{lll}
   \multicolumn{3}{c}{Input parameters of \texttt{sfh2exp}}\\\hline\hline
  Parameter               &Unit&Description\\\hline
  \texttt{tau\_main}      &Myr                &e-folding time of the main stellar population model\\
  \texttt{tau\_burst}     &Myr                &e-folding time of the late starburst population model \\
  \texttt{f\_burst}       &--                 &Mass fraction of the late burst population\\
  \texttt{age}            &Myr                &Age of the main stellar population in the galaxy\\
  \texttt{burst\_age}     &Myr                &Age of the late burst\\
  \texttt{sfr\_0}         &M$_\odot$~yr$^{-1}$&Value of SFR at $t=0$ if normalise is False\\
  \texttt{normalise}      &--                 &Normalise the SFH to produce one solar mass\\\hline
 \end{tabular}
 \caption{Input parameters of the \texttt{sfh2exp} module.\label{tab:input_sfh2exp}}
\end{table*}

\begin{table*}[!htbp]
 \centering
 \begin{tabular}{lll}
   \multicolumn{3}{c}{Output parameters of \texttt{sfh2exp}}\\\hline\hline
  Parameter               &Unit               &Description\\\hline
  \texttt{sfh.sfr}        &M$_\odot$~yr$^{-1}$&Instantaneous SFR\\
  \texttt{sfh.sfr10Myrs}  &M$_\odot$~yr$^{-1}$&Average SFR over 10 Myr\\
  \texttt{sfh.sfr100Myrs} &M$_\odot$~yr$^{-1}$&Average SFR over 100 Myr\\
  \texttt{sfh.age}        &Myr               &Age of the main stellar population in the galaxy\\
  \texttt{sfh.integrated} &M$_\odot$         &Integral of the SFH\\
  \texttt{sfh.tau\_main}  &Myr               &e-folding time of the main stellar population model\\
  \texttt{sfh.tau\_burst} &Myr               &e-folding time of the late starburst population model\\
  \texttt{sfh.f\_burst}   &--                &Mass fraction of the late burst population\\
  \texttt{sfh.burst\_age} &Myr               &Age of the late burst in Myr\\\hline
 \end{tabular}
 \caption{Output parameters of the \texttt{sfh2exp} module.\label{tab:output_sfh2exp}}
\end{table*}

\FloatBarrier
\subsubsection{\texttt{sfhdelayed}}
\begin{table*}[!htbp]
 \centering
 \begin{tabular}{lll}
   \multicolumn{3}{c}{Input parameters of \texttt{sfhdelayed}}\\\hline\hline
  Parameter               &Unit&Description\\\hline
  \texttt{tau\_main}      &Myr                &e-folding time of the main stellar population model\\
  \texttt{age\_main}      &Myr                &Age of the main stellar population in the galaxy\\
  \texttt{tau\_burst}     &Myr                &e-folding time of the late starburst population model\\
  \texttt{age\_burst}     &Myr                &Age of the late burst\\
  \texttt{f\_burst}       &--                 &Mass fraction of the late burst population\\
  \texttt{sfr\_A}         &M$_\odot$~yr$^{-1}$&Value of SFR at $t=0$ if normalise is False\\
  \texttt{normalise}      &--                 &Normalise the SFH to produce one solar mass\\\hline
 \end{tabular}
 \caption{Input parameters of the \texttt{sfhdelayed} module.\label{tab:input_sfhdelayed}}
\end{table*}

\begin{table*}[!htbp]
 \centering
 \begin{tabular}{lll}
   \multicolumn{3}{c}{Output parameters of \texttt{sfhdelayed}}\\\hline\hline
  Parameter               &Unit               &Description\\\hline
  \texttt{sfh.sfr}        &M$_\odot$~yr$^{-1}$&Instantaneous SFR\\
  \texttt{sfh.sfr10Myrs}  &M$_\odot$~yr$^{-1}$&Average SFR over 10 Myr\\
  \texttt{sfh.sfr100Myrs} &M$_\odot$~yr$^{-1}$&Average SFR over 100 Myr\\
  \texttt{sfh.integrated} &M$_\odot$         &Integral of the SFH\\
  \texttt{sfh.age\_main}  &Myr                &Age of the main stellar population in the galaxy\\
  \texttt{sfh.tau\_main}  &Myr                &e-folding time of the main stellar population model\\
  \texttt{sfh.age\_burst} &Myr                &Age of the late burst\\
  \texttt{sfh.tau\_burst} &--                 &e-folding time of the late starburst population model\\
  \texttt{sfh.f\_burst}   &Myr                &Mass fraction of the late burst population\\\hline
 \end{tabular}
 \caption{Output parameters of the \texttt{sfhdelayed} module.\label{tab:output_sfhdelayed}}
\end{table*}

\FloatBarrier
\subsubsection{\texttt{sfhdelayedbq}}

\begin{table*}[!htbp]
 \centering
 \begin{tabular}{lll}
   \multicolumn{3}{c}{Input parameters of \texttt{sfhdelayedbq}}\\\hline\hline
  Parameter               &Unit&Description\\\hline
  \texttt{tau\_main}      &Myr                &e-folding time of the main stellar population model\\
  \texttt{age\_main}      &Myr                &Age of the main stellar population in the galaxy\\
  \texttt{age\_bq}        &Myr                &Age of the burst/quench\\
  \texttt{r\_sfr}         &--                 &Ratio of the SFR after/before \texttt{age\_bq}\\
  \texttt{sfr\_A}         &M$_\odot$~yr$^{-1}$&Value of SFR at $t=0$ if normalise is False\\
  \texttt{normalise}      &--                 &Normalise the SFH to produce one solar mass\\\hline
 \end{tabular}
 \caption{Input parameters of the \texttt{sfhdelayedbq} module.\label{tab:input_sfhdelayedbq}}
\end{table*}

\begin{table*}[!htbp]
 \centering
 \begin{tabular}{lll}
   \multicolumn{3}{c}{Output parameters of \texttt{sfhdelayedbq}}\\\hline\hline
  Parameter               &Unit               &Description\\\hline
  \texttt{sfh.sfr}        &M$_\odot$~yr$^{-1}$&Instantaneous SFR\\
  \texttt{sfh.sfr10Myrs}  &M$_\odot$~yr$^{-1}$&Average SFR over 10 Myr\\
  \texttt{sfh.sfr100Myrs} &M$_\odot$~yr$^{-1}$&Average SFR over 100 Myr\\
  \texttt{sfh.integrated} &M$_\odot$          &Integral of the SFH\\
  \texttt{sfh.age\_main}  &Myr                &Age of the main stellar population in the galaxy\\
  \texttt{sfh.tau\_main}  &Myr                &e-folding time of the main stellar population model\\
  \texttt{sfh.age\_bq}    &Myr                &Age of the burst/quench\\
  \texttt{sfh.r\_sfr}     &--                 &Ratio of the SFR after/before \texttt{age\_bq}\\\hline
 \end{tabular}
 \caption{Output parameters of the \texttt{sfhdelayedbq} module.\label{tab:output_sfhdelayedbq}}
\end{table*}

\FloatBarrier
\subsubsection{sfhperiodic}
\begin{table*}[!htbp]
 \centering
 \begin{tabular}{lll}
   \multicolumn{3}{c}{Input parameters of \texttt{sfhperiodic}}\\\hline\hline
  Parameter               &Unit&Description\\\hline
  \texttt{type\_bursts}   &--                 &Type of the individual star formation episodes. 0: exponential, 1: delayed, 2: rectangle\\
  \texttt{delta\_bursts}  &Myr                &Elapsed time between the beginning of each burst\\
  \texttt{tau\_bursts}    &Myr                &Duration (rectangle) or e-folding time of all short events\\
  \texttt{age}            &Myr                &Age of the main stellar population in the galaxy\\
  \texttt{sfr\_A}         &M$_\odot$~yr$^{-1}$&Multiplicative factor controlling the amplitude of SFR (valid for each event) if normalise is False\\
  \texttt{normalise}      &--                 &Normalise the SFH to produce one solar mass\\\hline
 \end{tabular}
 \caption{Input parameters of the \texttt{sfhperiodic} module.\label{tab:input_sfhperiodic}}
\end{table*}

\begin{table*}[!htbp]
 \centering
 \begin{tabular}{lll}
   \multicolumn{3}{c}{Output parameters of \texttt{sfhperiodic}}\\\hline\hline
  Parameter                 &Unit               &Description\\\hline
  \texttt{sfh.sfr}          &M$_\odot$~yr$^{-1}$&Instantaneous SFR\\
  \texttt{sfh.sfr10Myrs}    &M$_\odot$~yr$^{-1}$&Average SFR over 10 Myr\\
  \texttt{sfh.sfr100Myrs}   &M$_\odot$~yr$^{-1}$&Average SFR over 100 Myr\\
  \texttt{sfh.integrated}   &M$_\odot$          &Integral of the SFH\\
  \texttt{sfh.type\_bursts} &--                 &Type of the individual star formation episodes\\
  \texttt{sfh.delta\_bursts}&Myr                &Elapsed time between the beginning of each burst\\
  \texttt{sfh.tau\_bursts}  &Myr                &Duration (rectangle) or e-folding time of all short events\\\hline
 \end{tabular}
 \caption{Output parameters of the \texttt{sfhperiodic} module.\label{tab:output_sfhperiodic}}
\end{table*}

\FloatBarrier
\subsubsection{\texttt{sfh\_buat08}}
\begin{table*}[!htbp]
 \centering
 \begin{tabular}{lll}
   \multicolumn{3}{c}{Input parameters of \texttt{sfh\_buat08}}\\\hline\hline
  Parameter               &Unit&Description\\\hline
  \texttt{velocity}       &km~s$^{-1}$        &Rotational velocity of the galaxy\\
  \texttt{age}            &Myr                &Age of the oldest stars in the galaxy\\
  \texttt{normalise}      &--                 &Normalise the SFH to produce one solar mass\\\hline
 \end{tabular}
 \caption{Input parameters of the \texttt{sfh\_buat08} module.\label{tab:input_sfh_buat08}}
\end{table*}

\begin{table*}[!htbp]
 \centering
 \begin{tabular}{lll}
   \multicolumn{3}{c}{Output parameters of \texttt{sfh\_buat08}}\\\hline\hline
  Parameter                 &Unit               &Description\\\hline
  \texttt{sfh.sfr}          &M$_\odot$~yr$^{-1}$&Instantaneous SFR\\
  \texttt{sfh.sfr10Myrs}    &M$_\odot$~yr$^{-1}$&Average SFR over 10 Myr\\
  \texttt{sfh.sfr100Myrs}   &M$_\odot$~yr$^{-1}$&Average SFR over 100 Myr\\
  \texttt{sfh.integrated}   &M$_\odot$          &Integral of the SFH\\
  \texttt{sfh.velocity}     &km~s$^{-1}$        &Rotational velocity of the galaxy\\\hline
 \end{tabular}
 \caption{Output parameters of the \texttt{sfh\_buat08} module.\label{tab:output_sfh_buat08}}
\end{table*}

\FloatBarrier
\subsubsection{\texttt{sfhfromfile}}
\begin{table*}[!htbp]
 \centering
 \begin{tabular}{lll}
   \multicolumn{3}{c}{Input parameters of \texttt{sfhfromfile}}\\\hline\hline
  Parameter               &Unit&Description\\\hline
  \texttt{filename}       &-- &Name of the file containing the SFH\\
  \texttt{sfr\_column}    &-- &List of column indices of the SFR\\
  \texttt{age}            &Myr&Age in Myr at which the SFH will be looked at\\
  \texttt{normalise}      &-- &Normalise the SFH to produce one solar mass\\\hline
 \end{tabular}
 \caption{Input parameters of the \texttt{sfhfromfile} module.\label{tab:input_sfhfromfile}}
\end{table*}

\begin{table*}[!htbp]
 \centering
 \begin{tabular}{lll}
   \multicolumn{3}{c}{Output parameters of \texttt{sfhfromfile}}\\\hline\hline
  Parameter                 &Unit               &Description\\\hline
  \texttt{sfh.sfr}          &M$_\odot$~yr$^{-1}$&Instantaneous SFR\\
  \texttt{sfh.sfr10Myrs}    &M$_\odot$~yr$^{-1}$&Average SFR over 10 Myr\\
  \texttt{sfh.sfr100Myrs}   &M$_\odot$~yr$^{-1}$&Average SFR over 100 Myr\\
  \texttt{sfh.integrated}   &M$_\odot$          &Integral of the SFH\\
  \texttt{sfh.index}        &--                 &Index of the column\\\hline
 \end{tabular}
 \caption{Output parameters of the \texttt{sfhfromfile} module.\label{tab:output_sfhfromfile}}
\end{table*}

\FloatBarrier
\subsection{Stellar populations}
\subsubsection{\texttt{bc03}}
\begin{table*}[!htbp]
 \centering
 \begin{tabular}{lll}
   \multicolumn{3}{c}{Input parameters of \texttt{bc03}}\\\hline\hline
  Parameter               &Unit&Description\\\hline
  \texttt{imf}            &--  &Initial mass function: 0 (Salpeter) or 1 (Chabrier)\\
  \texttt{metallicity}    &--  &Metallicity\\
  \texttt{separation\_age}&Myr &Age of the separation between the young and the old star populations\\\hline
 \end{tabular}
 \caption{Input parameters of the \texttt{bc03} module.\label{tab:input_bc03}}
\end{table*}

\begin{table*}[!htbp]
 \centering
 \begin{tabular}{lll}
   \multicolumn{3}{c}{Output parameters of \texttt{bc03}}\\\hline\hline
  Parameter                 &Unit               &Description\\\hline
  \texttt{stellar.imf}            &--  &Initial mass function: 0 (Salpeter) or 1 (Chabrier)\\
  \texttt{stellar.metallicity}    &--  &Metallicity\\
  \texttt{stellar.old\_young\_separation\_age}&Myr &Age of the separation between the young and the old star populations\\
  \texttt{stellar.m\_star\_young}             &M$_\odot$&Stellar mass of the young population\\
  \texttt{stellar.m\_gas\_young}              &M$_\odot$&Gas mass of the young population\\
  \texttt{stellar.n\_ly\_young}               &Photons  &Number of Lyman continuum photons of the young population\\
  \texttt{stellar.lum\_ly\_young}             &W        &Luminosity of the Lyman continuum of the young population\\
  \texttt{stellar.lum\_young}                 &W        &Luminosity of the young population\\
  \texttt{stellar.m\_star\_old}               &M$_\odot$&Stellar mass of the old population\\
  \texttt{stellar.m\_gas\_old}                &M$_\odot$&Gas mass of the old population\\
  \texttt{stellar.n\_ly\_old}                 &Photons  &Number of Lyman continuum photons of the old population\\
  \texttt{stellar.lum\_ly\_old}               &W        &Luminosity of the Lyman continuum of the old population\\
  \texttt{stellar.lum\_old}                   &W        &Luminosity of the old population\\
  \texttt{stellar.m\_star}                    &M$_\odot$&Total stellar mass\\
  \texttt{stellar.m\_gas}                     &M$_\odot$&Total gas mass\\
  \texttt{stellar.n\_ly}                      &Photons  &Total number of Lyman continuum photons\\
  \texttt{stellar.lum\_ly}                    &W        &Total Lyman continuum luminosity\\
  \texttt{stellar.lum}                        &W        &Total luminosity\\
  \texttt{stellar.age\_m\_star}               &Myr      &Mass--weighted age\\\hline
 \end{tabular}
 \caption{Output parameters of the \texttt{bc03} module.\label{tab:output_bc03}}
\end{table*}

\FloatBarrier
\subsubsection{\texttt{m2005}}
\begin{table*}[!htbp]
 \centering
 \begin{tabular}{lll}
   \multicolumn{3}{c}{Input parameters of \texttt{m2005}}\\\hline\hline
  Parameter               &Unit&Description\\\hline
  \texttt{imf}            &--  &Initial mass function: 0 (Salpeter) or 1 (Chabrier)\\
  \texttt{metallicity}    &--  &Metallicity\\
  \texttt{separation\_age}&Myr &Age of the separation between the young and the old star populations\\\hline
 \end{tabular}
 \caption{Input parameters of the \texttt{m2005} module.\label{tab:input_m2005}}
\end{table*}

\begin{table*}[!htbp]
 \centering
 \begin{tabular}{lll}
   \multicolumn{3}{c}{Output parameters of \texttt{m2005}}\\\hline\hline
  Parameter                                   &Unit     &Description\\\hline
  \texttt{stellar.imf}                        &--       &Initial mass function: 0 (Salpeter) or 1 (Chabrier)\\
  \texttt{stellar.metallicity}                &--       &Metallicity\\
  \texttt{stellar.old\_young\_separation\_age}&Myr      &Age of the separation between the young and the old star populations\\
  \texttt{stellar.mass\_total\_young}         &M$_\odot$&Stellar mass of the young population\\
  \texttt{stellar.mass\_alive\_young}         &M$_\odot$&Alive stars mass of the young population\\
  \texttt{stellar.mass\_white\_dwarf\_young}  &M$_\odot$&White dwarfs mass of the young population\\
  \texttt{stellar.mass\_neutron\_young}       &M$_\odot$&Neutron stars mass of the young population\\
  \texttt{stellar.mass\_black\_hole\_young}   &M$_\odot$&Black home mass of the young population\\
  \texttt{stellar.lum\_old}                   &W         &Luminosity of the old population\\
  \texttt{stellar.mass\_total\_old}           &M$_\odot$&Stellar mass of the old population\\
  \texttt{stellar.mass\_alive\_old}           &M$_\odot$&Alive stars mass of the old population\\
  \texttt{stellar.mass\_white\_dwarf\_old}    &M$_\odot$&White dwarfs mass of the old population\\
  \texttt{stellar.mass\_neutron\_old}         &M$_\odot$&Neutron stars mass of the old population\\
  \texttt{stellar.mass\_black\_hole\_old}     &M$_\odot$&Black home mass of the old population\\
  \texttt{stellar.lum\_old}                   &W         &Luminosity of the old population\\
  \texttt{stellar.mass\_total}                &M$_\odot$&Total stellar mass\\
  \texttt{stellar.mass\_alive}                &M$_\odot$&Total alive stars mass\\
  \texttt{stellar.mass\_white\_dwarf}         &M$_\odot$&Total white dwarfs mass\\
  \texttt{stellar.mass\_neutron}              &M$_\odot$&Total neutron stars mass\\
  \texttt{stellar.mass\_black\_hole}          &M$_\odot$&Total black home mass\\
  \texttt{stellar.lum}                        &W         &Total luminosity\\\hline
 \end{tabular}
 \caption{Output parameters of the \texttt{m2005} module.\label{tab:output_m2005}}
\end{table*}

\FloatBarrier
\subsection{Nebular emission}
\subsubsection{\texttt{nebular}}
\begin{table*}[!htbp]
 \centering
 \begin{tabular}{lll}
   \multicolumn{3}{c}{Input parameters of \texttt{nebular}}\\\hline\hline
  Parameter               &Unit       &Description\\\hline
  \texttt{logU}           &--         &Ionisation parameter\\
  \texttt{f\_esc}         &--         &Fraction of Lyman continuum photons escaping the galaxy\\
  \texttt{f\_dust}        &--         &Fraction of Lyman continuum photons absorbed by dust\\
  \texttt{lines\_width}    &km~s$^{-1}$&Line width\\
  \texttt{emission}       &--         &Include nebular emission\\\hline
 \end{tabular}
 \caption{Input parameters of the \texttt{nebular} module.\label{tab:input_nebular}}
\end{table*}

\begin{table*}[!htbp]
 \centering
 \begin{tabular}{lll}
   \multicolumn{3}{c}{Output parameters of \texttt{nebular}}\\\hline\hline
  Parameter                        &Unit       &Description\\\hline
  \texttt{nebular.f\_esc}          &--         &Fraction of Lyman continuum photons escaping the galaxy\\
  \texttt{nebular.f\_dust}         &--         &Fraction of Lyman continuum photons absorbed by dust\\
  \texttt{dust.dust\_luminosity}   &W          &Luminosity absorbed by dust\\
  \texttt{nebular.lines\_width}     &km~s$^{-1}$&Line width\\
  \texttt{nebular.logU}            &--         &Ionisation parameter\\\hline
 \end{tabular}
 \caption{Output parameters of the \texttt{nebular} module.\label{tab:output_nebular}}
\end{table*}

\FloatBarrier
\subsection{Attenuation law}

\FloatBarrier
\subsubsection{\texttt{dustatt\_modified\_CF00}}
\begin{table*}[!htbp]
 \centering
 \begin{tabular}{lll}
   \multicolumn{3}{c}{Input parameters of \texttt{dustatt\_modified\_CF00}}\\\hline\hline
  Parameter                         &Unit  &Description\\\hline
  \texttt{Av\_ISM}                  &mag   &V-band attenuation in the interstellar medium\\
  \texttt{mu}                       &--    &Av\_ISM / (Av\_BC+Av\_ISM)\\
  \texttt{slope\_ISM}               &--    &Power law slope of the attenuation in the ISM\\
  \texttt{slope\_BC}                &--    &Power law slope of the attenuation in the birth clouds\\
  \texttt{filters}                  &--    &Filters for which the attenuation will be computed and added to the SED information dictionary\\\hline
 \end{tabular}
 \caption{Input parameters of the \texttt{dustatt\_modified\_CF00} module.\label{tab:input_dustatt_modified_CF00}}
\end{table*}

\begin{table*}[!htbp]
 \centering
 \begin{tabular}{lll}
   \multicolumn{3}{c}{Output parameters of \texttt{dustatt\_modified\_CF00}}\\\hline\hline
  Parameter                         &Unit  &Description\\\hline
  \texttt{attenuation.Av\_ISM}      &mag   &V-band attenuation in the interstellar medium\\
  \texttt{attenuation.Av\_BC}       &mag   &V-band attenuation in the birth clouds\\
  \texttt{attenuation.mu}           &--    &Av\_ISM / (Av\_BC+Av\_ISM)\\
  \texttt{attenuation.slope\_ISM}   &--    &Power law slope of the attenuation in the ISM\\
  \texttt{attenuation.slope\_BC}    &--    &Power law slope of the attenuation in the birth clouds\\
  \texttt{dust.luminosity}          &W     &Luminosity absorbed by the dust\\
  \texttt{attenuation.[filters]}    &--     &Attenuation in \texttt{filters}\\\hline
 \end{tabular}
 \caption{Input parameters of the \texttt{dustatt\_modified\_CF00} module.\label{tab:output_dustatt_modified_CF00}}
\end{table*}

\FloatBarrier
\subsubsection{\texttt{dustatt\_modified\_starburst}}
\begin{table*}[!htbp]
 \centering
 \begin{tabular}{lll}
   \multicolumn{3}{c}{Input parameters of \texttt{dustatt\_modified\_starburst}}\\\hline\hline
  Parameter                         &Unit  &Description\\\hline
  \texttt{E\_BV\_lines}             &mag   &E(B-V)l, the colour excess of the nebular lines light for both the young and old population\\
  \texttt{E\_BV\_factor}            &--    &Reduction factor to apply on \textsc{E\_BV\_lines} to compute E(B-V)s the stellar continuum attenuation\\
  \texttt{uv\_bump\_wavelength}     &nm    &Central wavelength of the UV bump\\
  \texttt{uv\_bump\_width}          &nm    &Width (FWHM) of the UV bump\\
  \texttt{uv\_bump\_amplitude}      &--    &Amplitude of the UV bump. For the Milky Way: 3\\
  \texttt{powerlaw\_slope}          &--    &Slope delta of the power law modifying the attenuation curve\\
  \texttt{Ext\_law\_emission\_lines}&--    &Extinction law to use for attenuating the emission lines flux.\\&& Possible values are: 1, 2, 3. 1: MW, 2: LMC, 3: SMC\\
  \texttt{Rv}                       &--    &Ratio of total to selective extinction, A\_V / E(B-V), for the extinction curve applied to\\&& emission lines. Standard value is 3.1 for MW using CCM89, but can be changed.\\&& For SMC and LMC using Pei92 the value is automatically set to 2.93 and 3.16 respectively,\\&& no matter the value you write.\\
  \texttt{filters}                  &--    &Filters for which the attenuation will be computed and added to the SED information dictionary\\\hline
 \end{tabular}
 \caption{Input parameters of the \texttt{dustatt\_modified\_starburst} module.\label{tab:input_dustatt_modified_starburst}}
\end{table*}

\begin{table*}[!htbp]
 \centering
 \begin{tabular}{lll}
   \multicolumn{3}{c}{Output parameters of \texttt{dustatt\_modified\_starburst}}\\\hline\hline
  Parameter                                &Unit  &Description\\\hline
  \texttt{dust.luminosity}                 &W     &Luminosity absorbed by the dust\\
  \texttt{attenuation.[filters]}           &--    &Attenuation in \texttt{filters}\\
  \texttt{attenuation.E\_BV\_lines}        &mag   &E(B-V)l, the colour excess of the nebular lines light for both the\\&& young and old population\\
  \texttt{attenuation.E\_BVs}              &mag   &E(B-V)s, the colour excess of the stellar light for both the young and old population\\
  \texttt{attenuation.E\_BV\_factor}       &--    &Reduction factor to apply on \texttt{E\_BV\_lines} to compute E(B-V)s the stellar continuum\\&& attenuation\\
  \texttt{attenuation.uv\_bump\_wavelength}&nm    &Central wavelength of the UV bump\\
  \texttt{attenuation.uv\_bump\_width}     &nm    &Width (FWHM) of the UV bump\\
  \texttt{attenuation.uv\_bump\_amplitude} &--    &Amplitude of the UV bump. For the Milky Way: 3\\
  \texttt{attenuation.powerlaw\_slope}     &--    &Slope delta of the power law modifying the attenuation curve\\
  \texttt{attenuation.filters}             &--    &Filters for which the attenuation will be computed and added to\\&& the SED information dictionary\\\hline
 \end{tabular}
 \caption{Output parameters of the \texttt{dustatt\_modified\_starburst} module.\label{taboutnput_dustatt_modified_starburst}}
\end{table*}

\FloatBarrier
\subsection{Dust emission}
\subsubsection{\texttt{dale2014}}
\begin{table*}[!htbp]
 \centering
 \begin{tabular}{lll}
   \multicolumn{3}{c}{Input parameters of \texttt{dale2014}}\\\hline\hline
  Parameter                         &Unit  &Description\\\hline
  \texttt{fracAGN}                  &mag   &AGN fraction\\
  \texttt{alpha}                    &--    &Alpha slope\\\hline
 \end{tabular}
 \caption{Input parameters of the \texttt{dale2014} module.\label{tab:input_dale2014}}
\end{table*}

\begin{table*}[!htbp]
 \centering
 \begin{tabular}{lll}
   \multicolumn{3}{c}{Output parameters of \texttt{dale2014}}\\\hline\hline
  Parameter                         &Unit  &Description\\\hline
  \texttt{dust.luminosity}          &W     &Dust luminosity (only added where there are no stellar populations)\\
  \texttt{agn.fracAGN\_dale2014}    &mag   &AGN fraction\\
  \texttt{dust.alpha}               &--    &Alpha slope\\\hline
 \end{tabular}
 \caption{Output parameters of the \texttt{dale2014} module.\label{tab:output_dale2014}}
\end{table*}

\FloatBarrier
\subsubsection{\texttt{dl2007}}
\begin{table*}[!htbp]
 \centering
 \begin{tabular}{lll}
   \multicolumn{3}{c}{Input parameters of \texttt{dl2007}}\\\hline\hline
  Parameter                         &Unit  &Description\\\hline
  \texttt{qpah}                     &--    &Mass fraction of PAH\\
  \texttt{umin}                     &Habing&Minimum radiation field\\
  \texttt{umax}                     &Habing&Maximum radiation field\\
  \texttt{gamma}                    &--    &Fraction illuminated from Umin to Umax\\\hline
 \end{tabular}
 \caption{Input parameters of the \texttt{dl2007} module.\label{tab:input_dl2007}}
\end{table*}

\begin{table*}[!htbp]
 \centering
 \begin{tabular}{lll}
   \multicolumn{3}{c}{Input parameters of \texttt{dl2007}}\\\hline\hline
  Parameter                         &Unit  &Description\\\hline
  \texttt{dust.luminosity}          &W     &Dust luminosity (only added where there are no stellar populations)\\
  \texttt{dust.qpah}                &--    &Mass fraction of PAH\\
  \texttt{dust.umin}                &Habing&Minimum radiation field\\
  \texttt{dust.umax}                &Habing&Maximum radiation field\\
  \texttt{dust.gamma}               &--    &Fraction illuminated from Umin to Umax\\
  \texttt{dust.mass}                &kg    &Dust mass\\\hline
 \end{tabular}
 \caption{Output parameters of the \texttt{dl2007} module.\label{tab:output_dl2007}}
\end{table*}

\FloatBarrier
\subsubsection{\texttt{dl2014}}
\begin{table*}[!htbp]
 \centering
 \begin{tabular}{lll}
   \multicolumn{3}{c}{Input parameters of \texttt{dl2014}}\\\hline\hline
  Parameter                         &Unit  &Description\\\hline
  \texttt{qpah}                     &--    &Mass fraction of PAH\\
  \texttt{umin}                     &Habing&Minimum radiation field\\
  \texttt{alpha}                    &--    &Powerlaw slope dU/dM propto U\^{}alpha\\
  \texttt{gamma}                    &--    &Fraction illuminated from Umin to Umax\\\hline
 \end{tabular}
 \caption{Input parameters of the \texttt{dl2007} module.\label{tab:input_dl2014}}
\end{table*}

\begin{table*}[!htbp]
 \centering
 \begin{tabular}{lll}
   \multicolumn{3}{c}{Input parameters of \texttt{dl2014}}\\\hline\hline
  Parameter                         &Unit  &Description\\\hline
  \texttt{dust.luminosity}          &W     &Dust luminosity (only added where there are no stellar populations)\\
  \texttt{dust.qpah}                &--    &Mass fraction of PAH\\
  \texttt{dust.umin}                &Habing&Minimum radiation field\\
  \texttt{dust.alpha}               &--    &Powerlaw slope dU/dM propto U\^{}alpha\\
  \texttt{dust.gamma}               &--    &Fraction illuminated from Umin to Umax\\
  \texttt{dust.mass}                &kg    &Dust mass\\\hline
 \end{tabular}
 \caption{Output parameters of the \texttt{dl2007} module.\label{tab:output_dl2014}}
\end{table*}

\FloatBarrier
\subsubsection{\texttt{casey2012}}
\begin{table*}[!htbp]
 \centering
 \begin{tabular}{lll}
   \multicolumn{3}{c}{Input parameters of \texttt{casey2012}}\\\hline\hline
  Parameter                         &Unit  &Description\\\hline
  \texttt{temperature}              &K     &Temperature of the dust\\
  \texttt{beta}                     &--    &Emissivity index of the dust\\
  \texttt{alpha}                    &--    &Mid-infrared powerlaw slope\\\hline
 \end{tabular}
 \caption{Input parameters of the \texttt{casey2012} module.\label{tab:input_casey2012}}
\end{table*}

\begin{table*}[!htbp]
 \centering
 \begin{tabular}{lll}
   \multicolumn{3}{c}{Input parameters of \texttt{casey2012}}\\\hline\hline
  Parameter                         &Unit  &Description\\\hline
  \texttt{dust.luminosity}          &W     &Dust luminosity (only added where there are no stellar populations)\\
  \texttt{dust.temperature}         &K     &Temperature of the dust\\
  \texttt{dust.beta}                &--    &Emissivity index of the dust\\
  \texttt{dust.alpha}               &--    &Mid-infrared powerlaw slope\\\hline
 \end{tabular}
 \caption{Output parameters of the \texttt{casey2012} module.\label{tab:output_casey2012}}
\end{table*}

\FloatBarrier
\subsection{Synchrotron radio emission}
\subsubsection{synchrotron}
\begin{table*}[!htbp]
 \centering
 \begin{tabular}{lll}
   \multicolumn{3}{c}{Input parameters of \texttt{synchrotron}}\\\hline\hline
  Parameter                         &Unit  &Description\\\hline
  \texttt{qir}                      &--    &The value of the FIR/radio correlation coefficient\\
  \texttt{alpha}                    &--    &The slope of the power-law synchrotron emission\\\hline
 \end{tabular}
 \caption{Input parameters of the \texttt{synchrotron} module.\label{tab:input_synchrotron}}
\end{table*}

\begin{table*}[!htbp]
 \centering
 \begin{tabular}{lll}
   \multicolumn{3}{c}{Output parameters of \texttt{synchrotron}}\\\hline\hline
  Parameter                         &Unit  &Description\\\hline
  \texttt{radio.qir}                &--    &The value of the FIR/radio correlation coefficient\\
  \texttt{radio.alpha}              &--    &The slope of the power-law synchrotron emission\\\hline
 \end{tabular}
 \caption{Output parameters of the \texttt{synchrotron} module.\label{tab:output_synchrotron}}
\end{table*}

\FloatBarrier
\subsection{Active nucleus}
\subsubsection{\texttt{fritz2006}}

\begin{table*}[!htbp]
 \centering
 \begin{tabular}{lll}
   \multicolumn{3}{c}{Input parameters of \texttt{fritz2006}}\\\hline\hline
  Parameter                         &Unit  &Description\\\hline
  \texttt{r\_ratio}                 &--    &Ratio of the maximum to minimum radii of the dust torus\\
  \texttt{tau}                      &--    &Optical depth at 9.7 microns\\
  \texttt{beta}                     &--    &Beta\\
  \texttt{gamma}                    &--    &Gamma\\
  \texttt{opening\_angle}           &degree&Full opening angle of the dust torus \citep[Fig. 1 of][]{fritz2006a}\\
  \texttt{psy}                      &degree&Angle between equatorial axis and line of sight\\
  \texttt{fracAGN}                   &--    &AGN fraction\\\hline
 \end{tabular}
 \caption{Input parameters of the \texttt{fritz2006} module.\label{tab:input_fritz2006}}
\end{table*}

\begin{table*}[!htbp]
 \centering
 \begin{tabular}{lll}
   \multicolumn{3}{c}{Output parameters of \texttt{fritz2006}}\\\hline\hline
  Parameter                         &Unit  &Description\\\hline
  \texttt{agn.r\_ratio}             &--    &Ratio of the maximum to minimum radii of the dust torus\\
  \texttt{agn.tau}                  &--    &Optical depth at 9.7 microns\\
  \texttt{agn.beta}                 &--    &Beta\\
  \texttt{agn.gamma}                &--    &Gamma\\
  \texttt{agn.opening\_angle}       &degree&Full opening angle of the dust torus \citep[Fig. 1 of][]{fritz2006a}\\
  \texttt{agn.psy}                  &degree&Angle between equatorial axis and line of sight\\
  \texttt{agn.fracAGN}               &--    &AGN fraction\\\hline
 \end{tabular}
 \caption{Output parameters of the \texttt{fritz2006} module.\label{tab:output_fritz2006}}
\end{table*}

\FloatBarrier
\subsection{Physical property measurement}
\subsubsection{\texttt{restframe\_parameters}}

\begin{table*}[!htbp]
 \centering
 \begin{tabular}{lll}
   \multicolumn{3}{c}{Input parameters of \texttt{restframe\_parameters}}\\\hline\hline
  Parameter                         &Unit  &Description\\\hline
  \texttt{beta\_calz94}             &--    &UV slope measured in the same way as in \cite{calzetti1994a}\\
  \texttt{D4000}                    &--    &D4000 break using the \cite{balogh1999a} definition\\
  \texttt{IRX}                      &--    &IRX computed from the GALEX FUV filter and the dust luminosity\\
  \texttt{EW\_lines}                &--    &Central wavelength of the emission lines for which to compute the equivalent width\\
  \texttt{luminosity\_filters}      &--    &Filters for which the rest-frame luminosity will be computed\\
  \texttt{colours\_filters}         &--    &Rest-frame colours to be computed\\\hline
 \end{tabular}
 \caption{Input parameters of the \texttt{restframe\_parameters} module.\label{tab:input_restframe_parameters}}
\end{table*}

\begin{table*}[!htbp]
 \centering
 \begin{tabular}{lll}
   \multicolumn{3}{c}{Output parameters of \texttt{restframe\_parameters}}\\\hline\hline
  Parameter                             &Unit  &Description\\\hline
  \texttt{param.beta\_calz94}                  &--        &UV slope measured in the same way as in \cite{calzetti1994a}\\
  \texttt{param.D4000}                         &--        &D4000 break using the \cite{balogh1999a} definition\\
  \texttt{param.IRX}                           &--        &IRX computed from the GALEX FUV filter and the dust luminosity\\
  \texttt{param.EW([line])}                    &nm        &Central wavelength of the emission lines for which to compute the\\&& equivalent width\\
  \texttt{param.restframe\_Lnu([filter]})      &W~m$^{-2}$&Filters for which the rest-frame luminosity will be computed\\
  \texttt{param.restframe\_[filter1]-[filter2]}&AB mag    &Rest-frame colours to be computed\\\hline
 \end{tabular}
 \caption{Output parameters of the \texttt{restframe\_parameters} module.\label{tab:output_restframe_parameters}}
\end{table*}

\FloatBarrier
\subsection{Intergalactic medium}
\subsubsection{\texttt{redshifting}}

\begin{table*}[!htbp]
 \centering
 \begin{tabular}{lll}
   \multicolumn{3}{c}{Input parameters of \texttt{redshifting}}\\\hline\hline
  Parameter                         &Unit  &Description\\\hline
  \texttt{redshift}                 &--    &Redshift to apply to the galaxy\\\hline
 \end{tabular}
 \caption{Input parameters of the \texttt{redshifting} module.\label{tab:input_redshifting}}
\end{table*}

\begin{table*}[!htbp]
 \centering
 \begin{tabular}{lll}
   \multicolumn{3}{c}{Input parameters of \texttt{redshifting}}\\\hline\hline
  Parameter                         &Unit  &Description\\\hline
  \texttt{universe.redshift}                 &--    &Redshift to apply to the galaxy\\
  \texttt{universe.luminosity\_distance}     &m     &Luminosity distance at redshift\\
  \texttt{universe.age}                      &Myr   &Age of the universe at redshift\\\hline
 \end{tabular}
 \caption{Output parameters of the \texttt{redshifting} module.\label{tab:output_redshifting}}
\end{table*}

\FloatBarrier
\section{Usage of \texttt{CIGALE} executables}

\begin{table*}[!htbp]
 \centering
 \begin{tabular}{lll}
  \hline\hline
  Executable              &Command                                           &Description\\\hline
  \texttt{pcigale}        &\texttt{init}                                     &Creates a skeleton configuration file\\
                          &\texttt{genconf}                                  &Fills the configuration file according to the modules indicated\\
                          &\texttt{check}                                    &Performs basic checks and computes the number of models\\
                          &\texttt{run}                                      &Runs \texttt{CIGALE}\\\hline
  \texttt{pcigale-plots}  &\texttt{sed [-{}-nologo -{}-type mJy, lum]}       &Plots the SED in units of mJy or W\\
                          &\texttt{pdf}                                      &Plots the PDF of the analysed physical properties\\
                          &\texttt{chi2}                                     &Plots the $\chi^2$ of the analysed physical properties\\
                          &\texttt{mock}                                     &Plots the comparison between mock and true physical properties\\\hline
  \texttt{pcigale-filters}&\texttt{list}                                     &Lists all the filters in the database\\
                          &\texttt{add file1, [file2, \ldots]}               &Adds one or more filters in the database\\
                          &\texttt{del filtername, [filtername2, \ldots]}    &Deletes one or more filters from the database\\
                          &\texttt{plot [filtername1, filtername2, \ldots]} &Plots one or more filters\\\hline
   \end{tabular}
 \caption{Usage of the \texttt{CIGALE} executable 1. to generate and fit models to observations (\texttt{CIGALE}), 2. plot the SED, the $\chi^2$, and the PDF (\texttt{pcigale-plots}), and 3. manage the filters in the database and plot them (\texttt{pcigale-filters}).\label{tab:scripts}}
\end{table*}

\FloatBarrier
\section{Example of a \texttt{pcigale.ini} configuration file\label{sec:pcigale.ini}}
\lstdefinelanguage{Ini}
{
    basicstyle=\ttfamily\small,
    columns=fullflexible,
    morecomment=[s][\color{Orchid}\bfseries]{[}{]},
    morecomment=[l]{\#},
    morecomment=[l]{;},
    commentstyle=\color{gray}\ttfamily,
    morekeywords={},
    otherkeywords={=,:},
    keywordstyle={\color{green}\bfseries}
}
\begin{lstlisting}[language=Ini]
# File containing the input data. The columns are 'id' (name of the
# object), 'redshift' (if 0 the distance is assumed to be 10 pc),
# 'distance' (Mpc, optional, if present it will be used in lieu of the
# distance computed from the redshift), the filter names for the fluxes,
# and the filter names with the '_err' suffix for the uncertainties. The
# fluxes and the uncertainties must be in mJy for broadband data and in
# W/m2 for emission lines. This file is optional to generate the
# configuration file, in particular for the savefluxes module.
data_file = brown14-with-dist.fits

# Optional file containing the list of physical parameters. Each column
# must be in the form module_name.parameter_name, with each line being a
# different model. The columns must be in the order the modules will be
# called. The redshift column must be the last one. Finally, if this
# parameter is not empty, cigale will not interpret the configuration
# parameters given in pcigale.ini. They will be given only for
# information. Note that this module should only be used in conjonction
# with the savefluxes module. Using it with the pdf_analysis module will
# yield incorrect results.
parameters_file = 

# Avaiable modules to compute the models. The order must be kept.
# SFH:
# * sfh2exp (double exponential)
# * sfhdelayed (delayed SFH with optional exponential burst)
# * sfhdelayedbq (delayed SFH with optional constant burst/quench)
# * sfhfromfile (arbitrary SFH read from an input file)
# * sfhperiodic (periodic SFH, exponential, rectangle or delayed)
# SSP:
# * bc03 (Bruzual and Charlot 2003)
# * m2005 (Maraston 2005)
# Nebular emission:
# * nebular (continuum and line nebular emission)
# Dust attenuation:
# * dustatt_modified_CF00 (modified Charlot & Fall 2000 attenuation law)
# * dustatt_modified_starburst (modified starburst attenuaton law)
# Dust emission:
# * casey2012 (Casey 2012 dust emission models)
# * dale2014 (Dale et al. 2014 dust emission templates)
# * dl2007 (Draine & Li 2007 dust emission models)
# * dl2014 (Draine et al. 2014 update of the previous models)
# * themis (Themis dust emission models from Jones et al. 2017)
# AGN:
# * fritz2006 (AGN models from Fritz et al. 2006)
# Radio:
# * radio (synchrotron emission)
# Restframe parameters:
# * restframe_parameters (UV slope, IRX-beta, D4000, EW, etc.)
# Redshift+IGM:
# * redshifting (mandatory, also includes the IGM from Meiksin 2006)
sed_modules = sfhdelayed, bc03, nebular, dustatt_modified_starburst, dale2014,
              restframe_parameters, redshifting

# Method used for statistical analysis. Available methods: pdf_analysis,
# savefluxes.
analysis_method = pdf_analysis

# Number of CPU cores available. This computer has 96 cores.
cores = 48

# Bands to consider. To consider uncertainties too, the name of the band
# must be indicated with the _err suffix. For instance: FUV, FUV_err.
bands = galex.FUV, galex.FUV_err, galex.NUV, galex.NUV_err, sdss.up,
         sdss.up_err, sdss.gp, sdss.gp_err, sdss.rp, sdss.rp_err, sdss.ip,
         sdss.ip_err, sdss.zp, sdss.zp_err, 2mass.J, 2mass.J_err, 2mass.H,
         2mass.H_err, 2mass.Ks, 2mass.Ks_err, WISE1, WISE1_err,
         spitzer.irac.ch1, spitzer.irac.ch1_err, spitzer.irac.ch2,
         spitzer.irac.ch2_err, WISE2, WISE2_err, spitzer.irac.ch3,
         spitzer.irac.ch3_err, spitzer.irac.ch4, spitzer.irac.ch4_err, WISE3,
         WISE3_err, WISE4, WISE4_err, spitzer.mips.24, spitzer.mips.24_err

# Properties to be considered. All properties are to be given in the
# rest frame rather than the observed frame. This is the case for
# instance the equivalent widths and for luminosity densities.
properties = 


# Configuration of the SED creation modules.
[sed_modules_params]
  
  [[sfhdelayed]]
    # e-folding time of the main stellar population model in Myr.
    tau_main = 1, 500, 1000, 2000, 3000, 4000, 5000, 6000, 7000, 8000
    # Age of the main stellar population in the galaxy in Myr. The precision
    # is 1 Myr.
    age_main = 13000
    # e-folding time of the late starburst population model in Myr.
    tau_burst = 5, 10, 25, 50, 100
    # Age of the late burst in Myr. The precision is 1 Myr.
    age_burst =  5, 10, 25, 50, 100, 200, 350, 500, 750, 1000
    # Mass fraction of the late burst population.
    f_burst = 0, 0.0001, 0.0005, 0.001, 0.005, 0.01, 0.05, 0.1, 0.25
    # Value of SFR at t = 0 in M_sun/yr.
    sfr_A = 1.0
    # Normalise the SFH to produce one solar mass.
    normalise = True
  
  [[bc03]]
    # Initial mass function: 0 (Salpeter) or 1 (Chabrier).
    imf = 1
    # Metalicity. Possible values are: 0.0001, 0.0004, 0.004, 0.008, 0.02,
    # 0.05.
    metallicity = 0.02
    # Age [Myr] of the separation between the young and the old star
    # populations. The default value in 10^7 years (10 Myr). Set to 0 not to
    # differentiate ages (only an old population).
    separation_age = 10
  
  [[nebular]]
    # Ionisation parameter
    logU = -3.0
    # Fraction of Lyman continuum photons escaping the galaxy
    f_esc = 0.0
    # Fraction of Lyman continuum photons absorbed by dust
    f_dust = 0.0
    # Line width in km/s
    lines_width = 300.0
    # Include nebular emission.
    emission = True
  
  [[dustatt_modified_starburst]]
    # E(B-V)l, the colour excess of the nebular lines light for both the
    # young and old population.
    E_BV_lines = 0.005, 0.01, 0.025, 0.05, 0.075, 0.10, 0.15, 0.20, 0.25, 0.30,
                   0.35, 0.40, 0.45, 0.50, 0.55, 0.60, 0.65, 0.70 
    # Reduction factor to apply on E_BV_lines to compute E(B-V)s the stellar
    # continuum attenuation. Both young and old population are attenuated
    # with E(B-V)s.
    E_BV_factor = 0.25, .5, .75
    # Central wavelength of the UV bump in nm.
    uv_bump_wavelength = 217.5
    # Width (FWHM) of the UV bump in nm.
    uv_bump_width = 35.0
    # Amplitude of the UV bump. For the Milky Way: 3.
    uv_bump_amplitude = 0., 1., 2., 3.
    # Slope delta of the power law modifying the attenuation curve.
    powerlaw_slope = -.5, -.4, -.3, -.2, -.1, 0.
    # Extinction law to use for attenuating the emissio  n lines flux.
    # Possible values are: 1, 2, 3. 1: MW, 2: LMC, 3: SMC. MW is modelled
    # using CCM89, SMC and LMC using Pei92.
    Ext_law_emission_lines = 1
    # Ratio of total to selective extinction, A_V / E(B-V), for the
    # extinction curve applied to emission lines.Standard value is 3.1 for
    # MW using CCM89, but can be changed.For SMC and LMC using Pei92 the
    # value is automatically set to 2.93 and 3.16 respectively, no matter
    # the value you write.
    Rv = 3.1
    # Filters for which the attenuation will be computed and added to the
    # SED information dictionary. You can give several filter names
    # separated by a & (don't use commas).
    filters = V_B90 & FUV
  
  [[dale2014]]
    # AGN fraction. It is not recommended to combine this AGN emission with
    # the of Fritz et al. (2006) models.
    fracAGN = 0.0
    # Alpha slope. Possible values are: 0.0625, 0.1250, 0.1875, 0.2500,
    # 0.3125, 0.3750, 0.4375, 0.5000, 0.5625, 0.6250, 0.6875, 0.7500,
    # 0.8125, 0.8750, 0.9375, 1.0000, 1.0625, 1.1250, 1.1875, 1.2500,
    # 1.3125, 1.3750, 1.4375, 1.5000, 1.5625, 1.6250, 1.6875, 1.7500,
    # 1.8125, 1.8750, 1.9375, 2.0000, 2.0625, 2.1250, 2.1875, 2.2500,
    # 2.3125, 2.3750, 2.4375, 2.5000, 2.5625, 2.6250, 2.6875, 2.7500,
    # 2.8125, 2.8750, 2.9375, 3.0000, 3.0625, 3.1250, 3.1875, 3.2500,
    # 3.3125, 3.3750, 3.4375, 3.5000, 3.5625, 3.6250, 3.6875, 3.7500,
    # 3.8125, 3.8750, 3.9375, 4.0000
    alpha = 1.0, 1.5, 2.0, 2.5, 3.0, 3.5, 4.0
  
  [[restframe_parameters]]
    # UV slope measured in the same way as in Calzetti et al. (1994).
    beta_calz94 = True
    # D4000 break using the Balogh et al. (1999) definition.
    D4000 = False
    # IRX computed from the GALEX FUV filter and the dust luminosity.
    IRX = True
    # Central wavelength of the emission lines for which to compute the
    # equivalent width. The half-bandwidth must be indicated after the '/'
    # sign. For instance 656.3/1.0 means oth the nebular line and the
    # continuum are integrated over 655.3-657.3 nm.
    EW_lines = 
    # Filters for which the rest-frame luminosity will be computed. You can
    # give several filter names separated by a & (don't use commas).
    luminosity_filters = 
    # Rest-frame colours to be computed. You can give several colours
    # separated by a & (don't use commas).
    colours_filters = 
  
  [[redshifting]]
    # Redshift of the objects. Leave empty to use the redshifts from the input
    # file.
    redshift = 0.


# Configuration of the statistical analysis method.
[analysis_params]
  # List of the physical properties to estimate. Leave empty to analyse
  # all the physical properties (not recommended when there are many
  # models).
  variables = stellar.m_star, dust.luminosity, attenuation.FUV, sfh.sfr100Myrs,
                param.IRX, param.beta_calz94
  # If true, save the best SED for each observation to a file.
  save_best_sed = True
  # If true, for each observation and each analysed property, save the raw
  # chi2. It occupies ~15 MB/million models/variable.
  save_chi2 = False
  # If true, for each object check whether upper limits are present and
  # analyse them.
  lim_flag = False
  # If true, for each object we create a mock object and analyse them.
  mock_flag = True
  # When redshifts are not given explicitly in the redshifting module,
  # number of decimals to round the observed redshifts to compute the grid
  # of models. To disable rounding give a negative value. Do not round if
  # you use narrow-band filters.
  redshift_decimals = 2
  # Number of blocks to compute the models and analyse the observations.
  # If there is enough memory, we strongly recommend this to be set to 1.
  blocks = 1
\end{lstlisting}
\clearpage
\twocolumn

\end{document}